\documentclass[11pt,a4paper]{article}
\pdfoutput=1
\usepackage{amssymb,amsmath,amsfonts, mathtools, mathrsfs}
\usepackage[utf8]{inputenc} % per usare le lettere accentate come à,è etc...
\usepackage[dvipsnames]{xcolor}

\newif\ifnatbibsort\natbibsorttrue

\relax

\ifnatbibsort\RequirePackage[numbers,sort&compress]{natbib}\else\RequirePackage[numbers,compress]{natbib}\fi
\RequirePackage[colorlinks=true
,urlcolor=blue
,anchorcolor=blue
,citecolor=blue
,filecolor=blue
,linkcolor=blue
,menucolor=blue
,pagecolor=blue
,linktocpage=true
,pdfproducer=medialab
,pdfa=true
]{hyperref}

\usepackage{graphicx}
\usepackage{caption}
\usepackage{tensor}
\usepackage{subfigure}
\usepackage{enumerate}
\usepackage{dsfont}
\usepackage{verbatim}
\usepackage{amsfonts,euscript,amssymb,stmaryrd,braket}
\usepackage{tikz}
\usetikzlibrary{arrows,decorations.markings,patterns}
\usepackage{slashed}

\def\clock{{\count0=\time
		\divide\count0 60
		\ifnum\count0<10 0\fi\the\count0
		\multiply\count0 -60 \advance\count0 \time
		:\ifnum\count0<10 0\fi \the\count0
}}
\newcommand{\timestamp}{{\small\vbox{\hbox{\tt\jobname.tex}
			\hbox{\the\day/\the\month/\the\year, \clock}}}}

\newcommand{\bea}{\begin{eqnarray}}
\newcommand{\eea}{\end{eqnarray}}

\newcommand{\be}{\begin{equation}}
\newcommand{\ee}{\end{equation}}

\makeatletter
\let\old@startsection=\@startsection
\let\oldl@section=\l@section
\renewcommand{\@startsection}[6]{\old@startsection{#1}{#2}{#3}{#4}{#5}{#6\mathversion{bold}}}
\renewcommand{\l@section}[2]{\oldl@section{\mathversion{bold}#1}{#2}}
\makeatother

\numberwithin{equation}{section}
\usepackage{color}

\def \RR {{\mathbb R}}

\def\ri {{\rm i}}
\def\rd {{\rm d}}
\def\e {{\rm e}}
\def\prt {{\partial}}

\begin{document}
	\renewcommand{\thefootnote}{\arabic{footnote}}

	\overfullrule=0pt
	\parskip=2pt
	\parindent=12pt
	\headheight=0in \headsep=0in \topmargin=0in \oddsidemargin=0in

	\vspace{ -3cm} \thispagestyle{empty} \vspace{-1cm}
	\begin{flushright} 
		\footnotesize
		\textcolor{red}{\phantom{print-report}}
	\end{flushright}

\begin{center}
	\vspace{.0cm}

	{\Large\bf \mathversion{bold}
	Modular Hamiltonians for the massless Dirac field
	}
	\\
	\vspace{.25cm}
	\noindent
	{\Large\bf \mathversion{bold}
	in the presence of a boundary}

	\vspace{0.8cm} {
		Mihail Mintchev$^{\,a,}$\footnote[1]{mintchev@df.unipi.it}
		and Erik Tonni$^{\,b,}$\footnote[2]{erik.tonni@sissa.it}
	}
	\vskip  0.7cm
	
	\small
	{\em
		$^{a}\,$Dipartimento di Fisica, Universit\'a di Pisa and INFN Sezione di Pisa, \\
		largo Bruno Pontecorvo 3, 56127 Pisa, Italy
		\vskip 0.05cm
		$^{b}\,$SISSA and INFN Sezione di Trieste, via Bonomea 265, 34136, Trieste, Italy 
	}
	\normalsize

\end{center}

\vspace{0.3cm}
\begin{abstract} 

We study the modular Hamiltonians of an interval for the massless Dirac fermion on the half-line.
The most general boundary conditions ensuring the global energy conservation lead to consider
two phases, where either the vector or the axial symmetry is preserved. 
In these two phases we derive the corresponding modular Hamiltonian in explicit form.
Its density involves a bi-local term localised in two points of the interval, one conjugate to the other. 
The associated modular flows 
are also established. Depending on the phase,
they mix fields with different chirality or charge that follow different modular trajectories. Accordingly, 
the modular flow preserves either the vector or the axial symmetry.
We compute the two-point correlation functions along the modular flow and show that they satisfy the Kubo-Martin-Schwinger 
condition in both phases. 
The entanglement entropies are also derived. 
\end{abstract}

\vspace{1cm}

\begin{center}
{\it Dedicated to the loving memory of Ios\'e Scalfi}
\end{center}

\newpage

%%%%%%%%%%%%%%%%%%%%%%%%%%%%%%%%%%%%%
\tableofcontents
%%%%%%%%%%%%%%%%%%%%%%%%%%%%%%%%%%%%%

\newpage
%%%%%%%%%%%%%%%%%%%%%%%%%%%%%%%%%%%%%%%%%%%
\section{Introduction}
\label{sec_intro}

Entanglement quantifiers are crucial quantities to explore
quantum field theories, quantum gravity models, condensed matter systems and quantum information theory. 

The geometric entanglement between complementary spatial regions is studied 
by considering a quantum system in a state described by its density matrix $\rho$
and a spatial bipartition $A\cup B$ of the entire space.
Assuming that the Hilbert space can be factorised as $\mathcal{H} = \mathcal{H}_A \otimes \mathcal{H}_B$,
the reduced density matrix $\rho_A \equiv \textrm{Tr}_{{}_{\mathcal{H}_B}} \rho\equiv \textrm{Tr}_{\!{}_{B}} \rho$ 
of the region $A$ is obtained by tracing out the degrees of freedom corresponding to the complementary region. 
The reduced density matrix $\rho_A$ is both hermitean and positive semidefinite, hence
it can be written as $\rho_A \propto e^{-K_A}$, 
where the proportionality constant guarantees the normalisation condition $\textrm{Tr}_{\!{}_A} \rho_A =1$.
The hermitean operator $K_A$ is the modular Hamiltonian (also known as entanglement Hamiltonian) of the region $A$
\cite{Haag-book}.
The spectrum of the modular Hamiltonian provides interesting quantities like e.g. the entanglement entropy. 
%%%

The modular Hamiltonian $K_A$ allows to introduce a family of unitary operators 
$U(\tau)=e^{-\textrm{i} \tau K_A}$, parameterised by  $\tau\in \RR$,
that generates a flow $\mathcal{O}(\tau) \equiv U(\tau) \,\mathcal{O}\,U(-\tau)$ for any operator $\mathcal{O}$ localised in $A$,
which is called modular flow of the operator $\mathcal{O}$ \cite{takesaki-book}.
The modular flow of $\mathcal{O}$ satisfies
$ \textrm{Tr}_{\!{}_A} \!\big(\rho_A \,\mathcal{O}\big) 
=   \textrm{Tr}_{\!{}_A}\!\big(\rho_A  \,\mathcal{O}(\tau)\big)$.
%%%
The modular flow provides the intrinsic internal dynamics 
induced by the reduced density matrix of a subsystem.

The modular Hamiltonian is known analytically in terms of the fundamental fields of the model in few cases.
A seminal result due to Bisognano and Wichmann \cite{Bisognano:1975ih,Bisognano:1976za} states
that the modular Hamiltonian of half-space $x>0$
for a Lorentz invariant quantum field theory in its vacuum is given by the 
boost generator in the $x$-direction.
Other analytic expressions of modular Hamiltonians have been found for conformal field theories
by combining this crucial result with the conformal symmetry.
For a conformal field theory in generic spacetime dimension and in its ground state, 
this analysis provides the modular Hamiltonian of a spherical region \cite{Hislop:1981uh,Brunetti:1992zf,Casini:2011kv}.
In $1+1$ dimensions, where the conformal symmetry has infinitely many generators, 
the Bisognano-Wichmann modular Hamiltonian and the conformal transformations allow to find
the modular Hamiltonians in some other cases of physical interest \cite{Wong:2013gua, Cardy:2016fqc},
including time-dependent scenarios  \cite{Cardy:2016fqc}
and spatially inhomogeneous systems \cite{Tonni:2017jom}.
All these modular Hamiltonians are local because the corresponding densities are local operators.

Very few analytic expressions of modular Hamiltonians are available in the literature
that cannot be found through the result of Bisognano and Wichmann and the conformal symmetry,
even in conformal field theories. 
An important example is the modular Hamiltonian of the 
$1+1$ dimensional massless Dirac fermion in the ground state
for the union of disjoint intervals on the infinite line. 
By employing the modular Hamiltonian on the lattice obtained by Peschel \cite{Peschel:2003rdm},
this modular Hamiltonian has been written by Casini and Huerta \cite{Casini:2009vk},
who found also the corresponding modular flow for the Dirac field. 
This modular Hamiltonian is very interesting because,
beside a local term, it contains also a bi-local operator 
that induces a mixing along the modular flow  between 
the field evaluated in two different points. 
The correlators of the Dirac field along the modular flow generated 
by this modular Hamiltonian have been obtained in \cite{Longo:2009mn}, 
verifying that they satisfy the Kubo-Martin-Schwinger (KMS) condition \cite{Haag-book}.
The validity of this condition guarantees the uniqueness of the modular flow \cite{takesaki-book}.
Further analyses of this modular Hamiltonian and of its modular flow are reported in 
\cite{Rehren:2012wa,  Hollands:2019hje}.
%%%

Physical boundaries heavily influence the entanglement of complementary spatial regions.
For instance, in two-dimensional boundary conformal field theories \cite{Cardy:1984bb, Cardy:1986gw, Cardy:1989ir}, 
the Affleck-Ludwig boundary entropy \cite{Affleck:1991tk} occurs in the entanglement entropy \cite{Calabrese:2004eu}.
The role of a physical boundary has been studied also through algebraic quantum field theory methods 
\cite{Longo:2004fu, Longo:2010we, Witten:2018zxz}.

A class of modular Hamiltonians in two-dimensional conformal field theories can be  found through 
boundary conformal field theory techniques \cite{Cardy:2016fqc} 
and this approach allows to explain various features of some entanglement spectra 
\cite{Lauchli:2013jga, Cardy:2016fqc, Tonni:2017jom, Alba:2017bgn,DiGiulio:2019lpb,Surace:2019mft,Roy:2020frd}.
The modular Hamiltonians of more complicated configurations can be found 
by employing  techniques based on the specific model.
The typical example is the massless Dirac fermion,
where the modular Hamiltonians of an arbitrary number of disjoint intervals on the infinite line
when the entire system is in its ground state \cite{Casini:2009vk} 
and of an interval on the circle when the system is in a thermal state \cite{Blanco:2019xwi, Fries:2019ozf, Hollands:2019hje}
have been found. Other interesting related studies are \cite{Klich:2015ina, Klich:2017qmt, Arias:2018tmw}.

In this manuscript we consider the massless Dirac fermion in its ground state on the half-line $x>0$.
Imposing boundary conditions at $x=0$ that guarantee the energy conservation,
two inequivalent models (phases) are allowed. 
These two phases are characterised by different conservation laws.
In particular, either the charge or the helicity is preserved but not both of them.
Instead, for the massless Dirac field on the line both these symmetries are preserved. 
Furthermore, on the half-line the two components of the massless Dirac field are coupled through the boundary condition. 
%%%

In this model, we study the modular Hamiltonians $K_A$ of an interval $A$ on the half-line.
Analytic expressions for $K_A$ in terms of the components of the Dirac field are obtained in both phases.
Beside the expected local term, also a bi-local term occurs 
which involves fields evaluated in two conjugate points within the interval.
This bi-local term breaks either the vector or the axial symmetry. 
This is the main difference with respect to the bi-local term in the modular Hamiltonian
of two disjoint equal intervals on the line for the massless Dirac fermion in the ground state found in \cite{Casini:2009vk},
which preserves both these symmetries. 

The modular flows of the Dirac field generated by these modular Hamiltonians are also obtained. 
Depending on the phase, the modular flow mixes fields with either different chirality or different charge 
evaluated in conjugate points.
The characteristic symmetry of each phase is preserved along the corresponding modular flow. 
We find the two-point correlators of the fields along the modular flow that satisfy the KMS condition.

The outline of the manuscript is as follows.
In Sec.\;\ref{sec_dirac_fermion} we discuss the massless Dirac fermion on the half-line
and its correlation functions.
In Sec.\;\ref{sec_eh} we derive the modular Hamiltonians $K_A$ of an interval in the two phases.
The corresponding entanglement entropies are computed in Sec\;\ref{sec-ee}.
The modular flows of the components of the Dirac field and  their correlators 
are discussed  in Sec.\;\ref{sec_mod_flow} and in Sec.\;\ref{sec_correlators} respectively.
Some interesting limiting regimes are considered in Sec.\;\ref{sec_limiting-regimes}.
In Sec\;\ref{sec-mod-ev-part1} we study $K_A$ and its flow in the spacetime. 
The results are summarized in Sec.\;\ref{sec_conclusions}.
In the Appendices\;\ref{app_2intervals}, \ref{app-shift-operator} and \ref{app-bilocal-eh} 
we provide some technical details, the derivations of some expressions reported in the main text and further analyses,
including some results  for the massless Dirac fermion in spacetimes that are invariant under spatial translations.

%\newpage
%%%%%%%%%%%%%%%%%%%%%%%%%%%%%%%%%%%%%%%%%%%

\section{Dirac fermions on the half-line}
\label{sec_dirac_fermion}

In this section we summarise the main properties of massless Dirac fermions on the half-line 
$\RR_+ \equiv [0,\infty)$, fixing also the notation adopted throughout the manuscript.

\subsection{Bulk dynamics}

The massless Dirac field $\psi (t,x)$ is the following doublet made by the two complex fields
\be
\label{psi-doublet}
\psi (t,x)=
\bigg(
\begin{array}{c}\psi_1(t,x) \\ \psi_2(t,x) \\ \end{array}
\bigg)\,.
\ee
In the bulk of the half-line, it satisfies the scale invariant equation of motion
\begin{equation}
\label{dirac-eom}
(\gamma_t \partial_t - \gamma_x \partial_x)\psi (t,x) = 0 
\;\;\qquad\;\;  x>0
\end{equation} 
where 
\be
\gamma_t = 
\bigg(\begin{array}{cc}
0\; & 1 \\ 1 \; &  0
\end{array}  \bigg)
\;\;\qquad\;\;
\gamma_x 
= 
\bigg(\begin{array}{cc}
0 \; & -1\\ 1 \; &  0
\end{array}  \bigg)\,.
\end{equation}  

Denoting the Hermitean conjugation through the asterisk,
the energy-momentum tensor for the Dirac field (\ref{psi-doublet}) can be written in terms of 
the following components\footnote{The remaining components can be obtained from the symmetry and tracelessness conditions.}
\begin{eqnarray} 
\label{endens}
T_{tt}(t,x) &=& \frac{\ri}{2} \left [(\prt_x \psi^*) \,\gamma_t \gamma_x \,\psi -
\psi^* \,\gamma_t \gamma_x \,(\prt_x \psi) \right ]\! (t,x) 
\\
\rule{0pt}{.7cm}
T_{xt}(t,x) &=&  \frac{\ri}{2} \left [(\prt_t \psi^*) \, \gamma_t \gamma_x \, \psi -
\psi^*\, \gamma_t\gamma_x \,(\prt_t \psi) 
\right ] \! (t,x) \,.
\label{encurr}
\end{eqnarray} 
The equation of motion (\ref{dirac-eom}) implies the local energy conservation of the energy-momentum tensor
\begin{equation} 
\prt_t T_{tt}(t,x) - \prt_x T_{xt}(t,x) = 0
\;\;\qquad\;\;  x>0\,.
\label{encons}
\end{equation}

%%%
The bulk dynamics is invariant both under the vector phase transformation
\begin{equation}
\psi_r (t,x) 
\; \longmapsto \;
\e^{\ri \, \theta_{\textrm{\tiny v}}} \,\psi_r(t,x) 
\;\;\qquad\;\; 
\theta_{\textrm{\tiny v}} \in [0,2\pi)
\label{v1}
\end{equation}
and under the axial phase transformation
\begin{equation}
\psi_r (t,x) 
\; \longmapsto \;
\e^{\ri (-1)^r\, \theta_{\textrm{\tiny a}}}\, \psi_r(t,x)
\;\;\qquad\;\; 
\theta_{\textrm{\tiny a}} \in [0,2\pi)\,.
\label{a1}
\end{equation}

Denoting by $j_t(t,x)$ and $j_x(t,x)$ the components of the current corresponding to the vector $U(1)$ symmetry
and by  $k_t(t,x)$ and $k_x(t,x)$ the components of the current corresponding to the axial $U(1)$ symmetry, we have that
\begin{equation} 
j_t(t,x) = k_x(t,x) = [\psi^* \psi ](t,x) 
\;\;\qquad\;\; 
j_x(t,x) = k_t(t,x) = [\psi^* \gamma_t\gamma_x \psi] (t,x) \,.
\label{e2}
\end{equation}
These currents are locally conserved when the equation of motion (\ref{dirac-eom}) holds, namely
\begin{equation}
\prt_t j_t(t,x) -\prt_x j_x(t,x) = 0
\;\;\qquad\;\;
\prt_t k_t(t,x) -\prt_x k_x(t,x) = 0
\;\;\qquad\;\;
 x>0
\label{currcons}
\end{equation} 
and they describe the electric and helical transport in the system.

\subsection{Boundary conditions}

In order to fully determine the dynamics on the half-line $\RR_+$, the boundary condition at $x=0$ must be imposed.
This choice deeply influences the symmetry content of the model. 

We adopt the most general boundary conditions that ensure global energy conservation.
This means that the vanishing of the energy flow through the boundary is imposed, namely \cite{Cardy:1984bb, Cardy:1986gw, Cardy:1989ir}
\begin{equation}
\label{eflow}
T_{xt}(t,0) = 0
\;\;\qquad\;\;  
t \in \RR \,.
\end{equation} 
This boundary condition can be satisfied in two ways: either
\be
\label{bc5}
\psi_1 (t, 0) = \e^{\ri \alpha_{\textrm{\tiny v}}}\, \psi_2 (t,0)
\;\;\qquad\;\;
\alpha_{\textrm{\tiny v}} \in [0,2\pi)
\;\;\qquad\;\;  
t \in \RR 
\ee
or
\be
\label{bc6}
\psi_1 (t, 0) = \e^{-\ri \alpha_{\textrm{\tiny a}}} \,\psi^*_2 (t,0) 
\;\;\qquad\;\;
\alpha_{\textrm{\tiny a}} \in [0,2\pi)
\;\;\qquad\;\;  
t \in \RR 
\ee
which provide a scale invariant coupling of fields with different chirality at $x=0$.

We remark that $\alpha_{\textrm{\tiny v}} $ and $\alpha_{\textrm{\tiny a}}$ parametrise  
all self-adjoint extensions \cite{ReedSimon-book} of the Hamiltonian 
\begin{equation} 
h =  \bigg(\begin{array}{cc}\ri \,\prt_x & 0\\ 0 & -\,\ri \,\prt_x    \\ \end{array} \bigg)
\;\;\qquad\;\;
x\in \RR_+
\label{bc2}
\end{equation}
which is obtained by rewriting the equation of motion (\ref{dirac-eom}) in the form 
\begin{equation}
\ri \, \prt_t \psi(t,x) = h\, \psi (t,x) \,.
\label{bc1}
\end{equation} 

The boundary conditions (\ref{bc5}) and (\ref{bc6}) lead to different conservation laws for the system.
In particular, the boundary condition (\ref{bc5}) preserves the vector symmetry and breaks the axial one,
while the opposite holds in the model where (\ref{bc6}) is imposed. 
Thus, on the half-line the basic physical requirement of energy conservation (i.e. the self-adjointness of the Hamiltonian) 
implies that either the vector symmetry or the axial symmetry is necessarily broken.
Instead, for the massless Dirac fermion on the infinite line both these symmetries are preserved. 
At quantum level this feature provides two different phases for the massless Dirac fermion on the half-line,
which are characterised by either the conservation of the charge or by the conservation of the helicity
(but not both of them) \cite{Liguori:1997vd}. 
Throughout this manuscript
we will refer to these two inequivalent models respectively as vector phase and axial phase.

%%%%%%%%%%%%%%%%%%%%%%%%%%%%%%%%%%
\subsection{Quantization}

The quantum fields $\psi_r(t,x)$ in (\ref{psi-doublet}) satisfy the equation of motion (\ref{dirac-eom}),
the following the equal-time anti-commutation relations 
\bea
\label{car1}
& &
\big[\, \psi_{r_1}(t,x_1)\, ,\, \psi^*_{r_2}(t,x_2,)\,\big]_+ 
= \,
\delta_{r_1r_2}\, \delta (x_1-x_2)
\\
\rule{0pt}{.45cm}
\label{car2}
& &
\big[\, \psi_{r_1}(t,x_1)\, ,\, \psi_{r_2}(t,x_2)\, \big]_+
= \,
\big[\, \psi^*_{r_1}(t,x_1)\, ,\, \psi^*_{r_2}(t,x_2)\, \big]_+ =\,0
\eea
and also a boundary condition that is either (\ref{bc5}) or (\ref{bc6}).
These fields can be described in terms of 
two mutually anticommuting algebras ${\cal A}_+$ and ${\cal B}_+$ generated respectively by 
\begin{equation} 
\label{car}
\big\{a(k)\,,\, a^\ast(k)  \big\}
\;\;\qquad\;\;
 \big\{b (k)\,,\, b^\ast(k)  \big\}
 \;\;\qquad\;\;
 k\geqslant 0
\end{equation} 
which satisfy the canonical anticommutation relations.

\subsubsection{Vector phase} 
\label{subsec-vec-phase}

In the vector phase, where the boundary condition (\ref{bc5}) is imposed. 
We denote the components of the Dirac field by
\be
\label{lambda-doublet}
\lambda(t,x)=
\bigg(
\begin{array}{c}\lambda_1(x+t) \\ \lambda_2(x-t) \\ \end{array} 
\bigg) 
\;\;\qquad\;\; x \in \RR_+  \;\quad\;  t\in \RR 
\ee
can which be written as \cite{Mintchev:2011mx}
\bea
\label{npsi1hl}
\lambda_1(x+t) 
&=&
\int_0^{\infty} \! 
\Big[\, a(k) \,\e^{-\ri k (x+t)} + \e^{\ri \alpha_{\textrm{\tiny v}}}\, b^\ast(k) \, \e^{\ri k (x+t)} \,\Big] \frac{\rd k}{2\pi} 
\\
\rule{0pt}{.7cm}
\label{npsi2hl}
\lambda_2(x-t) 
&=& 
\int_0^{\infty} \!
\Big[ \,\e^{-\ri \alpha_{\textrm{\tiny v}}}\, a(k) \,\e^{\ri k (x-t)} + b^\ast(k) \,\e^{-\ri k (x-t)} \, \Big] \, \frac{\rd k}{2\pi} \,.
\eea
Notice that each component $\lambda_r$ depends on the angle $\alpha_{\textrm{\tiny v}}$ and 
involves the same generators of ${\cal A}_+$ and ${\cal B}_+$, 
which is not the case of the Dirac fermion on the line\footnote{On the infinite line $\RR$,
two copies of ${\cal A}_+$ and ${\cal B}_+$ are needed to write the components of the Dirac field as
\bea
\label{psi1-line}
\psi_1(x+t) 
&=&
\int_{0}^{\infty} \! 
\Big[\, a_1(k) \,\e^{-\ri k (x+t)} + b_1^\ast(k) \, \e^{\ri k (x+t)} \,\Big] \frac{\rd k}{2\pi} 
\\
\rule{0pt}{.55cm}
\label{psi2-line}
\psi_2(x-t) 
&=& 
\int_{0}^{\infty} \! 
\Big[\, a_2(k) \,\e^{\ri k (x-t)} + b_2^\ast(k) \, \e^{-\ri k (x-t)} \,\Big] \frac{\rd k}{2\pi} \,.
\eea
}.

For the vacuum expectation values of the fields in (\ref{npsi1hl}) and (\ref{npsi2hl}) 
in the Fock representation of ${\cal A}_+$ and ${\cal B}_+$, one finds
\bea
\label{n11}
\langle \lambda_1(x_1+t_1)\,\lambda_1^*(x_2+t_2)\rangle 
&=&
C (t_{12}+ x_{12})
\\
\label{n22}
\langle \lambda_2(x_1-t_1)\,\lambda_2^*(x_2-t_2)\rangle &=&  C(t_{12}-x_{12}) 
\\
\label{n12}
\langle \lambda_1(x_1+t_1)\,\lambda_2^*(x_2-t_2)\rangle &=& \e^{\ri \alpha_{\textrm{\tiny v}}} \, C (t_{12}+\hat{x}_{12})
\\
\label{n21}
\langle \lambda_2(x_1-t_1)\,\lambda_1^*(x_2+t_2)\rangle &=& \e^{-\ri \alpha_{\textrm{\tiny v}}}  \,C(t_{12}- \hat{x}_{12})
\eea
where we have introduced 
\be
\label{not2}
t_{12} \equiv t_1-t_2
\;\;\qquad\;\;
x_{12} \equiv x_1-x_2
\;\;\qquad\;\;
\hat{x}_{12} \equiv x_1+x_2 
\ee
and 
\be
\label{not2}
C(\zeta) 
\,\equiv\, 
\frac{1}{2\pi \ri (\zeta - \ri \varepsilon)} 
\,=\, \frac{1}{2\pi \ri} \left[\,{\rm P.V.}\,\frac{1}{\zeta} + \ri \pi \, \delta(\zeta) \,\right]
\;\;\qquad\;\;
\varepsilon >0 \,.
\ee
In this phase the axial symmetry (\ref{a1}) is broken, hence the nontrivial correlation functions (\ref{n12}) and (\ref{n21}) 
that mix different helicities are allowed.

We find it convenient to collect the correlation functions
(\ref{n11}), (\ref{n22}), (\ref{n12}) and (\ref{n21}) at equal times $t_1=t_2\equiv t$ into the following matrix 
\bea
\label{vcm}
\boldsymbol{C}(x,y;\alpha_{\textrm{\tiny v}}) 
&\equiv&
\Bigg(
\begin{array}{cc} 
\langle \lambda_1 (x+t) \, \lambda_1^*(y+t)\rangle \;\; & \langle \lambda_1(x+t)\,\lambda_2^*(y-t)\rangle 
\\ 
\rule{0pt}{.5cm}
\langle \lambda_2(x-t)\, \lambda_1^*(y+t)\rangle \;\; & \langle \lambda_2(x-t)\, \lambda_2^*(y-t)\rangle  
\end{array} 
\Bigg) 
\nonumber \\
&=&
\Bigg(\,
\begin{array}{cc} 
C(x-y) \; & \e^{\ri \alpha } \,C(x+y)
\\ 
\rule{0pt}{.5cm}
\e^{-\ri \alpha }\, C(-x-y)  \; & C(-x+y)
\end{array} 
\Bigg) .
\eea

It is worth discussing the meaning of the phase factor $\e^{\ri \alpha_{\textrm{\tiny v}}}$ in (\ref{bc5}) and 
the possibility to absorb it through the field redefinition 
\be 
\lambda_1(x+t) \longmapsto \lambda_1(x+t) 
\;\;\qquad\;\;
\lambda_2(x-t) \longmapsto e^{-\ri \alpha_{\textrm{\tiny v}}} \lambda_2(x-t)
\label{bc10}
\ee
which  leaves invariant both the equations of motion (\ref{dirac-eom}) and the equal-time 
canonical relations (\ref{car1}) and (\ref{car2}). 
This leads us to describe the boundary condition at $x \to \infty$.

The boundary condition at $x \to \infty$ can be studied by introducing the following family of states 
\be 
\Phi (h) = 
\left( \,\int_0^\infty 
 h(p) \,a^*(p) \,\frac{\rd p}{2\pi}\, \right) \Omega
\;\;\qquad \;\;
h \in {\cal D}(\RR_+) 
\label{Phi}
\ee
where $\Omega$ is the Fock vacuum and ${\cal D}(\RR_+)$ is the space of $C^\infty$
functions with compact support in $\RR_+$. It is well known that 
the Fourier transform 
\be 
\hat h (y) 
= 
\int_{-\infty}^\infty   h(p) \, \e^{-ipy}\,\frac{\rd p}{2\pi}
= 
\int_0^\infty h(p)\, \e^{-ipy} \,\frac{\rd p}{2\pi} 
\label{ft}
\ee
is a smooth function that is rapidly decreasing as $|y| \to \infty$. 
We take $h \in {\cal D}(\RR_+)$ such that   
\be 
\hat h (0) = \int_0^\infty  h(p)\, \frac{\rd p}{2\pi}\, \neq 0 \,.
\label{test}
\ee
Within the family of one-particle states (\ref{Phi}) defined by (\ref{Phi}) and (\ref{test}), one can 
%$\{\Phi(h)\, :\, h\in {\cal D}(\RR_+)\}$, 
evaluate the following expectation values
\be
\langle \Omega |\lambda_1(x+t) |\Phi (h) \rangle = {\hat h}(t+x) 
\;\;\qquad\;\;
\langle \Omega |\lambda_2(x-t) |\Phi (h) \rangle = e^{-\ri \alpha_{\textrm{\tiny v}}}{\hat h}(t-x) 
\label{bc11-12}
\ee
which imply 
\be 
\lim_{x\to \infty \atop{x\not=-t}} \langle \Omega |\lambda_1(x+t) |\Phi (h) \rangle = 
\lim_{x\to \infty \atop{x\not=t}} \langle \Omega |\lambda_2(x-t) |\Phi (h) \rangle = 0 \,.
\label{lim1}
\ee
However, considering the limit along the components of the light cone, given by $t=-x$ and $t=x$, one finds 
\be
\lim_{x\to \infty \atop{x =-t}} \langle \Omega |\lambda_1(x+t) |\Phi (h) \rangle 
= 
{\hat h} (0) \neq 0
\;\;\qquad\;\;
\lim_{x\to \infty \atop{x =t}} \langle \Omega |\lambda_2(x-t) |\Phi (h) \rangle 
=  
e^{-\ri \alpha_{\textrm{\tiny v}}}{\hat h}(0) \neq 0\,.
\label{lim2-3}
\ee
Comparing (\ref{bc5}) and (\ref{lim2-3}), we conclude that
the phase $e^{\ri \alpha_{\textrm{\tiny v}}}$ cannot be absorbed in $\lambda_2$ as in (\ref{bc10})
at $x=0$ and at $x=\infty$ simultaneously.

Finally we observe that the angle $\alpha_{\textrm{\tiny v}}$ has a simple physical interpretation in 
the context of scattering theory: the boundary induces a non-trivial one-body scattering matrix, 
which describes the particle reflection at $x=0$. The angle $\alpha_{\textrm{\tiny v}}$ defines the 
scattering phase shift of an outgoing particle with respect to an incoming one. 

\newpage

\subsubsection{Axial phase}

The axial phase implements the boundary condition (\ref{bc6}). 
Denoting by $\chi_r(t,x)$ the components of the Dirac field in this phase throughout the manuscript, we have that
these fields can be decomposed through the  generators of $\mathcal{A}_+$ and $\mathcal{B}_+$
introduced in Sec.\,\ref{subsec-vec-phase}.
These decompositions read
\bea
\label{npsisc1} 
\chi_1(x+t) 
&=&
\int_0^{\infty} \!
\left[\, \e^{-\ri \alpha_{\textrm{\tiny a}}} \,b(k) \, \e^{- \ri k (x+t)} + a^*(k) \, \e^{\ri k (x+t)}\, \right] \frac{\rd k}{2\pi} 
\\
\rule{0pt}{.7cm}
\label{npsisc2} 
\chi_2(x-t) 
&=& 
\int_0^{\infty} \!
\left[\, \e^{-\ri \alpha_{\textrm{\tiny a}}} \, a(k) \, \e^{\ri k (x-t)} + b^\ast(k) \, \e^{-\ri k (x-t)} \,\right]\frac{\rd k}{2\pi} 
\eea
which depend on the angle $\alpha_{\textrm{\tiny a}}$. 
The corresponding correlation functions on the vacuum are
\bea
\label{nsc11}
\langle \chi_1^*(x_1+t_1)\, \chi_1(x_2+t_2) &=& C (t_{12}+ x_{12})
\\
\label{nsc22}
\langle \chi_2(x_1-t_1)\, \chi_2^*(x_2-t_2)\rangle &=& C(t_{12}-x_{12}) 
\\
\label{nsc12}
\langle \chi_1^*(x_1+t_1)\, \chi_2^*(x_2-t_2)\rangle &=& \e^{\ri \alpha_{\textrm{\tiny a}}} \,C (t_{12}+\hat{x}_{12})
\\
\label{nsc21}
\langle \chi_2(x_1-t_1)\, \chi_1(x_2+t_2)\rangle &=& \e^{-\ri \alpha_{\textrm{\tiny a}}} \, C (t_{12}-\hat{x}_{12}) \,.
\eea
Notice that the violation of the electric charge conservation is manifest in  (\ref{nsc12}) and (\ref{nsc21}).
%%%%
Comparing (\ref{n11}), (\ref{n22}), (\ref{n12}) and (\ref{n21}) 
with (\ref{nsc11}), (\ref{nsc22}), (\ref{nsc12}) and (\ref{nsc21})  respectively, 
we observe that at equal times $t_1=t_2\equiv t$ we can write
\be
\label{acm}
\Bigg(
\begin{array}{cc} \langle \chi_1^*(x+t)\, \chi_1(y+t)\rangle \;\; & \langle \chi_1^*(x+t)\, \chi_2^*(y-t)\rangle 
\\
\rule{0pt}{.5cm}
\langle \chi_2(x-t)\, \chi_1(y+t) \rangle \;\; & \langle \chi_2(x-t)\, \chi_2^*(y-t)\rangle \\ \end{array} 
\Bigg)
= \,
\boldsymbol{C}(x,y;\alpha_{\textrm{\tiny a}})\,.
\ee
Thus, the matrices (\ref{vcm}) and (\ref{acm}) coincide, except for the angles $\alpha_{\textrm{\tiny v}}$ and $\alpha_{\textrm{\tiny a}}$. 

The discussion made in Sec.\,\ref{subsec-vec-phase} about the meaning of the phase $\e^{\ri \alpha_{\textrm{\tiny a}}}$
in (\ref{bc6}) and the possibility to reabsorb it through a redefinition of the fields (\ref{chi}) can be easily adapted 
also to this phase, arriving at the same conclusion.

Since in the vector phase both components of the doublet $\lambda (t,x)$ defined in (\ref{psi-doublet}) 
transform in the same way under the vector phase transformations (\ref{v1}), 
in the axial phase we find it convenient to introduce the following doublet
\be
\label{chi}
\chi (t,x) = 
\bigg(\begin{array}{c} \chi_1^*(x+t) \\  \chi_2(x-t)  \end{array} \bigg) 
\ee
whose components transform in the same way under the axial phase transformations (\ref{a1}). 

At this point, in order to treat both phases in a unified way, we introduce the doublet 
\be
\label{fn1}
\psi (t,x) = 
\bigg(\begin{array}{c} \psi_1 (x+t) \\  \psi_2(x-t)  \end{array} \bigg) \qquad \qquad 
\psi(t,x) 
\equiv 
\left\{
\begin{array}{ll} 
\lambda (t,x) \;\;\;\;\; &   \textrm{vector phase} 
\\
\rule{0pt}{.5cm}
\chi (t,x) & \textrm{axial phase}
\end{array} 
\right.
\ee
with $\lambda $ and $\chi$ given by (\ref{lambda-doublet}) and (\ref{chi}) respectively. With this notation 
the boundary conditions (\ref{bc5}) and (\ref{bc6}) take the form 
\be
\label{bcnew}
\psi_1(t) = \e^{\ri \alpha}\, \psi_2(-t) \qquad \qquad 
\alpha  
\equiv 
\left\{
\begin{array}{ll}
\alpha_{\rm v} \;\;\;\; & {\rm vector\; phase} 
\\
\rule{0pt}{.5cm}
\alpha_{\rm a} & {\rm axial\; phase}\,.
\end{array}
\right.
\ee

%\newpage
%%%%%%%%%%%%%%%%%%%%%%%%%%%%%%%%%%%%%%%%%%%
\section{Modular Hamiltonians of an interval on the half-line}
\label{sec_eh}

The massless Dirac fermion is described by a quadratic field theory, hence 
the modular Hamiltonian of an interval $A=[a,b] \subset \RR_+$ on the half-line 
for a massless Dirac fermion in its ground state
can be written in the following quadratic form
\be
\label{K_A lambda}
K_A \,= 
\int_A \int_A 
:\! \psi^* (0,x) \,
\boldsymbol{H}_A(x,y)\, 
\psi(0,y)\!:
\rd x  \,\rd y 
\ee
where $:\cdots :$ denotes the normal product in the oscillator algebras ${\cal A}_+$ and ${\cal B}_+$
and  $\psi(t,x)$ is the two-components field in (\ref{fn1}).  

The kernel ${\boldsymbol H}_A(x,y)$ in (\ref{K_A lambda})  is the $2\times 2$ matrix 
given by \cite{Peschel:2003rdm, EislerPeschel:2009review,Casini:2009sr}
\be
\label{eh-matrix-peschel}
\boldsymbol{H}_A(x,y) \,=\, \log \! \big(\boldsymbol{C}_A(x,y;\alpha)^{-1} - \mathbb{I}\,\big) 
\;\;\qquad\;\;
x,y \in A
\ee
where $\mathbb{I}$ is the identity matrix and $\boldsymbol{C}_A$ is the reduced correlation functions matrix,
obtained by restricting the correlation functions matrix
(given by (\ref{vcm}) in the vector phase and by (\ref{acm}) in the axial phase) to the interval $A$.

In order to obtain an explicit expression for ${\boldsymbol H}_A$, the spectral problem 
associated to the reduced correlation functions matrix $\boldsymbol{C}_A$ must be solved.
This means that we have to find the eigenvalues $\sigma_s$ and the eigenfunctions $\Phi_{s,p}(x) $ such that
\be
\label{sp0} 
\int_a^b 
\boldsymbol{C}_A(x,y;\alpha )\, \Phi_{s,p}(y) \, \rd y
=
\sigma_s\, \Phi_{s,p}(x) 
\;\;\qquad\;\;
x \in A
\ee
where $s$ and $p$ are two parameters specified below.

\subsection{The spectral problem}
\label{sec_spectral_problem}

In order to solve the spectral problem (\ref{sp0}) 
for the massless Dirac field on the half-line when the subsystem is the interval $A=[a,b] \subset \RR_+$,
let us first consider the  auxiliary spectral problem corresponding to 
the massless Dirac field in its ground state on the infinite line and where the bipartition of the line is given by 
the two disjoint equal intervals $A_{\textrm{\tiny sym}} \equiv [-b,-a] \cup [a,b] \subset \RR$ and by its complement on the line. 
%%%
This auxiliary spectral problem reads
\be
\label{sp1} 
\int_{A_{\textrm{\tiny sym}}} \!\!
C(x-y )\, \phi_{s,p}(y) \, \rd y
\,= \,\sigma_s\, \phi_{s,p}(x) 
\;\;\qquad\;\;
x\in A_{\textrm{\tiny sym}}
\ee
where  $C(x-y)$ is the distribution defined in (\ref{not2}).
The solution of this spectral problem can be found by specialising to $A_{\textrm{\tiny sym}}$
the solution of the spectral problem corresponding to an arbitrary number of disjoint intervals of generic lengths on the line,
which has been found in \cite{Casini:2009vk} by employing \cite{Musk-book}.
%%%
The explicit form of the spectral data $\{\sigma_s, \, \phi_{s,p} \,|\,  s \in \RR,\, p=1,2\}$ for (\ref{sp1}) has been reported in the Appendix\;\ref{app_2intervals}
(see (\ref{sigma-s-def}) and (\ref{phi_s form disjoint})).

The spectral problem (\ref{sp0}) can be solved through the auxiliary spectral problem (\ref{sp1}) by first observing that the latter one can be rewritten as follows
\be
\label{sp2} 
\int_a^b 
C(x-y )\, \phi_{s,p}(y) \, \rd y \,+ \int_a^b C(x+y )\, \phi_{s,p}(-y) \, \rd y
\,=\,
\sigma_s\, \phi_{s,p}(x) 
\;\;\qquad \;\;
x\in A_{\textrm{\tiny sym}} \,.
\ee
Since $A_{\textrm{\tiny sym}} $ is symmetric under reflection with respect to the origin at $x=0$, 
we can conclude that (\ref{sp2}) holds also for $x \mapsto -x$. 
This observation provides another identity that can be combined with (\ref{sp2}).
This leads to write the solution of (\ref{sp0}) in terms of $\phi_{s,p}$ and $\alpha$ as follows
\be
\label{sp3}
\Phi_{s,p}(x) 
\equiv 
\Bigg(
\begin{array}{c}
\Phi^{(1)}_{s,p}(x)
\\
\rule{0pt}{.5cm}
\Phi^{(2)}_{s,p}(x)
\end{array} 
\Bigg) =
\Bigg(
\begin{array}{c}
e^{\textrm{i} \alpha} \,\phi_{s,p}(x)
\\
\rule{0pt}{.5cm}
\phi_{s,p}(-x)
\end{array} 
\Bigg)\,.
\ee

We stress  that the function $\phi_{s,p}$ in (\ref{sp3}) is the solution of the spectral problem (\ref{sp1}) in  $A_{\textrm{\tiny sym}} $. 
The completeness and orthonormality of 
the system $\{\phi_{s,p}(x)  :  x\in A_{\textrm{\tiny sym}}, s\in \RR, p=1,2\}$  
implies that (\ref{sp3}) form a complete set of orthonormal eigenfunctions in $A$, namely
\be
\sum_{p=1}^2\int_{-\infty}^\infty  \overline \Phi^{(i)}_{s,p}(x) \Phi^{(j)}_{s,p}(y) \, \rd s = \delta_{ij}\, \delta(x-y)
\;\qquad\;
\sum_{i=1}^2\int_a^b \overline \Phi^{(i)}_{s,p}(x) \Phi^{(i)}_{r,q}(x)\, \rd x = \delta_{pq}\, \delta(s-r)
\ee
where the overline indicates the complex conjugation.

\subsection{The kernel ${\boldsymbol H}_A$}

The solution of the spectral problem (\ref{sp0})  leads us to write the 
reduced correlation functions matrix through its spectral representation
\be
\label{C_A spectral rep}
\boldsymbol{C}_A(x,y;\alpha ) = 
\sum_{p=1}^2 \int_{-\infty}^{+\infty} \!\!  \sigma_s\, \Phi_{s,p}(x)\, \Phi^\ast_{s,p}(y) \, \rd s \,.
\ee
The relation (\ref{eh-matrix-peschel}) tells us that  $\boldsymbol{C}_A$ and $\boldsymbol{H}_A$ share the same eigenfunctions
and that, since the eigenvalues $\sigma_s$ of $\boldsymbol{C}_A$ are  (\ref{sigma-s-def}),
the eigenvalues of $\boldsymbol{H}_A$ are given by $-2\pi s = \log(1/\sigma_s - 1) $ with $s\in \RR$.
Thus, the spectral representation of the kernel $\boldsymbol{H}_A$ reads
\be
\label{H_A spectral rep}
\boldsymbol{H}_A(x,y)
\,=\,
- \,2\pi \sum_{p=1}^2\int_{-\infty}^{+\infty} \!\! s \; \Phi_{s,p}(x)\, \Phi_{s,p}^\ast (y) \, \rd s\,.
\ee
Expressing the eigenfunctions (\ref{sp3}) through the explicit form of $\phi_{s,p}$ given in (\ref{phi_s form disjoint}), we find
\be
\label{H_A-matrix-s-integral}
\Phi_{s,p}(x)\, \Phi_{s,p}^\ast (y)
\,=\,
\Bigg(\!
\begin{array}{cc}
m(x,y) \, e^{-\textrm{i} s[w(x) - w(y)]}
&
\;\;-\,e^{\textrm{i} \alpha} \, m(x,-y) \, e^{-\textrm{i} s[w(x) - w(-y)]}
\\
\rule{0pt}{.5cm}
- e^{-\textrm{i} \alpha} \, m(-x,y)\, e^{-\textrm{i} s[w(-x) - w(y)]}
&
\;\; m(-x,-y)\, e^{-\textrm{i} s[w(-x) - w(-y)]}
\end{array} 
\,\Bigg)
\ee
where we have introduced 
\be
\label{w-function-def}
w(x) \,\equiv\,  
\log \!\left [ \frac{(x+b)(x-a)}{(x+a)(b-x)} \right ]
\ee
that plays an important role throughout our analysis,
and $m(x,y)$ is defined in terms of the functions $m_p(x)$ in (\ref{m-1-2-def}) as follows
\be
\label{mtilde-def-split}
m(x,y) 
\equiv 
\sum_{p=1}^2 m_p(x)\, m_p(y) 
=
\frac{(b-a) \,(   x\, y + a\, b  )}{\pi \, \sqrt{(b^2 - x^2)(x^2 - a^2)\, (b^2 - y^2)(y^2 - a^2)}}\,.
\ee

The integration in (\ref{H_A spectral rep}) can be performed by observing that
\be
\label{E1}
\int_{-\infty}^\infty s\, \e^{-\ri s [w(\eta_x x)-w(\eta_y y)]}\, \rd s 
\,=\,
\ri \pi \left[\, \frac{\eta_x}{w^\prime(x)}\,\prt_x - \frac{\eta_y}{w^\prime(y)}\,\prt_y \, \right] \delta \big(w(\eta_x x)-w(\eta_y y)\big)
\ee
where $\eta_{x}, \eta_y \in \{-1, +1\}$.
The support of the Dirac delta function in (\ref{E1}) is given by the zeros of the following function
\be
z_w(\eta_x x,\eta_y y) 
\equiv
w(\eta_x x) - w(\eta_y y) \,.
\ee

A crucial role  is played by the point $\tilde{x}$ conjugate to $x$ defined as
\be
\label{xtilde}
\tilde{x} \equiv \frac{a\, b}{x} \,.
\ee
Notice that, if $x\in A$, then also $\tilde{x} \in A$.

The integration in (\ref{H_A spectral rep}) of the diagonal elements of the matrix (\ref{H_A-matrix-s-integral}) 
provides Dirac delta functions localised where $z_w(x,y) =0 $ and $z_w(-x,-y) =0 $.
These equations are solved by $y=x$ and $y=-\tilde{x}$, but only the former solution is allowed
because the latter one does not belong to $\RR_+$ when $x\in A$.
Thus, for these integrals we have to use
\be
\label{w10}
\int_{-\infty}^\infty s\, \e^{-\ri s [w(x)-w(y)]}\, \rd s 
\,=\, 
-\int_{-\infty}^\infty s\, \e^{-\ri s [w(-x)-w(-y)]}\, \rd s
\,=\,
 \frac{\ri \pi }{w^\prime (x) \,w^\prime ( y)} \left (\prt_x - \prt_y \right )\delta (x-y)
\ee
which tells us that the diagonal elements in (\ref{H_A-matrix-s-integral}) lead to local terms in the modular Hamiltonian. 
%%%
Instead, the off-diagonal elements of (\ref{H_A-matrix-s-integral}) give Dirac delta functions 
localised on the zeros of $z_w(x,-y)$ and $ z_w(-x,y)$. 
They are $y=\tilde{x}$ and $y=-x$, but only the former solution is allowed because $-x \notin \RR_+$ when $x\in A$.
Thus,  for the integration  in (\ref{H_A spectral rep})  of the off-diagonal elements of (\ref{H_A-matrix-s-integral}) we need
\bea
\label{w11}
& & \hspace{-1cm}
\int_{-\infty}^\infty s\, \e^{-\ri s [w(x)-w(-y)]}\, \rd s \,=\, - \int_{-\infty}^\infty s\, \e^{-\ri s [w(-x)-w(y)]}\, \rd s \, =
\\
& & \hspace{5cm}
=
\frac{\ri \pi }{2\, w^\prime (x) \,w^\prime ( y)} 
\left[\, \frac{ab}{y^2} \, \prt_x \delta (x-\tilde{y}) +  \frac{ab}{x^2}\, \prt_y \delta (y-{\tilde{x}})\, \right] 
\nonumber
\eea
which tells us that these terms give origin to bi-local terms in the modular Hamiltonian.

The above discussion suggests to separate the diagonal terms and the off-diagonal terms in (\ref{H_A-matrix-s-integral}).
This leads to write the kernel (\ref{H_A spectral rep}) as follows
\be
\label{H_A-final}
\boldsymbol{H}_A(x,y) 
=
\boldsymbol{H}^{\textrm{\tiny loc}}_A(x,y)
+
\boldsymbol{H}^{\textrm{\tiny bi-loc}}_A(x,y) 
\ee 
where $\boldsymbol{H}^{\textrm{\tiny loc}}_A$ and $\boldsymbol{H}^{\textrm{\tiny bi-loc}}_A$
give origin respectively to the local terms and the bi-local terms in the modular Hamiltonian. 
More explicitly, the matrices in the r.h.s. of (\ref{H_A-final}) are
\be
\label{HA-kernel-local}
\boldsymbol{H}^{\textrm{\tiny loc}}_A(x,y)
\equiv 
- 2\pi\,\textrm{i}  \; M_+(x,y)\;
\bigg(
\begin{array}{cc}
(\partial_x - \partial_y) \,\delta(x-y) & 0
\\
\rule{0pt}{.3cm}
0 & -\,(\partial_x - \partial_y) \,\delta(x-y)
\end{array} 
 \bigg) 
 \ee
and 
 \be
\label{HA-kernel-bilocal}
\boldsymbol{H}^{\textrm{\tiny bi-loc}}_A(x,y)
\equiv 
- 2\pi\,\textrm{i}  \; M_-(x,y)\;
\bigg(
\begin{array}{cc}
0
&
e^{\textrm{i} \alpha} \, D(x,-y)
\\
\rule{0pt}{.3cm}
-e^{-\textrm{i} \alpha} \,D(x,-y)
&
0
\end{array} 
\bigg) 
\ee
where we have introduced
\be
M_\pm(x,y)
\,\equiv\,
\frac{(x\, y\pm a\,b)\, \sqrt{(b^2 - x^2)(x^2 - a^2)\, (b^2 - y^2)(y^2 - a^2)}}{4(b-a)\,(a\, b + x^2)\,(a\, b + y^2)}
\ee
and
\be
D(x,-y) 
\,\equiv \,
\frac{a b}{y^2} \, \partial_x \delta(x-\tilde{y}) +  \frac{a b}{x^2} \, \partial_y \delta(y-\tilde{x}) 
\ee
that satisfies the relation $D(-x,y) = - D(x,-y)$.

The decomposition (\ref{H_A-final}) leads to write the modular Hamiltonian (\ref{K_A lambda}) as follows
\be
\label{K_A-decomposition}
K_A  \,= \, K_A^{\textrm{\tiny loc}} + K_A^{\textrm{\tiny bi-loc}} 
\ee 
where $K_A^{\textrm{\tiny loc}}$ is a local operator, while $K_A^{\textrm{\tiny bi-loc}} $ is a bi-local operator,
which are discussed in Sec.\,\ref{sec-local-term} and Sec.\,\ref{sec-bilocal-term} respectively.

Setting either $\alpha = \alpha_{\textrm{\tiny v}}$ or $\alpha = \alpha_{\textrm{\tiny a}}$ 
in the above expressions, we obtain the corresponding results respectively for the vector phase and for the axial phase.

\subsubsection{Local term}
\label{sec-local-term}

The local term in (\ref{K_A-decomposition}) reads
\be
\label{K_A lambda local}
K_A^{\textrm{\tiny loc}} 
\equiv
\int_a^b  \!\! \int_a^b \!
:\! \psi^*(0,x) 
\,\boldsymbol{H}^{\textrm{\tiny loc}}_A(x,y) \;
\psi(0,y)\! :
 \rd x \, \rd y
\ee
where the kernel $\boldsymbol{H}^{\textrm{\tiny loc}}_A(x,y) $ has been defined in  (\ref{HA-kernel-local}).

The Dirac delta functions occurring in $\boldsymbol{H}^{\textrm{\tiny loc}}_A(x,y) $ guarantee that (\ref{K_A lambda local}) is a local operator
because it can be written as an integral over $A$ of fields evaluated at the same point. 
This integral can be found by first plugging (\ref{HA-kernel-local}) into (\ref{K_A lambda local}) and then integrating by parts. 
Exploiting the fact that $ [ \partial_x M_+(x,y) - \partial_y M_+(x,y) ] \,|_{x=y} =\, 0 $,  we find that (\ref{K_A lambda local}) becomes
\be
\label{K_A-local-def}
K_A^{\textrm{\tiny loc}} 
\,=\,
2\pi 
\int_a^b \!
\beta_{\textrm{\tiny loc}}(x) \, T_{tt}(0,x)\, \rd x 
\ee
where the operator $T_{tt}(t,x)$ is the normal ordered version of the energy density (\ref{endens}), namely
\be
\label{T00-lambda-def}
T_{tt}(t,x) 
\,\equiv\,
\,\frac{\textrm{i}}{2}
:\! \!
\Big[ \Big ((\partial_x \psi^\ast_1)\, \psi_1 - 
\psi^\ast_1\, (\partial_x \psi_1) \Big )(x+t)
- \Big((\partial_x \psi^\ast_2)\,  \psi_2 - \psi^\ast_2\, (\partial_x \psi_2)\Big) (x-t)
\Big]\!\! : 
\ee
and the weight function reads
\be
\label{beta-loc-def}
\beta_{\textrm{\tiny loc}}(x) 
\,\equiv\,
 2M_+(x,x) 
=
\frac{1}{w'(x)} 
=
\frac{(b^2 - x^2)\,(x^2 - a^2)}{2\,(b-a)\, (a\,b +x^2)}
\ee
which satisfies 
\be
\label{beta-loc-xtilde}
\beta_{\textrm{\tiny loc}}(\tilde{x}) = \frac{ab}{x^2}\;\beta_{\textrm{\tiny loc}}(x)
\qquad \Longleftrightarrow \qquad
\frac{\beta_{\textrm{\tiny loc}}(\tilde{x})}{\tilde{x}} =  \frac{\beta_{\textrm{\tiny loc}}(x)}{x}\,.
\ee

\subsubsection{Bi-local term}
\label{sec-bilocal-term}

The bi-local term in the decomposition (\ref{K_A-decomposition}) of the modular Hamiltonian (\ref{K_A lambda}) reads
\be
\label{K_A lambda bi-local}
K_A^{\textrm{\tiny bi-loc}} 
\equiv
\int_a^b  \!\! \int_a^b \!
:\! \psi^*(0,x) 
\,\boldsymbol{H}^{\textrm{\tiny bi-loc}}_A(x,y) \;
\psi(0,y)\! :
 \rd x \, \rd y
\ee
where the kernel $\boldsymbol{H}^{\textrm{\tiny bi-loc}}_A(x,y)$ is defined in (\ref{HA-kernel-bilocal}).

The operator (\ref{K_A lambda bi-local}) is bi-local because
the Dirac delta functions in $\boldsymbol{H}^{\textrm{\tiny bi-loc}}_A(x,y)$ 
allow to write it as an integral over $A$ of fields evaluated at different (conjugated) points.
%%%
This integral can be computed by plugging (\ref{HA-kernel-bilocal}) into (\ref{K_A lambda bi-local}) first and then integrating by parts.
Since $M_-(x,ab/x) = 0$, we find that (\ref{K_A lambda bi-local}) becomes
\be
\label{K_A-bilocal-def}
K_A^{\textrm{\tiny bi-loc}}  
\,=\,
2\pi 
\int_a^b \!
\beta_{\textrm{\tiny bi-loc}}(x) \, T_{\textrm{\tiny bi-loc}}(0,x, \tilde{x};\alpha) \, \rd x
\ee
where we have introduced the following bi-local operator 
\bea
\label{T-bilocal-def}
T_{\textrm{\tiny bi-loc}}(t,x, y ;\alpha) 
 &\equiv &
\frac{\textrm{i}}{2}\;
\bigg\{ \,
e^{\textrm{i} \alpha}
\!:\!\!\Big[\, \psi^\ast_1(y+t) \,  \psi_2(x-t) + \psi^\ast_1(x+t) \,  \psi_2(y-t) \Big]\!\!: 
\\
& & \hspace{.8cm}
- \;
e^{-\textrm{i} \alpha}
\!:\!\! \Big[\, \psi^\ast_2(y-t) \,  \psi_1(x+t) + \psi^\ast_2(x-t) \,  \psi_1(y+t) \Big] \!\!: \!
\bigg\}
\nonumber
\eea
and the weight function is 
\be
\label{beta-biloc-def}
\beta_{\textrm{\tiny bi-loc}}(x) 
\equiv
\frac{a\, b \, (b^2-x^2) \, (x^2-a^2)}{2\,(b-a)\, x\,(a\, b + x^2)^2} 
=
\frac{\beta_{\textrm{\tiny loc}}(\tilde{x}) }{x + \tilde{x}}
\ee
which satisfies 
\be
\beta_{\textrm{\tiny bi-loc}}(\tilde{x}) = \frac{x^2}{a\, b}\, \beta_{\textrm{\tiny bi-loc}}(x)
\qquad \Longleftrightarrow \qquad
\tilde{x} \, \beta_{\textrm{\tiny bi-loc}}(\tilde{x}) = x\, \beta_{\textrm{\tiny bi-loc}}(x)\,.
\ee 

The final form for the modular Hamiltonian of the interval $A\subset \RR_+$ is  (\ref{K_A-decomposition}),
where $K_A^{\textrm{\tiny loc}}$ and $K_A^{\textrm{\tiny bi-loc}}$ are the operators 
given by  (\ref{K_A-local-def}) and (\ref{K_A-bilocal-def}) respectively.

\subsubsection{Modular Hamiltonians in the vector and axial phases}
\label{sec-eh-phases}

The explicit expressions of the modular Hamiltonians (\ref{K_A-decomposition}) 
for the interval $A\subset \RR_+$ on the half-line when the Dirac field is in the ground state 
can be written by first using (\ref{K_A-local-def}) and (\ref{K_A-bilocal-def}) and then
performing the substitutions $(\psi, \alpha) \mapsto (\lambda, \alpha_{\textrm{\tiny v}})$ for the vector phase 
and $(\psi, \alpha) \mapsto (\chi, \alpha_{\textrm{\tiny a}})$ for the axial phase
into the operators (\ref{T00-lambda-def}) and (\ref{T-bilocal-def}),
where $\lambda $ and $\chi$ have been defined in (\ref{lambda-doublet}) and in (\ref{chi}) respectively. 

In the local term (\ref{K_A-local-def}), taking into account that the fermion fields anticommute under the normal product, we observe that 
$T_{tt}$ in (\ref{T00-lambda-def}) has the same form when expressed in terms of $\lambda$ or $\chi$. 
This is a consequence of the fact that $T_{tt}$ is invariant under both the vector and the axial transformations 
given in (\ref{v1}) and (\ref{a1}) respectively. 

In the bi-local term (\ref{K_A-bilocal-def}), 
the bi-local operator (\ref{T-bilocal-def}) in the vector phase reads
\bea
\label{T-bilocal-vec}
T^{\textrm{\tiny vector}}_{\textrm{\tiny bi-loc}}(t,x, y ;\alpha_{\textrm{\tiny v}}) 
 &\equiv &
\frac{\textrm{i}}{2}\;
\bigg\{ \,
e^{\textrm{i} \alpha_{\textrm{\tiny v}}}
\!:\!\!\Big[\, \lambda^\ast_1(y+t) \,  \lambda_2(x-t) + \lambda^\ast_1(x+t) \,  \lambda_2(y-t)\, \Big]\!\!: 
\nonumber
\\
& & \hspace{.8cm}
- \;
e^{-\textrm{i} \alpha_{\textrm{\tiny v}}}
\!:\!\! \Big[\, \lambda^\ast_2(y-t) \,  \lambda_1(x+t) + \lambda^\ast_2(x-t) \,  \lambda_1(y-t)\, \Big] \!\!: \!
\bigg\}
\eea
while in the axial phase it becomes
\bea
\label{T-bilocal-ax}
T^{\textrm{\tiny axial}}_{\textrm{\tiny bi-loc}}(t,x, y ;\alpha_{\textrm{\tiny a}}) 
 &\equiv &
\frac{\textrm{i}}{2}\;
\bigg\{ \,
e^{\textrm{i} \alpha_{\textrm{\tiny a}}}
\!:\!\!\Big[\, \chi_1(y+t) \,  \chi_2(x-t) + \chi_1(x+t) \,  \chi_2(y-t)\, \Big]\!\!: 
\nonumber
\\
& & \hspace{.8cm}
- \;
e^{-\textrm{i} \alpha_{\textrm{\tiny a}}}
\!:\!\! \Big[\, \chi^\ast_2(y-t) \,  \chi^\ast_1(x+t) + \chi^\ast_2(x-t) \,  \chi^\ast_1(y+t) \,\Big] \!\!: \!
\bigg\}\,.
\eea
These expressions depend explicitly on the angle determining the boundary condition at $x=0$.

It is  instructive to compare these modular Hamiltonians to the modular Hamiltonian of 
two disjoint equal intervals $A_{\textrm{\tiny sym}}$ on the line, 
obtained as a special case of the modular Hamiltonian of the union of a generic number of
disjoint intervals on the line found in \cite{Casini:2009vk}.
Also the modular Hamiltonian of $A_{\textrm{\tiny sym}}\subset \RR$ can be written as the sum of a local term and a bi-local one
\be
\label{K_Asym-decomposition}
K_{A_{\textrm{\tiny sym}}}  =  K_{A_{\textrm{\tiny sym}}}^{\textrm{\tiny loc}} + K_{A_{\textrm{\tiny sym}}}^{\textrm{\tiny bi-loc}} 
\ee 
where
\be
\label{K_Asym-terms}
K_{A_{\textrm{\tiny sym}}}^{\textrm{\tiny loc}} 
=
2\pi 
\int_{A_{\textrm{\tiny sym}}} \!\!
\beta_{\textrm{\tiny loc}}(x) \, T_{tt}(0,x)\, \rd x 
\;\;\qquad\;\;
K_{A_{\textrm{\tiny sym}}}^{\textrm{\tiny bi-loc}} 
=
2\pi 
\int_{A_{\textrm{\tiny sym}}} \!\!
\beta_{\textrm{\tiny bi-loc}}(x) \, T_{\textrm{\tiny bi-loc}}(0,x, -\tilde{x}) \, \rd x \,.
\ee
In these expressions $T_{tt}(t,x)$ is the normal ordered energy density (\ref{T00-lambda-def}),
the weight functions $\beta_{\textrm{\tiny loc}}(x)$ and $\beta_{\textrm{\tiny bi-loc}}(x)$ are 
(\ref{beta-loc-def}) and (\ref{beta-biloc-def}) respectively
and the bi-local operator $T_{\textrm{\tiny bi-loc}}(t,x, y)$ occurring in the bi-local term is defined as follows
\bea
\label{T-bilocal-2int}
T_{\textrm{\tiny bi-loc}}(t,x, y) 
 &\equiv &
\frac{\textrm{i}}{2}\;
\bigg\{ \,
\!:\!\!\Big[\, \psi^\ast_1(x+t) \,  \psi_1(y+t) - \psi^\ast_1(y+t) \,  \psi_1(x+t) \, \Big]\!\!: 
\nonumber
\\
& & \hspace{.8cm}
+ \,
\!:\!\! \Big[\,  \psi^\ast_2(x-t) \,  \psi_2(y-t) - \psi^\ast_2(y-t) \,  \psi_2(x-t) \, \Big] \!\!: \!
\bigg\}
\eea
where we remark that the fields in this operator are given by (\ref{psi1-line}) and (\ref{psi2-line}).

In the local term (\ref{K_A-local-def}),
the weight function and the functional dependence on the fermion fields in the integrand 
are the same occurring in the local term $K_{A_{\textrm{\tiny sym}}}^{\textrm{\tiny loc}}$
of the modular Hamiltonian of $A_{\textrm{\tiny sym}}$ on the line,
that is  the first expression in (\ref{K_Asym-terms}).
However, we stress that the fermionic fields on the half-line are given by
(\ref{npsi1hl}), (\ref{npsi2hl}), (\ref{npsisc1}) and (\ref{npsisc2}),
which depend explicitly on the angles $\alpha_{\textrm{\tiny v}}$ and $\alpha_{\textrm{\tiny a}}$ 
characterising the boundary condition at $x=0$.

As for the bi-local terms, comparing (\ref{K_A-bilocal-def}) and the second expression in (\ref{K_Asym-terms}), 
one observes that, while the same weight function occurs in the integrands, 
the corresponding bi-local operators are very different.
Indeed, we have that
(a) the operator (\ref{T-bilocal-2int}) is invariant under both vector (\ref{v1}) and axial (\ref{a1}) transformations, 
while this is not the case for (\ref{T-bilocal-vec}) and (\ref{T-bilocal-ax}) which preserve separately only the vector and axial 
symmetry;
(b) the point conjugate to $x$ in the integrand of $K_{A_{\textrm{\tiny sym}}}^{\textrm{\tiny bi-loc}} $ is $-\tilde{x}$, 
which belongs to the opposite interval in $A_{\textrm{\tiny sym}}$ with respect to $x$, 
while in the integrand of (\ref{K_A-bilocal-def})
both $x$ and its conjugate point $\tilde{x}$ belong to the interval $A$;
hence also the self-conjugate point $x=\sqrt{a \,b}$ (where $x=\tilde{x}$) occurs;
(c) the bi-local operators (\ref{T-bilocal-vec}) and (\ref{T-bilocal-ax}) depend explicitly on the boundary conditions through the angular parameter characterising the corresponding phase.

\section{Entanglement entropies}
\label{sec-ee}

The R\'enyi entropies are defined through
the moments of the reduced density matrix $\textrm{Tr}_{\!{}_A} \rho_A^n$ for integers $n \geqslant 2$ 
as 
\be
\label{renyi-def}
S_A^{(n)} \equiv \frac{1}{1-n}\, \log\!\big[ \textrm{Tr}_{\!{}_A} \rho_A^n\big]\,.
\ee
They provide the entanglement entropy $S_A$ by means of the following analytic continuation
of the integer parameter $n$ 
\be
S_A \equiv \lim_{n \to 1} S_A^{(n)}
=
-\, \partial_n \big( \textrm{Tr}_{\!{}_A} \rho_A^n \big) \big|_{n=1}\;.
\ee

In a two dimensional conformal field theory,
the moments of the reduced density matrix have been computed as the correlation functions of
the branch point twist fields $\mathcal{T}_n$ \cite{Calabrese:2004eu}.

When the interval $A=[0,b]$ is adjacent to the boundary of the half-line and the entire system is in the ground state,
$\textrm{Tr} \rho_A^n  = \langle \mathcal{T}_n(b)  \rangle \propto (2b/\epsilon)^{-\Delta_n}$ 
is the one-point function of the twist field, where $\epsilon$ is the ultraviolet cutoff and 
$\Delta_n$ is the conformal dimension of the twist field given by 
\be
\Delta_n \equiv \frac{c}{12} \left( n -\frac{1}{n}\right)
\ee
which is proportional to the central charge $c$ of the model.

When the interval $A=[a,b]$ on the half-line is not adjacent to the boundary, 
$\textrm{Tr} \rho_A^n$ can be found as the two-point function of the twist fields. 
Combining the R\'enyi entropies of two disjoint equal intervals on the line 
\cite{Caraglio:2008pk,Furukawa:2008uk,  Calabrese:2009ez, Calabrese:2010he} 
with the method of the images, one obtains
\be
\label{tr-rhoA-cft}
\textrm{Tr}_{\!{}_A} \rho_A^n 
= 
\langle \mathcal{T}_n(a) \, \mathcal{T}^\ast_n\! (b)  \rangle
=
c_n\! \left[ \frac{(a+b)^2\, \epsilon^2}{4\,a \,b\, (b-a)^2} \right]^{\Delta_n} \! \mathcal{F}_n(r)
\;\;\qquad\;\;
r \equiv \frac{(b-a)^2}{(b+a)^2}
\ee
where $r$ is the cross ratio of the endpoints of the interval and of their images, while $c_n$ is a constant.
The function $\mathcal{F}_n(r)$ depends on the full operator content of the boundary conformal field theory,
hence it encodes also the conformal boundary state, in a highly non-trivial way. 
These one-point and two-point correlation functions of twist fields allow to construct the following ultraviolet finite combinations
\be
\label{uv-finite-ratio}
S^{(n)}_{[0,a]} + S^{(n)}_{[0,b]} - S^{(n)}_{[a,b]}
\,=\,
\frac{1}{n-1}\, \log R_n
\;\;\qquad\;\;
R_n \equiv
\frac{\textrm{Tr}_{{}_{[a,b]}} \rho_{[a,b]}^n}{\big( \textrm{Tr}_{{}_{[0,a]}} \rho_{[0,a]}^n \big) \big( \textrm{Tr}_{{}_{[0,b]}} \rho_{[0,b]}^n \big)}\,.
\ee

A relevant quantity encoding some properties of the boundary is the Affleck-Ludwig boundary entropy $\log(g)$ \cite{Affleck:1991tk}.
Considering a conformal field theory in its ground state either on the half-line or on the line, 
the Affleck-Ludwig boundary entropy
can be found by combining the R\'enyi entropies $S_{[0,\ell]}^{(n)}$ of an interval of length $\ell$ adjacent to the boundary of the half line
and the R\'enyi entropies $\widetilde{S}_{[0,\ell]}^{(n)}$ of an interval of length $2\ell$ on the line as follows
\cite{Calabrese:2004eu, Cardy:2016fqc, Affleck:2009aa, Schollwok:2006aaa} 
\be
\label{log-g-from-S}
S_{[0,\ell]}^{(n)} - \frac{1}{2}\, \widetilde{S}_{[0,2\ell]}^{(n)}  = \log (g)\,.
\ee

The massless Dirac field is a conformal field theory with $c=1$. 
In this model, 
the R\'enyi entropies and the entanglement entropy 
can be evaluated also through the reduced correlation functions matrix respectively as 
\cite{EislerPeschel:2009review,Casini:2009sr}
\be
\label{renyi-fer}
S_A^{(n)} =
%\log\!\big[ \textrm{Tr} \rho_A^n \big] =  
\frac{1}{1-n}\;\textrm{tr} \big[  g_n(\boldsymbol{C}_A) \big]
 \;\;\qquad\;\;
 g_n(\gamma) \equiv \log\!\big[\gamma^n+(\mathbb{I}-\gamma)^n \big]
\ee
and 
\be
\label{SA-function-CA}
S_A  =  \textrm{tr} \big[ g(\boldsymbol{C}_A)\big]
 \;\;\qquad\;\;
 g(\gamma) \equiv
   -\, \gamma \log \gamma - (\mathbb{I}-\gamma) \log(\mathbb{I}-\gamma)
\ee
where $\mathbb{I}$ is the $2\times 2$ identity matrix, $\gamma$ is a generic $2\times 2$ matrix and 
$g(\gamma) \equiv -\partial_n g_{n}(\gamma)|_{n=1}$.
%%%%%

By employing the spectral representation of $\boldsymbol{C}_A$ in (\ref{SA-function-CA}),
for the entanglement entropies $S_A^{(n)}$ with integer $n\geqslant 1$,
which are given by $S_A^{(1)} \equiv S_A$ and (\ref{renyi-fer}) for $n\geqslant 2$, we find 
\be
\label{S_A-final-integral nomal}
S_A^{(n)}
=
\sum_{p=1}^2\,
\int_{A_{\epsilon}} 
\int_{-\infty}^{+\infty} \!\!  g_n(\sigma_s)\; 
\textrm{tr} \big[ \Phi_{s,p}(x)\,\Phi^\ast_{s,p}(x)\big]
\, \rd s
\, \rd x
\ee
where $\sigma_s$ is  (\ref{sigma-s-def}),
$g_n(\sigma)$ is defined in  (\ref{SA-function-CA}) for $n=1$ and in  (\ref{renyi-fer}) for $n\geqslant 2$,
and $A_{\epsilon} \equiv (a+\epsilon,b-\epsilon)$ with $\epsilon \to 0^+$ has been introduced to regularise the integral.
By using (\ref{phi_s form disjoint}) and (\ref{sp3}), we find
\be
\label{trace-phi-phi}
\textrm{tr} \big[ \Phi_{s,p}(x)\,\Phi^\ast_{s,p}(x)\big]
=
| \phi_{s,p}(x)|^2 + |\phi_{s,p}(-x)|^2
=
m_p(x)^2 + m_p(-x)^2
\ee
where the explicit expressions of $m_p(x)$ are given in  (\ref{m-1-2-def}).
Notice that (\ref{trace-phi-phi}) is independent of both the parameters $\alpha$ and $s$.
The independence of $s$ leads to the factorisation of the two integrals in (\ref{S_A-final-integral nomal});
hence we can evaluate them separately.
By using $\int_{-\infty}^{+\infty} g_n(\sigma_s)\, \rd s = \tfrac{\pi(n+1)}{12\,n}$;
for the entanglement entropies we obtain
\be
\label{ent-ent-bdy}
S_A^{(n)}
=
\frac{n+1}{12\, n}\,  \log\! \bigg[ \,\frac{(2b/\epsilon-1)(2a/\epsilon+1)([b-a]/\epsilon-1)^2}{(b+a)^2/\epsilon^2-1} \,\bigg] 
=
\frac{n+1}{6\, n}\,  \log\! \bigg[ \,\frac{2 \,\sqrt{a\,b}\,(b-a)}{(a+b) \,\epsilon} \, \bigg] + O(\epsilon) \,.
\ee
We remark that this expression is independent of $\alpha$.

Plugging (\ref{tr-rhoA-cft}) into (\ref{renyi-def}) and comparing the resulting expression with (\ref{ent-ent-bdy}),
we find that
\be
\mathcal{F}_n(r) = 1
\ee 
identically 
in both the phases and  for any choice of the boundary condition parameter. 
A similar simplification has been observed also in \cite{Casini:2009vk} 
for the massless Dirac fermion in its ground state on the line when the subsystem is the union of disjoint intervals,
and further explored in \cite{Headrick:2012fk,Coser:2013qda, Coser:2015dvp}.
Lattice results in the XX chain with open boundary conditions for the bipartition that we are considering 
have been discussed in \cite{Fagotti:2010cc}.

The entanglement entropies for an interval $A=[0,b]$ adjacent to the boundary
are given by (\ref{S_A-final-integral nomal}) with $A_\epsilon \equiv (0, b-\epsilon)$.
The result is
\be
\label{renyi-adj}
S_A^{(n)} =
\frac{n+1}{12\, n}\,  \log (2b/\epsilon-1)
=
\frac{n+1}{12\, n}\,  \Big( \log(b/\epsilon)  + \log 2 \Big) + O(\epsilon)\,.
\ee

From (\ref{ent-ent-bdy}) and (\ref{renyi-adj}), we find that the UV finite combination 
(\ref{uv-finite-ratio}) in this model reads
\be
 S^{(n)}_{[0,a]} + S^{(n)}_{[0,b]} - S^{(n)}_{[a,b]}
\,=\,
\frac{n+1}{12\, n}\, \log\!\Bigg(\frac{\big[(a+b)^2 -\epsilon^2\big]\, (2a-\epsilon)}{(b-a-\epsilon)^2\,(2a+\epsilon)} \Bigg) 
=
\frac{n+1}{6\, n}\, \log\!\bigg(\frac{b+a}{b-a} \bigg) + O(\epsilon)\,.
\ee

In order to evaluate the Affleck-Ludwig boundary entropy for the massless Dirac field through (\ref{log-g-from-S}),
we have to consider the entanglement entropies of an interval $A=[a,b]$ of length $\ell = b-a$ on the line, 
when the entire system is in its ground state. 
For the massless Dirac field,
the underlying spectral problem is solved by the eigenvalues (\ref{sigma-s-def}) and by the following eigenfunctions 
\cite{Casini:2009vk, EislerPeschel:2009review}
\be
\label{eigenfunc-1int}
\phi_s(x) = \sqrt{\frac{b-a}{2\pi (b-x)(x-a)}}\; \e^{- \textrm{i} s w(x)}
\;\;\qquad\;\;
w(x) \equiv  \log\! \bigg( \frac{x-a}{b-x} \bigg)
\qquad
x\in A\,.
\ee
The corresponding entanglement entropies are 
\be
\label{S_A-int-line}
\widetilde{S}_A^{(n)}
=\,
2
\int_{A_{\epsilon}} 
\int_{-\infty}^{+\infty} \!\!  g_n(\sigma_s)\; 
 \big| \phi_{s}(x) \big|^2
\, \rd s
\, \rd x
=
\frac{n+1}{6\, n}\, \log\! \left( \frac{b-a}{\epsilon} - 1\right)
=
\frac{n+1}{6\, n}\, \log\! \left( \frac{\ell}{\epsilon} \right) +O(\epsilon)
\ee
where $A_{\epsilon} = (a+\epsilon, b-\epsilon)$ and the factor $2$ occurs because the two components of the Dirac field give the 
same contribution. 
%%%
By using (\ref{renyi-adj}) and (\ref{S_A-int-line}), we can specialise (\ref{log-g-from-S}) to our case, 
finding $g=1$, 
which has been obtained also in \cite{Affleck:1991tk,Smith:2019jnh} through different methods.

%\newpage
%%%%%%%%%%%%%%%%%%%%%%%%%%%%%%%%%%%%%%%%%%%
\section{Modular flows of the Dirac field}
\label{sec_mod_flow}

The reduced density matrix $\rho_A \propto e^{-K_A}$ generates the one-parameter family of unitary operators given by
$\{ \rho_A^{\textrm{i}\tau} :  \tau \in \RR\}$
that defines an automorphism on the operator algebra known as modular flow \cite{Haag-book}.

The modular flow of the two components of the Dirac field are defined as 
\be
\label{mod-evolution-lambda}
 \psi_r(\tau,x)
 \,\equiv\,
\rho_A^{\textrm{i} \tau} \, \psi_r(x)\, \rho_A^{-\textrm{i} \tau} 
 \,=\,
e^{-\textrm{i} \tau K_A} \, \psi_r(x)\, e^{\textrm{i} \tau K_A} 
\;\; \qquad \;\;
x\in A
\;\; \qquad \;\;
r\in \{1,2\}
\ee
where the initial configuration $\psi_r(x)$ in the right hand side 
is defined by (\ref{fn1}) at  $t=0$. 
The fields (\ref{mod-evolution-lambda}) can be found by solving the following differential equations
\be
\label{partial differential equation-mod-evolution}
\textrm{i}\,\frac{ d\psi_r(\tau,x)}{d\tau}  = \big[\,K_A \,, \psi_r(\tau,x)\,\big]_{-}
\;\; \qquad \;\;
x\in A
\;\; \qquad \;\;
r\in \{1,2\}\,.
\ee

The modular Hamiltonian $K_A$ can be decomposed as in (\ref{K_A-decomposition});
hence the r.h.s. of (\ref{partial differential equation-mod-evolution}) is the sum of two terms
determined  by the local operator $K_A^{\textrm{\tiny loc}} $ in (\ref{K_A-local-def})
and by the bi-local operator $K_A^{\textrm{\tiny bi-loc}}$ in (\ref{K_A-bilocal-def}).
Both these contributions can be evaluated by means of the equal-time anticommutators (\ref{car1}) and (\ref{car2}). 
As for the term provided by $K_A^{\textrm{\tiny loc}} $ in (\ref{K_A-local-def}),
by using the expression of $T_{tt}$ in (\ref{T00-lambda-def}) one obtains
\be
\label{mod-flow-comm-local}
\big[\,K_A^{\textrm{\tiny loc}}   \,, \psi_r(\tau,y)\,\big]_{-}
=\,
2\pi \, \textrm{i}\; (-1)^{r-1} \, B_{\textrm{\tiny loc}}(y) \, \psi_r(\tau,y)
\ee
where  the differential operator $B_{\textrm{\tiny loc}}$ is defined in terms of the weight function (\ref{beta-loc-def}) as follows
\be
\label{B-func-def}
B_{\textrm{\tiny loc}}(y) 
\equiv
 \beta_{\textrm{\tiny loc}}(y)\, \partial_y + \frac{1}{2}\, \partial_y \beta_{\textrm{\tiny loc}}(y)\,.
\ee
The contribution of the bi-local operator $K_A^{\textrm{\tiny bi-loc}}$ in (\ref{K_A-bilocal-def})
to the r.h.s. of (\ref{partial differential equation-mod-evolution})
can be found by exploiting the expression (\ref{T-bilocal-def}). The result reads
\be
\label{KA-bilocal-commutator}
\big[\,K_A^{\textrm{\tiny bi-loc}}   , \psi_r(\tau,y)\,\big]_{-}
=\,
2\pi\,\textrm{i}\; (-1)^{r} \, e^{- \textrm{i}(-1)^r\alpha}\;
\beta_{\textrm{\tiny bi-loc}}(y) \,\psi_k(\tau,\tilde{y}) 
\;\;\qquad\;\;
k \neq r
\ee
where the weight function (\ref{beta-biloc-def}) occurs and $\tilde{y}$ is 
the point conjugate to $y$.

The two partial differential equations in (\ref{partial differential equation-mod-evolution}) 
are coupled because both the components of the Dirac field occur in (\ref{KA-bilocal-commutator}).
In order to study the resulting  system of partial differential equations,
we find it convenient to introduce the following doublet of fields
\be
\label{Lambda-doublet-def}
\Psi(\tau,x)
\equiv
\bigg(
\begin{array}{c}
\psi_1(\tau,x) \\  \rule{0pt}{.3cm} \psi_2(\tau,\tilde{x}) 
\end{array}  \bigg) \,.
\ee
By using (\ref{mod-flow-comm-local}) and (\ref{KA-bilocal-commutator}) with $y=x$ and $y = \tilde{x}$, 
combined with (\ref{partial differential equation-mod-evolution}), 
in terms of the doublet (\ref{Lambda-doublet-def}) 
we obtain the following system of four partial differential equations 
\be
\label{mod-flow-4-matrix-bdy}
\frac{d}{d\tau}
\bigg(
\begin{array}{c}
\Psi(\tau, x) \\ \Psi(\tau, \tilde{x}) 
\end{array}  
\bigg) 
=\, 
2\pi \,\big[\, \boldsymbol{B}(x) \oplus \boldsymbol{B}(\tilde{x})\,\big] 
\bigg(
\begin{array}{c}
\Psi(\tau, x) \\ \Psi(\tau, \tilde{x}) 
\end{array}  
\bigg) 
\ee
where the $4\times 4$ block diagonal matrix within the square brackets is a differential operator 
expressed through the $2\times 2$ matrix differential operator defined as
\be
\boldsymbol{B}(x)
\equiv
\bigg(
\begin{array}{cc}
B_{\textrm{\tiny loc}}(x) & - \,e^{\textrm{i} \alpha}  \beta_{\textrm{\tiny bi-loc}}(x) 
\\
e^{-\textrm{i} \alpha}  \beta_{\textrm{\tiny bi-loc}}(\tilde{x}) & - \,B_{\textrm{\tiny loc}}(\tilde{x}) 
\end{array} 
\bigg) 
\ee
in terms of the differential operator $B_{\textrm{\tiny loc}}(x) $ introduced in (\ref{B-func-def}).
The solution of the system of partial differential equations in (\ref{mod-flow-4-matrix-bdy}) can be found by solving 
\be
\label{system-partial differential equation-Lambda-2comps}
\frac{d}{d\tau}\, \Psi(\tau, x) = 2\pi \, \boldsymbol{B}(x)\, \Psi(\tau, x) \qquad x\in A
\ee
with the initial condition 
\be
\label{initialcond}
\Psi(0,x)
= \bigg(
\begin{array}{c}
\psi_1(x) \\  \rule{0pt}{.3cm} \psi_2(\tilde{x})  
\end{array}  \bigg) \,.
\ee
This system can be solved by first performing a suitable transformation 
that decouples the two equations in (\ref{system-partial differential equation-Lambda-2comps}).
Each of the two resulting decoupled equations  
defines the flow of a one-parameter abelian group that can 
be determined through a standard technique.
All the details of this procedure are reported in Appendix\;\ref{app-shift-operator}.

\begin{figure}[t!]
\vspace{-.3cm}
\hspace{-.7cm}
%\begin{center}
\includegraphics[width=1.12\textwidth]{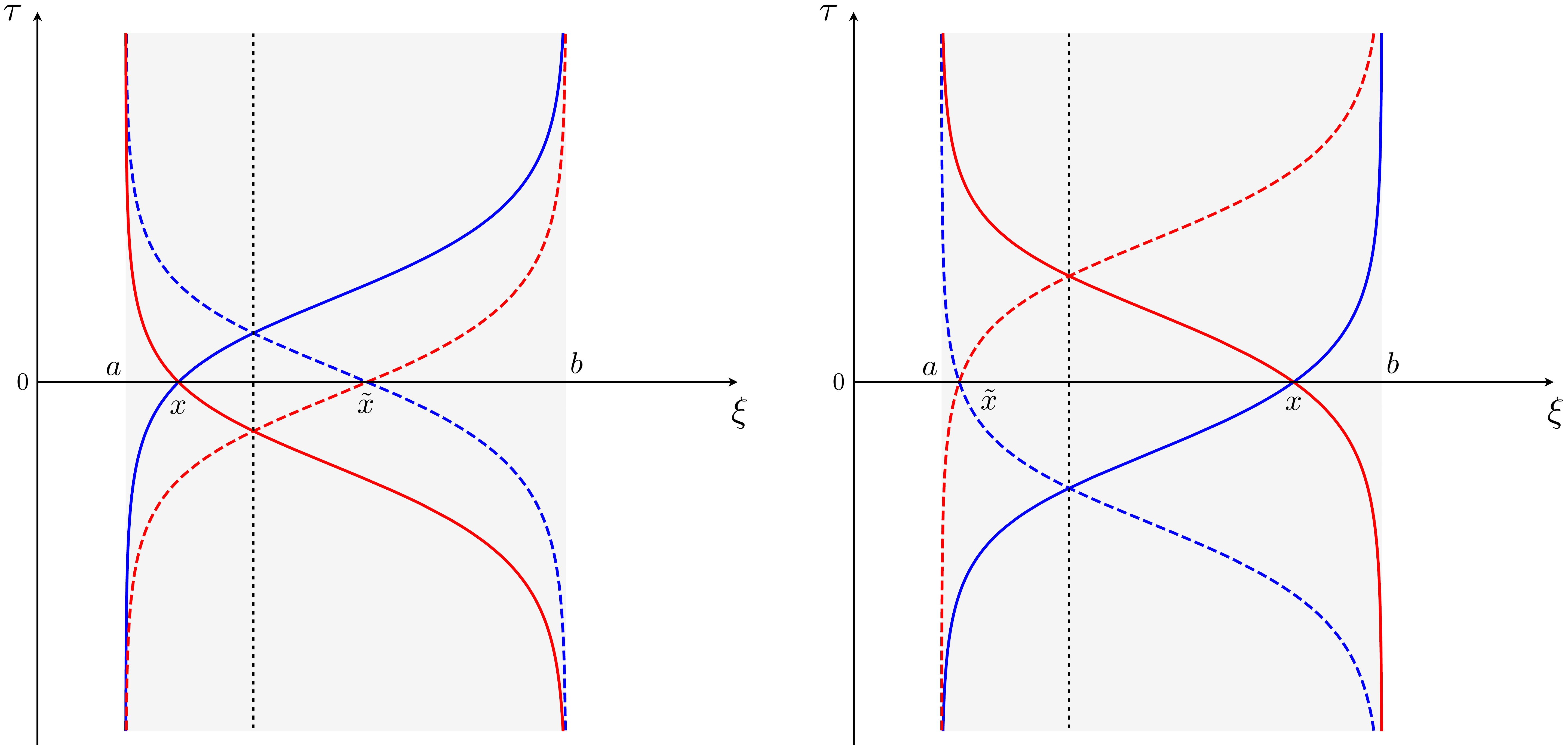}
% \end{center}
\vspace{-.2cm}
\caption{ 
Modular evolution of the arguments of the fields mixed by the modular flow in the r.h.s's of (\ref{psi-mod-flow-ab}),
for a point $x$ (solid lines) and its conjugate $\tilde{x}$ (dashed lines) at $\tau=0$.
The blue and the red curves correspond to
the first and the second equation in (\ref{psi-mod-flow-ab}) respectively.
The vertical dotted line identifies the point $\sqrt{ab}$.
In the left panel $x< \sqrt{ab}$, while $x> \sqrt{ab}$ in the right panel.
}
\label{figure-xi}
\end{figure}

The solution of (\ref{system-partial differential equation-Lambda-2comps}) and  (\ref{initialcond}) 
is one of the main results of this manuscript. It reads
\be
\label{psi-mod-flow-ab}
\left\{ \begin{array}{l}
\displaystyle
\psi_1(\tau,x)
=
\left[
P(\xi;  x) 
\left(
 \big( a \,b + x \,\xi \big) \,\psi_1(\xi) 
- \frac{a\,b}{\xi}\; e^{\textrm{i}\alpha} \big(\xi - x\big)\, \psi_2(ab/\xi)
\right)
\right]
\! \bigg|_{\xi = \xi(\tau,x)}
\\
\rule{0pt}{.85cm}
\displaystyle
\psi_2(\tau,x)
=
\left[
P(\xi;  x) 
\left(
 \big( a \,b + x \,\xi \big) \,\psi_2(\xi) 
- \frac{a\,b}{\xi}\; e^{-\textrm{i}\alpha} \big(\xi - x\big)\, \psi_1(ab/\xi)
\right)\right]
\! \bigg|_{\xi = \xi(-\tau,x)}
\end{array}
\right.
\ee
where
\be
\label{P-function-def}
P(\xi;  x)
\equiv
\sqrt{\frac{\beta_{\textrm{\tiny loc}}(\xi)}{\beta_{\textrm{\tiny loc}}(x)\, (a \,b + x^2)(a\,b+\xi^2) }}
\ee
and  $\xi(\tau,x)$ can be defined in terms of the function $w(x)$ introduced in (\ref{w-function-def}) as follows
\be
\label{xi-def}
\xi(\tau,x) 
=
\frac{ (b-a)\big(e^{2\pi \tau + w(x)}-1\big) 
+
\sqrt{(b-a)^2\big(e^{2\pi \tau + w(x)}-1\big)^2 + 4 ab \,\big(e^{2\pi \tau + w(x)}+1\big)^2}
}{
2\,\big(1+e^{2\pi \tau + w(x)}\big)}\,.
\ee
This function describes the modular evolution (parameterised by $\tau\in \RR$) of any point $x \in A$.
It satisfies
\be
\label{x-tilde-xi-tilde}
\xi(0,x) =x \qquad 
\xi(\tau,x) \in (a,b)\qquad  
\textrm{and} \qquad
\xi(-\tau, \tilde{x}) = \frac{a\, b}{\xi(\tau,x)} \equiv \tilde{\xi}(\tau,x)\,.
%\equiv \tilde{\xi}(\tau, x) 
\ee
Notice that, since $x=\sqrt{ab}$ is self-conjugate under (\ref{xtilde}), the second expression in (\ref{x-tilde-xi-tilde})
implies that $\xi(\tau,\sqrt{ab}\,)\,\xi(-\tau,\sqrt{ab}\,) = a b$ for any $\tau \in \RR$.

In Fig.\,\ref{figure-xi}, 
we show the arguments of the fields  in the r.h.s.'s of (\ref{psi-mod-flow-ab})
in the half-plane parameterised by $(\xi, \tau)$, by employing (\ref{xi-def}).
We consider a point $x\in (a,b)$ at $\tau=0$, whose conjugate point is $\tilde{x}$.
The blue curves represent the arguments $\xi(\tau,x)$ and $ab/\xi(\tau,x)$ occurring in the first equation of (\ref{psi-mod-flow-ab}),
while the red curves correspond to $\xi(-\tau,x)$ and $ab/\xi(-\tau,x)$, 
which appear in the second equation of (\ref{psi-mod-flow-ab}).
At $\tau=0$, the solid and the dashed curves pass through $x$ and $\tilde{x}$ respectively. 
A solid curve and the corresponding dashed one having the same colour intersect at one of the points
whose coordinates are $(\xi , \tau) = (\sqrt{ab} \,, \pm \tau_0)$, 
depending on the position of $x$ with respect to $\sqrt{ab}$,
where $\tau_0$ is obtained by solving $\xi(\tau_0, x) =\sqrt{ab}$.
The value of $\tau_0$ can be determined from
(\ref{xi-def-general}), (\ref{int-disjoint-bdy-special-case}) and (\ref{w-function-def}),
finding $2\pi \tau_0 = | w(\sqrt{ab}\,) -w(x) | = | w(x) |$ because  $w(\sqrt{ab}\,) =0$.

We find it instructive to write the explicit expressions of the solution (\ref{psi-mod-flow-ab}) in the two inequivalent phases by using (\ref{fn1}).
In the vector phase, the modular flow of the massless Dirac field is
\be
\label{lambda-mod-flow-ab}
\left\{ \begin{array}{l}
\displaystyle 
\lambda_1(\tau,x)
=
\left[
P(\xi;  x) 
\left(
 \big( a \,b + x \,\xi \big) \,\lambda_1(\xi) 
- \frac{a\,b}{\xi}\; e^{\textrm{i}\alpha_{\textrm{\tiny v}}} \big(\xi - x\big)\, \lambda_2(ab/\xi)
\right)
\right]
\! \bigg|_{\xi = \xi(\tau,x)}
\\
\rule{0pt}{.85cm}
\displaystyle
\lambda_2(\tau,x)
=
\left[
P(\xi;  x) 
\left(
 \big( a \,b + x \,\xi \big) \,\lambda_2(\xi) 
- \frac{a\,b}{\xi}\; e^{-\textrm{i}\alpha_{\textrm{\tiny v}}} \big(\xi - x\big)\, \lambda_1(ab/\xi)
\right)\right]
\! \bigg|_{\xi = \xi(-\tau,x)} 
\end{array}
\right.
\,.
\ee
This modular flow mixes fields with different chiralities, but the electric charge is preserved.
The mixing is non-local because it involves a field in $\xi$ and another one in its conjugate point $ab/\xi$. 

In the axial phase, the modular flow reads
\be
\label{chi-mod-flow-ab}
\left\{ \begin{array}{l}
\displaystyle 
\chi_1(\tau,x)
=
\left[
P(\xi;  x) 
\left(
 \big( a \,b + x \,\xi \big) \,\chi_1(\xi) 
- \frac{a\,b}{\xi}\; e^{-\textrm{i}\alpha_{\textrm{\tiny a}}} \big(\xi - x\big)\, \chi^*_2(ab/\xi)
\right)
\right]
\! \bigg|_{\xi = \xi(\tau,x)}
\\
\rule{0pt}{.85cm}
\displaystyle
\chi_2(\tau,x)
=
\left[
P(\xi;  x) 
\left(
 \big( a \,b + x \,\xi \big) \,\chi_2(\xi) 
- \frac{a\,b}{\xi}\; e^{-\textrm{i}\alpha_{\textrm{\tiny a}}} \big(\xi - x\big)\, \chi^*_1(ab/\xi)
\right)\right]
\! \bigg|_{\xi = \xi(-\tau,x)} 
\end{array}
\right.
\,.
\ee
This bi-local modular flow mixes fields with different electric charge, but the chirality is preserved,
contrary to (\ref{lambda-mod-flow-ab}).

%\newpage
%%%%%%%%%%%%%%%%%%%%%%%%%%%%%%%%%%%%%%%%%%%
\section{Correlation functions along the modular flow}
\label{sec_correlators}

%\noindent

The correlation functions of the fermion fields $\lambda_r (\tau, x)$ and $\chi_r(\tau,x)$, 
evolving trough the modular flow generated by the modular Hamiltonian $K_A$, describe the quantum fluctuations 
along the modular ``time" $\tau$. They can be obtained 
by first employing either (\ref{lambda-mod-flow-ab}) or (\ref{chi-mod-flow-ab}) and then 
by writing the initial data in the Fock representation of the algebras ${\cal A}_+$ and ${\cal B}_+$,
i.e. using either (\ref{npsi1hl}) and (\ref{npsi2hl}) or (\ref{npsisc1}) and (\ref{npsisc2}) for the initial field configurations.

The derivation significantly simplifies by adopting the identity $w(\xi(\tau,x)) = 2\pi\, \tau +w(x)$ written in the exponential form
instead of the cumbersome explicit expression (\ref{xi-def}).

We find it convenient to introduce the following distribution 
\be
\label{Wfunc}
W(\tau; x, y) 
\equiv
\frac{e^{w(x)} - e^{w(y)}}{2\pi \textrm{i} (x - y)}\;
\frac{1}{e^{w(x)+\pi \tau} - e^{w(y)-\pi \tau} - \textrm{i} \varepsilon} 
\ee
where the first factor in the r.h.s. is regular at $x=y$; indeed, from (\ref{w-function-def}) we have
\be
\frac{e^{w(x)} - e^{w(y)}}{x - y}
\,=\,
\frac{2(b-a) \, (x \, y + a\, b)}{(b-x)(x + a)\, (b-y)(y + a)}\,.
\ee
We remark that (\ref{Wfunc}) satisfies
\be
\label{Widentity}
W(\tau\pm \textrm{i} \,; x, y) 
=
W(-\tau ; y, x) 
=
\overline{W(\tau ; x, y) }\,.
\ee
By using (\ref{w-function-def}), we find that for the interval $A \subset \RR_+$ the distribution (\ref{Wfunc}) becomes
\be
\label{W-mod-function-def}
W(\tau; x, y) 
=
\frac{(b-a)\, (a\,b + x\, y)}{\pi \textrm{i}\,\big[(x-a)(b+x)\, (b-y)(y+a)\, e^{\pi \tau} - (y-a)(b+y)\, (b-x)(x+a)\, e^{-\pi \tau} - \textrm{i}  \varepsilon\big]}\,.
\ee

The distribution (\ref{Wfunc}) represents the modular counterpart of (\ref{not2}). 
Indeed, analogously to the conventional time evolution, where all the correlators 
(\ref{n11}), (\ref{n22}), (\ref{n12}) and (\ref{n21})  are written in terms of the distribution (\ref{not2}), 
we find that all the non-vanishing correlators of the fields along the modular flow can be expressed 
 in terms of (\ref{Wfunc}).

After some algebra, we find that the eight non-vanishing two-point functions in the vector phase can be written as follows
\bea
\label{corr-11-mod}
\langle \lambda_1(\tau_1,x_1)\,\lambda_1^*(\tau_2,x_2)\rangle 
&=&
\langle \lambda^\ast_1(\tau_1,x_1)\,\lambda_1(\tau_2,x_2)\rangle 
=
W(\tau_{12};x_1,x_2) 
\\
\label{corr-22-mod}
\langle \lambda_2(\tau_1,x_1)\,\lambda_2^*(\tau_2,x_2)\rangle 
&=&
\langle \lambda^\ast_2(\tau_1,x_1)\,\lambda_2(\tau_2,x_2)\rangle 
=  
W(\tau_{12};x_2,x_1) 
\\
\label{corr-mixed1-mod}
\langle \lambda_1(\tau_1,x_1)\,\lambda_2^*(\tau_2,x_2)\rangle 
&=&
\overline{\langle \lambda_2(\tau_2,x_2)\,\lambda_1^*(\tau_1,x_1)\rangle}
= 
e^{\textrm{i} \alpha_{\textrm{\tiny v}}} W(\tau_{12};x_1,-x_2)
\\
\label{corr-mixed2-mod}
\langle \lambda_1^\ast(\tau_1,x_1)\,\lambda_2(\tau_2,x_2)\rangle 
&=&
\overline{ \langle \lambda_2^\ast(\tau_2,x_2)\,\lambda_1(\tau_1,x_1)\rangle }
=
e^{-\textrm{i} \alpha_{\textrm{\tiny v}}} \;
W(\tau_{12};x_1,-x_2)
\eea
where we have adopted the notation
\be 
\tau_{12} \equiv \tau_1 - \tau_2 \,.
\ee

The vacuum expectation values (\ref{corr-11-mod}), (\ref{corr-22-mod}), (\ref{corr-mixed1-mod}) and (\ref{corr-mixed2-mod})
satisfy some basic properties.
First, the Hilbert space structure of the theory provides the following relations
\bea
\langle \,\lambda^\ast_{r_1}(\tau_1,x_1)\, \lambda_{r_2}(\tau_2,x_2)\, \rangle
&=&
\overline {\langle \,\lambda^\ast_{r_2}(\tau_2,x_2)\, \lambda_{r_1}(\tau_1,x_1)\, \rangle }
\\
\langle \,\lambda_{r_1}(\tau_1,x_1)\, \lambda^*_{r_2}(\tau_2,x_2)\, \rangle
&=&
\overline {\langle \,\lambda_{r_2}(\tau_2,x_2)\, \lambda^*_{r_1}(\tau_1,x_1)\, \rangle }\,.
\eea
Second, they obey the modular equations of motion following from (\ref{system-partial differential equation-Lambda-2comps}) when $x_1 \neq x_2$.
For instance, we have that
\be
\label{modeq}
\left [\frac{1}{2\pi}\,\partial_{\tau_1} - B_{\textrm{\tiny loc}}(x_1) \right ]\langle \lambda_1(\tau_1,x_1)\,\lambda_1^*(\tau_2,x_2)\rangle 
= 
- \,e^{\textrm{i} \alpha_{\textrm{\tiny v}}} \beta_{\textrm{\tiny bi-loc}}(x_1)  \langle \lambda_2(\tau_1,{\tilde x}_1)\,\lambda_1^*(\tau_2,x_2)\rangle \,.
\ee
By exploiting (\ref{corr-11-mod}) and (\ref{corr-mixed1-mod}), 
this equation is equivalent to the following one satisfied by (\ref{Wfunc}) (in the limit $\varepsilon \to 0$)
\be
\label{Weq}
\left [\frac{1}{2\pi}\,\partial_{\tau} - B_{\textrm{\tiny loc}}(x) \right ]W(\tau;x,y) 
+
 \beta_{\textrm{\tiny bi-loc}}(x) \, W(\tau;-{\tilde x},y) 
\,=\,
0\,.
\ee
Notice that the partial differential equation (\ref{modeq}) involves both $x$ and $\tilde x$, 
keeping trace of the bi-local character of the modular evolution. 
We remark that equations similar to (\ref{modeq}) hold for all correlation functions in 
(\ref{corr-11-mod}), (\ref{corr-22-mod}), (\ref{corr-mixed1-mod}) and (\ref{corr-mixed2-mod}).
Their validity provides a valuable consistency check of the whole construction.

A fundamental property satisfied by the correlation functions 
(\ref{corr-11-mod}), (\ref{corr-22-mod}), (\ref{corr-mixed1-mod}) and (\ref{corr-mixed2-mod})
is the Kubo-Martin-Schwinger (KMS) condition \cite{Haag-book}
\bea
\label{KMS1}
\langle \,\lambda_{r_1}(\tau_1,x_1)\, \lambda_{r_2}^*(\tau_2+\tau+\textrm{i},x_2)\, \rangle
&=&
\langle \,\lambda_{r_2}^*(\tau_2+\tau,x_2)\,\lambda_{r_1}(\tau_1,x_1)\,  \rangle
\\
\label{KMS2}
\langle \,\lambda^\ast_{r_1}(\tau_1,x_1)\, \lambda_{r_2}(\tau_2+\tau+\textrm{i},x_2)\, \rangle
&=&
\langle \,\lambda_{r_2}(\tau_2+\tau,x_2)\,\lambda^\ast_{r_1}(\tau_1,x_1)\,  \rangle
\eea
where $r_1, r_2 \in \{1,2\}$,
whose validity is a consequence of the first equality in (\ref{Widentity}).
From these KMS relations, one infers that the expectation values 
(\ref{corr-11-mod}), (\ref{corr-22-mod}), (\ref{corr-mixed1-mod}) and (\ref{corr-mixed2-mod})
behave like thermal correlators with inverse temperature $\beta =1$ in our units. 
We remark that the KMS condition is a distinguishing property of the modular group
(see Theorem 1.2 in chapter VIII of \cite{takesaki-book});
hence their validity provides a strong check of our results.

We find worth observing that the mixed correlators (\ref{corr-mixed1-mod}) and  (\ref{corr-mixed2-mod}) vanish at the self-conjugate point 
$x_1 = x_2 = \sqrt {ab}$, as it an be verified by using (\ref{W-mod-function-def}).
This implies that at this point the states 
$\lambda_1(\tau_1,\sqrt{ab}\,) \,\Omega$ and $\lambda_2(\tau_2,\sqrt{ab}\,) \,\Omega$, where $\Omega $ is the Fock vacuum, are 
orthogonal for any $\tau_1,\tau_2 \in \RR$.

Notice that the two properties in (\ref{Widentity}) hold for a distribution $\mathcal{W}(\tau; x, y) $ of the form
\be
\label{Wfunc-g}
\mathcal{W}(\tau; x, y) 
=
\frac{g(x,y)}{2\pi \textrm{i} \,\big(e^{w(x)+\pi \tau} - e^{w(y)-\pi \tau} - \textrm{i} \varepsilon\big)} 
\ee
where $w(x)$ and $g(x,y)$ are real functions and $g(y,x) =g(x,y)$.
This class of distributions includes the ones occurring in the modular correlators of the massless Dirac field 
for various interesting cases like
the bipartition of the infinite line given by a generic number of disjoint intervals when the entire system is in its ground state \cite{Longo:2009mn,Hollands:2019hje} 
and the ones discussed in the Appendix\;\ref{app-examples-mf-trans}.

The above analysis can be adapted to write the corresponding quantities for the axial phase. 
In the axial phase, the correlators of the fields along the modular flow read
\bea
\label{corr-11-mod-ax}
\langle \chi_1(\tau_1,x_1)\,\chi^\ast_1(\tau_2,x_2)\rangle =
\langle \chi^\ast_1(\tau_1,x_1)\,\chi_1(\tau_2,x_2)\rangle 
&=&
W(\tau_{12} ; x_1 , x_2)
\\
\label{corr-22-mod-ax}
\langle \chi_2(\tau_1,x_1)\,\chi_2^*(\tau_2,x_2)\rangle 
=
\langle \chi^\ast_2(\tau_1,x_1)\,\chi_2(\tau_2,x_2)\rangle 
&=&  
W(\tau_{12} ; x_2 , x_1)
\\
\label{corr-mixed1-mod-ax}
\langle \chi^\ast_1(\tau_1,x_1)\,\chi_2^*(\tau_2,x_2)\rangle 
=
\overline{\langle \chi_2(\tau_2,x_2)\,\chi_1(\tau_1,x_1)\rangle}
&=& 
e^{\textrm{i} \alpha_{\textrm{\tiny a}}} \,  W(\tau_{12} ; x_1 , -x_2)
\\
\label{corr-mixed2-mod-ax}
\langle \chi_1(\tau_1,x_1)\,\chi_2(\tau_2,x_2)\rangle 
=
\overline{ \langle \chi_2^\ast(\tau_2,x_2)\,\chi_1^\ast(\tau_1,x_1)\rangle }
&=&
e^{-\textrm{i} \alpha_{\textrm{\tiny a}}} \,  W(\tau_{12} ; x_1 , -x_2)\,.
\eea
The crucial difference with respect to the vector phase is that
the mixed correlation functions (\ref{corr-mixed1-mod-ax}) and (\ref{corr-mixed2-mod-ax})
are not invariant under the vector phase transformation (\ref{v1}).

The correlators (\ref{corr-11-mod-ax}), (\ref{corr-22-mod-ax}), (\ref{corr-mixed1-mod-ax}) and (\ref{corr-mixed2-mod-ax})
satisfy the KMS condition, namely
\bea
\label{KMS1-ax}
\langle \,\chi_{r}(\tau_1,x_1)\, \chi_{r}^*(\tau_2+\tau+\textrm{i},x_2)\, \rangle
&=&
\langle \,\chi_{r}^*(\tau_2+\tau,x_2)\,\chi_{r}(\tau_1,x_1)\,  \rangle
\\
\label{KMS2-ax}
\langle \,\chi_{r_1}(\tau_1,x_1)\, \chi_{r_2}(\tau_2+\tau+\textrm{i},x_2)\, \rangle
&=&
\langle \,\chi_{r_2}(\tau_2+\tau,x_2)\,\chi_{r_1}(\tau_1,x_1)\,  \rangle
\\
\label{KMS3-ax}
\langle \,\chi^\ast_{r_1}(\tau_1,x_1)\, \chi^\ast_{r_2}(\tau_2+\tau+\textrm{i},x_2)\, \rangle
&=&
\langle \,\chi^\ast_{r_2}(\tau_2+\tau,x_2)\,\chi^\ast_{r_1}(\tau_1,x_1)\,  \rangle\,.
\eea

The correlators along the modular flow provide
the symmetry content along the modular evolution. 
%%%
The invariance of (\ref{corr-11-mod})-(\ref{corr-mixed2-mod}) and 
of (\ref{corr-11-mod-ax})-(\ref{corr-mixed2-mod-ax})
under translations in the modular parameter $\tau_i \longmapsto \tau_i + \tau_0$ implies the 
conservation of the modular ``energy". 
%%%
The invariance of (\ref{corr-11-mod})-(\ref{corr-mixed2-mod}) under vector transformations (\ref{v1}) 
and 
the invariance of (\ref{corr-11-mod-ax})-(\ref{corr-mixed2-mod-ax}) under axial transformations (\ref{a1}) 
lead to the existence respectively  of a conserved vector charge $Q_{\textrm{\tiny v}}$
and of a conserved axial charge $Q_{\textrm{\tiny a}}$ in the corresponding phases.

%\newpage
%%%%%%%%%%%%%%%%%%%%%%%%%%%%%%%%%%%%%%%%%%%
\section{Special bipartitions}
\label{sec_limiting-regimes}

In this section we consider some limiting regimes for the position of the interval $A =[a,b]$ on the half-line.
In all these limits, the modular flow of the Dirac field becomes local.

%%%%%%%%%%%%%%%%%%%%%%%%%%%%%%%%%
\subsection{Interval at large distance from the boundary}
\label{subsec_limit_distant}

The first case that we find worth considering is an interval of length $\ell$ at large distance from the boundary.
This limit can be performed by first setting $b=a+\ell$, $x=a+v$ with $v \in [0, \ell]$ and then sending $a\to \infty$.
In this limit, for the conjugate point we have
$\tilde{x} = a + \tilde{v} +O(1/a)$, where $\tilde{v} \equiv \ell - v$,
and the function (\ref{w-function-def}) simplifies to
\be
\label{w-function-def-large}
w(x) = 
\log \!\left(\frac{v}{\ell-v} \right).
\ee

The weight functions (\ref{beta-loc-def}) and (\ref{beta-biloc-def}) in the modular Hamiltonian become respectively 
\be
\label{betas-interval-far}
\beta_{\textrm{\tiny loc}}(x) =\beta_0(v) +O(1/a^2)
\;\;\qquad\;\;
\beta_{\textrm{\tiny bi-loc}}(x) =O(1/a)
 \;\;\qquad\;\;
\beta_0(v) \equiv \frac{v(\ell-v)}{\ell}
\ee
where one recognises that $\beta_0(v)$ is the weight function occurring in the modular Hamiltonian
of an interval of length $\ell$ in the infinite line (see (\ref{beta-loc-singleint-line})) \cite{Hislop:1981uh, Casini:2011kv}.

In order to get the modular flow of the Dirac field, one first observes that, in this limit, (\ref{xi-def}) becomes
\be
 \xi(\tau, x) 
% \,\equiv\, w^{-1}\big( w(x)+ 2\pi\,\tau\big)
\,=\,
a +  \zeta(\tau, v)  + O(1/a)
\;\;\qquad\;\;
\zeta(\tau, v) \equiv \frac{\ell\, v\, e^{2\pi \tau}}{\ell +(e^{2\pi \tau}-1) v}\,.
\ee
This allows to write the modular evolutions of the fields in this regime by specifying (\ref{ode-1-solution-bis-0}) to this case. 
The result reads
\be
\psi_1(\tau,x)  
=
\Big[
\sqrt{ \partial_v \zeta}  \; \psi_1(a+\zeta)
\Big]\!
 \Big|_{\zeta = \zeta(\tau,v)}
\;\;\qquad\;\;
 \psi_2(\tau,x)  
=
\Big[
\sqrt{ \partial_v \zeta}  \; \psi_2(a+\zeta)
\Big]\!
 \Big|_{\zeta = \zeta(-\tau,v)}.
\ee
where we used that $\tfrac{\beta_0(\zeta)}{\beta_0(v)} = \partial_v \zeta$.

The two-point functions of these fields is obtained by first observing that,
in this limiting regime, (\ref{W-mod-function-def}) becomes (we also introduce $y=a+z$)
\be
\label{W-function-interval-far}
W(\tau; x, y) 
=
\frac{\ell}{ 2\pi \textrm{i} \, \big[ (\ell-z)\, v\, e^{\pi \tau} - (\ell-v)\, z\, e^{-\pi \tau} - \textrm{i}  \varepsilon \big]}
\ee
and then employing this result in the correlators of Sec.\,\ref{sec_correlators} expressed through
this function.

%%%%%%%%%%%%%%%%%%%%%%%%%%%%%%%%%
\subsection{Interval adjacent to the boundary}
\label{subsec_limit_adjacent}

The case of an interval $A=[0,b]$ at the beginning of the half-line 
can be studied by taking the limit  $a\to 0$ in the above results. 

In this limit, the function (\ref{w-function-def}) becomes
\be
\label{w-func-adjacent}
 w(x) =  \log\!\bigg( \frac{x+b}{b-x} \bigg)
  \;\;\qquad\;\;
 x\in [0,b)
\ee
and for the weight functions (\ref{beta-loc-def}) and (\ref{beta-biloc-def}) we find
\be
\label{betas-a=0}
\beta_{\textrm{\tiny loc}}(x) \,\to\,\beta_0(x) 
\;\;\qquad\;\;
\beta_{\textrm{\tiny bi-loc}}(x) \,\to\, 0
\;\;\qquad\;\;
\beta_0(x) \equiv \frac{b^2-x^2}{2b}\,.
\ee
Thus, in this limit the bi-local operator in the modular Hamiltonian (\ref{K_A-decomposition}) vanishes 
and the remaining local term becomes the modular Hamiltonian found in \cite{Cardy:2016fqc} 
for an interval adjacent to the boundary of the half line,
specialised to the model that we are considering.

The modular flow of the Dirac field when the interval is adjacent to the boundary can be found by specifying
the analysis of the Appendix\;\ref{app-flow-local} to this case. 
In particular, $\beta_0(x) $ is the weight function obtained in (\ref{betas-a=0}).
By using (\ref{xi-def-general}),
the corresponding $ \xi(\tau, x)$ can be written in terms of $w(x)$ in (\ref{w-func-adjacent}),
where we remark that $x\in (0,b)$, hence the inverse function $w^{-1}(y)$ is defined when $y\geqslant 0$.
This result reads
\be
\label{xi-bdy}
 \xi(\tau, x) 
% \,\equiv\, w^{-1}\big( w(x)+ 2\pi\,\tau\big)
\,=\,
 b\; \frac{e^{2\pi \tau}  e^{w(x)} -1 }{e^{2\pi \tau}  e^{w(x)} + 1}
 \,=\,
 b\;\frac{x\, \cosh(\pi \tau) + b\, \sinh(\pi \tau)}{b\, \cosh(\pi \tau) + x\, \sinh(\pi \tau)} 
 \;\;\qquad\;\;
 \tau \leqslant \tau_0\,.
\ee
Notice that the limit $a \to 0$ of (\ref{xi-def}) gives (\ref{xi-bdy}) when $\tau \leqslant \tau_0$
and $\xi=0$ when $\tau \geqslant \tau_0$.

Alternatively, we can take the limit $a\to 0$ in the modular flow (\ref{psi-mod-flow-ab}), 
observing that the mixing terms vanish.
The modular flow of the Dirac field can be written through $\beta_0(x)$ in (\ref{betas-a=0}) as follows
\be
\label{psi12-mf-bdy}
\psi_1(\tau,x)  
=
\Big[
\sqrt{\partial_x \xi } \; \psi_1(\xi)
\Big]\!
 \Big|_{\xi = \xi(\tau,x)}
 \;\;\qquad\;\;
 \psi_2(\tau,x)  
=
\Big[
\sqrt{\partial_x \xi }  \; \psi_2(\xi)
\Big]\!
 \Big|_{\xi = \xi(-\tau,x)} \;.
\ee

\begin{figure}[t!]
\vspace{-.6cm}
\hspace{-.7cm}
%\begin{center}
\includegraphics[width=1.06\textwidth]{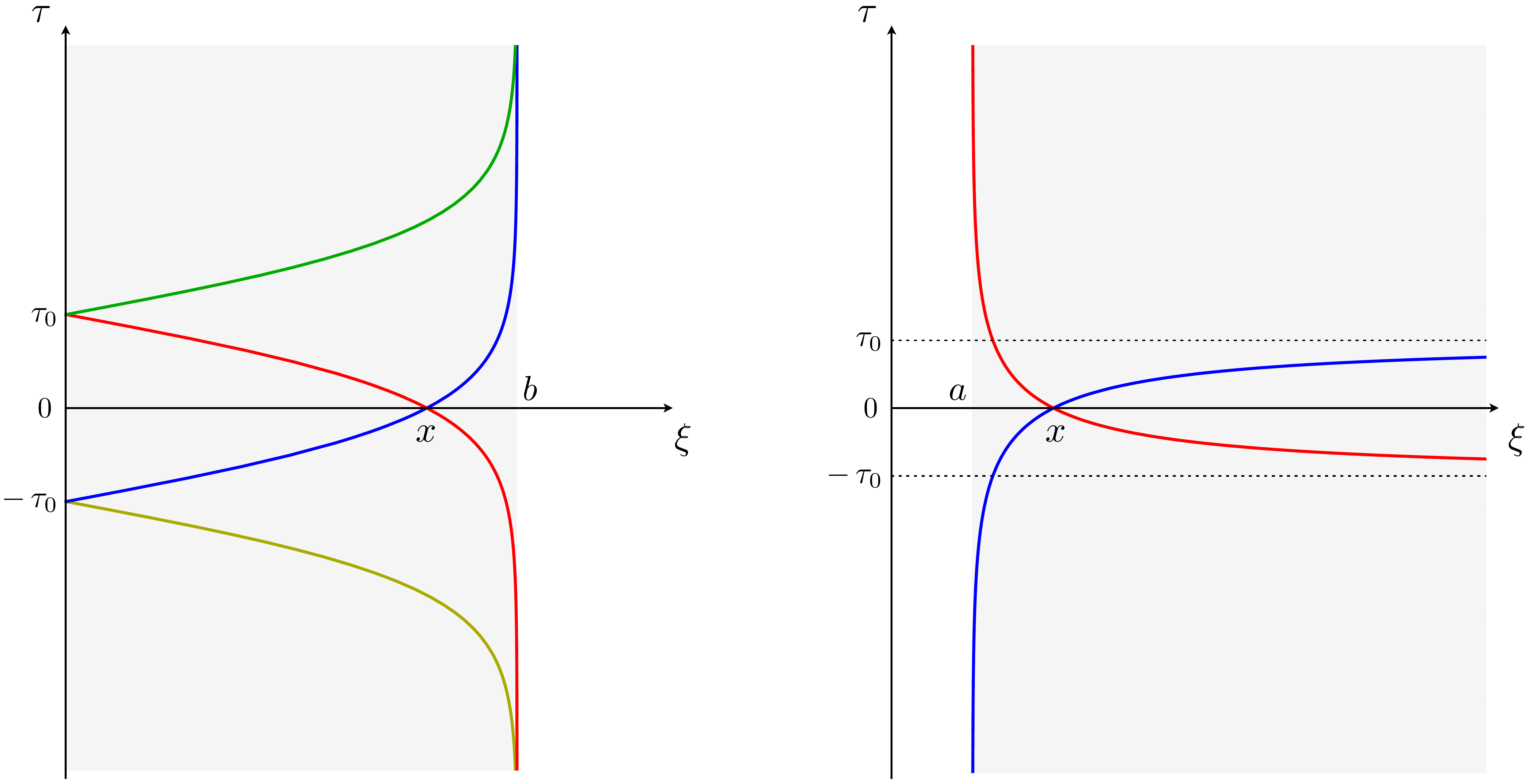}
% \end{center}
\vspace{-.25cm}
\caption{
Modular evolution through the arguments of the fields for a point $x$ at $\tau=0$
%(see the r.h.s. of \ref{psi-mod-flow-ab})
when $A=[0,b]$ (left panel, see (\ref{xi-bdy})) and when $A=[a,+\infty)$ (right panel, see (\ref{xi-semi-infinite})),
discussed in
Sec.\;\ref{subsec_limit_adjacent} and in Sec.\;\ref{subsec_limit_semi-infinite}
respectively.
}
\label{figure-limits}
\end{figure}

%In the left panel of Fig.\;\ref{figure-xi-bdy} we show the modular trajectories provided by 
Considering the arguments the fields in the r.h.s.'s of these expressions,
%The blue and the red curves that correspond respectively to the first and the second equation in (\ref{psi12-mf-bdy}),
we observe that the curves $\xi(\tau,x)$ and $\xi(-\tau,x)$ shown in the left panel of Fig.\,\ref{figure-limits}
intersect the boundary $\xi=0$ at $\tau=-\tau_0$ and $\tau=\tau_0$ respectively, where
$2\pi \tau_0 = | w(x) |$, with $w(x)$ given by (\ref{w-func-adjacent}). 
At these points either (\ref{bc5}) or (\ref{bc6}) holds, hence 
%the trajectories bounce and
the chirality of the corresponding field changes. 
Notice that $\xi \to b$ as $|\tau| \to \infty$.

The correlators of these fields can be found by first taking $a\to 0$ in (\ref{W-mod-function-def}), that gives
\be
\label{W-function-a=0}
W(\tau; x, y) 
=
\frac{b}{ \textrm{i}\pi \big[(x+b)(b-y)\, e^{\pi \tau} - (y+b)\, (b-x)\, e^{-\pi \tau} - \textrm{i}  \varepsilon\big]}
\ee
and then employing this result into the expressions in Sec.\,\ref{sec_correlators} written in terms of this function.

%\newpage
%%%%%%%%%%%%%%%%%%%%%%%%%%%%%%%%%
\subsection{Semi-infinite line separated from the boundary}
\label{subsec_limit_semi-infinite}

Another interesting bipartition of the half-line to consider is the one where 
the subsystem $A$ is the semi-infinite line separated from the boundary of the half-line at $x=0$, namely $A = [a, +\infty )$ with $a>0$.
This case corresponds to the limit $b \to +\infty $ of the results for the interval in the half-line.

In this limiting regime, the function (\ref{w-function-def}) becomes
\be
\label{w-func-semi-infinite}
 w(x) =  \log\! \bigg( \frac{x-a}{x+a} \bigg)
 \;\;\qquad\;\;
 x>a
\ee
and for the weight functions (\ref{beta-loc-def}) and (\ref{beta-biloc-def}),
occurring in the local and in the bi-local term of the  modular Hamiltonian (\ref{K_A-decomposition}),
we find respectively (see also \cite{Blanco:2013joa})
\be
\label{betas-b-infty}
\beta_{\textrm{\tiny loc}}(x) =\beta_{0}(x)  +O(1/b)
\qquad
\beta_{\textrm{\tiny bi-loc}}(x) =\frac{\beta_0(x)}{x} +O(1/b)
\qquad
\beta_{0}(x) \equiv
\frac{x^2-a^2}{2a} \,.
% \qquad
%b\to \infty
\ee
We remark that, despite the fact that $\beta_{\textrm{\tiny bi-loc}}(x) $ is non-vanishing in this regime, 
the modular Hamiltonian becomes local because $\tilde{x} \to \infty$ in this limit
and the fields $\{\psi_r(\tilde{x})\, :\, r=1,2\}$ occurring in the bi-local term (\ref{K_A-bilocal-def}) vanish (see (\ref{lim1})).

The function $\xi(\tau,x)$ can be constructed by using (\ref{w-prime-def}), (\ref{xi-def-general}) and (\ref{AB-special-case-beta0}), 
with the weight function $\beta_{0}(x)$ defined  in (\ref{betas-b-infty}). The result reads
\be
\label{xi-semi-infinite}
 \xi(\tau, x) 
% \,\equiv\, w^{-1}\big( w(x)+ 2\pi\,\tau\big)
\,=\,
 -\, a\, \frac{e^{w(x)+2\pi \tau} + 1 }{e^{w(x)+2\pi \tau} - 1}
 \,=\,
 a\, \frac{x\, \cosh(\pi \tau) -a\, \sinh(\pi \tau)}{a\, \cosh(\pi \tau) -x \, \sinh(\pi \tau)}\,.
\ee
This expression diverges when $\tau \to \tau_0$, where $\tau_0 = \tfrac{|w(x)|}{2\pi}$ and $w(x)$ is (\ref{w-func-semi-infinite}).

The  fields along the modular flow are obtained by specifying (\ref{ode-1-solution-bis-0}) to this case,
finding (\ref{psi12-mf-bdy}) with $\beta_0(x)$ and $\xi(\tau,x)$ 
given in (\ref{betas-b-infty}) and (\ref{xi-semi-infinite}) respectively, where $x>a$.
The limit $b\to +\infty$ of (\ref{psi-mod-flow-ab}) provides the same result. 

The arguments of the fields provided by the r.h.s. of (\ref{ode-1-solution-bis-0}) give the curves shown in the right panel of Fig.\;\ref{figure-limits}.
The blue and the red curves are $ \xi(\tau, x) $ and $ \xi(-\tau, x) $ respectively and they intersect at $(\xi , \tau)=(x,0)$.
Furthermore, they diverge as $\tau \to \tau_0$ or $\tau \to - \tau_0$ respectively,
where $2\pi \tau_0 = | w(x) |$, with $w(x)$ given by (\ref{w-func-semi-infinite}). 
The occurrence of $\tau_0$ is due to the presence of the boundary.
Indeed, for the semi-infinite line in the line \cite{Bisognano:1975ih,Bisognano:1976za},
one finds that $\xi(\tau,x) = x\, e^{2\pi \tau}$,
that diverges as $\tau \to +\infty$ (see also the corresponding comment in the Appendix\;\,\ref{app-examples-mf-trans}).
Thus, the internal dynamics of a semi-infinite line depends also on its complement.

The correlators of the fields along this modular flows can be written by employing the results 
discussed in Sec.\,\ref{sec_correlators} and observing that the function $W(\tau; x,y)$ in this regime 
can be obtained by plugging (\ref{w-func-semi-infinite}) into (\ref{Wfunc}), that gives
\be
\label{W-function-b=infty}
W(\tau; x, y) 
=
\frac{a}{ \pi \textrm{i}  \,\big[(x-a)(y+a)\, e^{\pi \tau} - (y-a)\, (x+a)\, e^{-\pi \tau} - \textrm{i}  \varepsilon\big]}\,.
\ee

Combining the modular Hamiltonian of this limiting regime and the one obtained 
in the Sec.\;\ref{subsec_limit_adjacent} for the interval adjacent to the boundary,
we can write the full modular Hamiltonian $K_{A\cup B}$ \cite{Haag-book} associated to
the bipartition $A \cup B$ of the half-line where $A = [0,\ell]$ is an interval of length $\ell$ adjacent to the boundary 
and $B = [\ell, +\infty)$ is its complement.
It reads
\be
\label{K-full-def}
K_{A\cup B} = K_A \otimes \boldsymbol{1}_B - \boldsymbol{1}_A \otimes K_B
\ee
where $\boldsymbol{1}_A$ and $\boldsymbol{1}_B$ denote the identity operators on $A$ and $B$ respectively, 
while
\be
\label{K-full-terms}
K_A = 2\pi \int_0^\ell \frac{\ell^2-x^2}{2\ell} \; T_{tt}(0,x)\, \rd x 
\;\;\qquad\;\;
K_B = 2\pi \int_\ell^\infty \frac{x^2-\ell^2}{2\ell}  \; T_{tt}(0,x)\, \rd x 
\ee
are the modular Hamiltonians respectively of the interval $A = [0,\ell]$ adjacent to the boundary 
and of its complement, that is a semi-infinite line at distance $\ell$ from the boundary.

Let us remark that, considering the system on the line, in its ground state
and the bipartition $A \cup B$ of the line where the finite subsystem is the interval $A =(-\ell, \ell)$,
the full modular Hamiltonian is (\ref{K-full-def}) where $K_A$ and $K_B$ are very similar 
to the operators reported in (\ref{K-full-terms})\footnote{The modular Hamiltonian $K_B$ can be found as the limit $b \to +\infty$ of $K_{A_{\textrm{\tiny sym}}}$.}.
Indeed, only the integration domains are different, which are $A =(-\ell, \ell)$ for $K_A$ and 
$B =(-\infty, -\ell) \cup(\ell, +\infty)$ for $K_B$, as expected. 

%\newpage

%%%%%%%%%%%%%%%%%%%%%%%%%%%%%%%%%%%%%%%%%%
\section{Modular evolution in the spacetime}
\label{sec-mod-ev-part1}

The modular evolution whose initial field configurations (at $\tau=0$)
are either (\ref{npsi1hl}) and (\ref{npsi2hl}) in the vector phase or (\ref{npsisc1}) and (\ref{npsisc2}) in the axial phase at $t=0$
has been described in Sec.\,\ref{sec_mod_flow}.
It is worth investigating the most general modular evolution where also the physical time $t$ is involved. 

For the massless Dirac field, the physical time can be included 
by exploiting the fact that each component $\psi_r$ of the Dirac field
depends only on one of the light-cone coordinates 
\be
\label{lc1}
u_\pm \equiv x\pm t  
\;\;\qquad\;\; 
x\geqslant 0 \, .  
\ee

In these coordinates and on the half line, $\psi_r$ satisfy the following anti-commutation relations 
\bea
\label{A1}
& &
[\psi_1 (u_+)\, ,\, \psi^*_1 (v_+)]_+ = \delta (u_+ - v_+) 
\hspace{1.4cm}
[\psi_2 (u_-)\, , \,\psi^*_2 (v_-)]_+ = \delta (u_- - v_-)  \qquad 
\\
\label{A2}
& &
[\psi_1 (u_+)\, ,\, \psi^*_2 (v_-)]_+=\e^{\ri \alpha}\delta (u_+ + v_-)
\hspace{.95cm}
[\psi_2 (u_-)\, ,\, \psi^*_1 (v_+)]_+=\e^{-\ri \alpha}\delta (u_- + v_+)\, .  
\eea

\begin{figure}[t!]
\vspace{-.7cm}
\hspace{1.6cm}
%\begin{center}
\includegraphics[width=.8\textwidth]{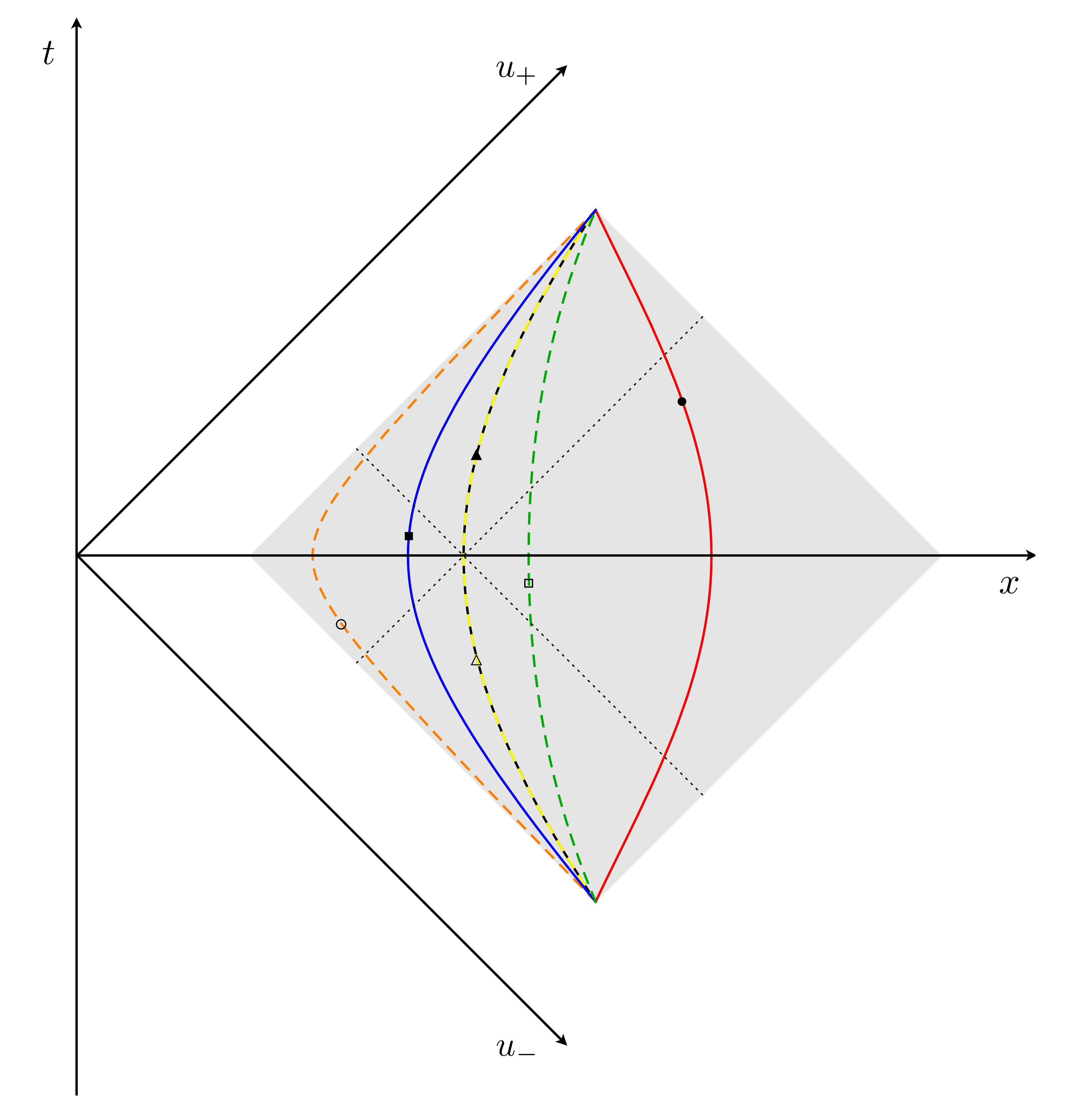}
% \end{center}
\vspace{-0cm}
\caption{ 
Three pairs of modular trajectories. 
Each pair is made by conjugate trajectories whose initial points at $\tau =0$ 
are indicated through the same symbol (filled or empty).
}
\label{figure-diamond-bdy}
\end{figure}

Here we consider two intervals $u_\pm \in [a\, , b]$, which parameterise the grey diamond $\mathcal{D}$ 
shown in Fig.\,\ref{figure-diamond-bdy}.
By employing  the spectral problem discussed in Sec.\;\ref{sec_spectral_problem} in the form 
\begin{equation} 
\frac{1}{2\pi \ri} 
\left(\!
\begin{array}{cc} 
\displaystyle \int_a^b \frac{\rd v_+}{u_+-v_+} \;\;\;& \displaystyle \e^{\ri \alpha } \! \int_a^b \frac{\rd v_-}{u_++v_-}
\\ 
\rule{0pt}{.9cm}
\displaystyle -\e^{-\ri \alpha}\int_a^b \frac{\rd v_+}{u_-+v_+} \;\;\;\;& \displaystyle -\int_a^b \frac{\rd v_-}{u_--v_-} 
\\ 
\end{array} \right)
\left (\begin{array}{c} \e^{\ri \alpha }\phi_{s,p}(v_+)
\\ 
\phi_{s,p}(-v_-)\end{array}\right ) = 
\sigma_s \left (\begin{array}{c} \e^{\ri \alpha }\phi_{s,p}(u_+)\\ \phi_{s,p}(-u_-)\end{array}\right )
\label{A3}
\end{equation}  
and repeating the analysis described in Sec.\,\ref{sec_eh},
we obtain the modular Hamiltonian $K_A$  in the coordinates $u_\pm$.
It can be decomposed again as
\be
\label{K_A-decomposition-lc}
K  \,= \, K^{\textrm{\tiny loc}} + K^{\textrm{\tiny bi-loc}} 
\ee 
where the local term is
\be
\label{K_A-local-def-lc}
K^{\textrm{\tiny loc}} 
\,=\,
2\pi  \int_a^b \! \beta_{\textrm{\tiny loc}}(u_+) \, T_{(1)}(u_+)\, \rd u_+
-
2\pi  \int_a^b \! \beta_{\textrm{\tiny loc}}(u_-) \, T_{(2)}(u_-)\, \rd u_-
\ee
with 
\be
T_{(r)}(u) 
\,\equiv\,
\,\frac{\textrm{i}}{2}
:\! \!
\Big[(\partial_u \psi^\ast_r)\, \psi_r - \psi^\ast_r\, (\partial_u \psi_r) \Big]\!\! : \! (u) 
\ee
while the bi-local term can be written by introducing the $\tilde{u}_\pm \equiv a b/ u_\pm$ conjugate to $u_\pm$ as follows 
\be
\label{K_A-bilocal-def-lc}
K^{\textrm{\tiny bi-loc}}  
\,=\,
2\pi \int_a^b \! \beta_{\textrm{\tiny bi-loc}}(u_+) \, T_{\textrm{\tiny bi-loc}}(u_+, \tilde{u}_+;\alpha) \, \rd u_+
+
2\pi \int_a^b \! \beta_{\textrm{\tiny bi-loc}}(u_-) \, T_{\textrm{\tiny bi-loc}}(u_-, \tilde{u}_-;\alpha) \, \rd u_-
\ee
which contains the bi-local hermitean operator defined as follows
\be
\label{T-bilocal-def-lc}
T_{\textrm{\tiny bi-loc}}(u, v ;\alpha) 
\equiv 
\frac{\textrm{i}}{2}\,
\!:\!\!\Big[\, e^{\textrm{i} \alpha}\, \psi^\ast_1(u) \,  \psi_2(v) - e^{-\textrm{i} \alpha} \,\psi^\ast_2(v) \,  \psi_1(u) \Big]\!\!: \,.
\ee
The weight functions $\beta_{\textrm{\tiny loc}}(z)$ and $\beta_{\textrm{\tiny bi-loc}}(z)$ 
are defined in (\ref{beta-loc-def}) and (\ref{beta-biloc-def}) respectively. 
Notice that, restricting to the slice defined by constant $t=0$, 
one finds that  (\ref{K_A-local-def-lc}) and (\ref{K_A-bilocal-def-lc}) become (\ref{K_A-local-def}) and (\ref{K_A-bilocal-def}) respectively.

The modular flow is defined by 
\bea
\label{lc-mod-evolution1}
\textrm{i}\,\frac{ d\psi_1(\tau,u_+)}{d\tau}  = \big[\,K \,, \psi_1(\tau,u_+)\,\big]_{-}
\;\; \qquad \;\; \psi_1(0,u_+) = \psi_1(u_+) 
\\
\rule{0pt}{.7cm}
\label{lc-mod-evolution2}
\textrm{i}\,\frac{ d\psi_2(\tau,u_-)}{d\tau}  = \big[\,K \,, \psi_2(\tau,u_-)\,\big]_{-}
\;\; \qquad \;\;
\psi_2(0,u_-) = \psi_2(u_-) 
\eea
where $u_\pm \in [a,b]$ and, according to (\ref{fn1}), the initial configurations $\psi_1(u_+)$ and $\psi_2(u_-)$ 
in the vector and axial phases are given by (\ref{npsi1hl})-(\ref{npsi2hl}) and (\ref{npsisc1})-(\ref{npsisc2}) respectively. 
Using that the mixed anti-commutators (\ref{A2}) vanish for $u_\pm, v_\pm \in [a,b]$, 
we can analyse (\ref{lc-mod-evolution1}) and (\ref{lc-mod-evolution2}) and solve them exactly as in Sec.\,\ref{sec_mod_flow}, finding
\begin{equation}
\label{lc4}
\left\{ \begin{array}{l}
\displaystyle
\psi_1(\tau,u_+)
=
\left[
P(\xi;  u_+) 
\left(
 \big( a \,b + \xi \,u_+ \big) \,\psi_1(\xi) 
- \frac{a\,b}{\xi}\; e^{\textrm{i}\alpha} \big(\xi - u_+\big)\, \psi_2(ab/\xi)
\right)
\right]
\! \bigg|_{\xi = \xi(\tau,u_+)}
\\
\rule{0pt}{.85cm}
\displaystyle
\psi_2(\tau,u_-)
=
\left[
P(\xi;  u_-) 
\left(
 \big( a \,b + \xi \,u_- \big) \,\psi_2(\xi) 
- \frac{a\,b}{\xi}\; e^{-\textrm{i}\alpha} \big(\xi - u_-\big)\, \psi_1(ab/\xi)
\right)\right]
\! \bigg|_{\xi = \xi(-\tau,u_-)}
\end{array}
\right.
\end{equation}
where $P(\xi;  u_\pm)$ has been defined in (\ref{P-function-def}).
At $t=0$ the expressions in (\ref{psi-mod-flow-ab}) are recovered. 

The solution (\ref{lc4}) for the modular flow of the Dirac field allows us to 
write the coordinates of the generic point along the modular trajectory in the spacetime
\be
\label{mod-trajec-1int}
x(\tau) = \frac{\xi(\tau, p_{+,0}) + \xi(-\tau, p_{-,0})}{2}
\;\;\qquad\;\;
t(\tau) = \frac{\xi(\tau, p_{+,0}) - \xi(-\tau, p_{-,0})}{2}
\;\;\qquad\;\;
\tau \in \RR
\ee
where $p_{\pm, 0}$ denote the light-cone coordinates of the initial point (at $\tau=0$) of the modular trajectory. 
In Fig.\,\ref{figure-diamond-bdy} we show some modular trajectories obtained through (\ref{mod-trajec-1int}).
A filled marker denotes an initial points $p_{\pm,0} = u_{\pm,0}$,
while the empty version of the same marker denotes the conjugate initial point such that $p_{\pm,0} = ab/u_{\pm,0}$.
These are the initial points of two conjugate modular trajectories (a solid curve and a dashed curve in Fig.\,\ref{figure-diamond-bdy})
that determine the modular evolution of the field in the spacetime. 
In Fig.\,\ref{figure-diamond-bdy} three pairs of modular trajectories are shown. 
Notice that the modular trajectory passing through the point whose light-cone coordinates are $u_\pm =\sqrt{ab}$
coincides with its conjugate trajectory (see the solid black curve and the yellow dashed curve in Fig.\,\ref{figure-diamond-bdy}).
Changing the initial point, the resulting pairs of modular trajectories span the entire grey diamond $\mathcal{D}$ in Fig.\,\ref{figure-diamond-bdy}.
Any modular trajectory begins in the lower vertex of $\mathcal{D}$ as $\tau \to -\infty$ and ends into the 
upper vertex of $\mathcal{D}$ as $\tau \to+\infty$.

The correlation functions along the modular flow
in the light cone coordinates can be obtained from (\ref{lc4}).
By adopting the notation $u_{r\pm} \in [a,b]$ with $r\in \{1,2\}$ for the two points, 
in the vector phase we find 
\bea
\label{corr-11-mod-lc}
\langle \lambda_1(\tau_1,u_{1+})\,\lambda_1^*(\tau_2,u_{2+})\rangle 
&=&
\langle \lambda^\ast_1(\tau_1,u_{1+})\,\lambda_1(\tau_2,u_{2+})\rangle 
=
W(\tau_{12};u_{1+},u_{2+}) 
\\
\label{corr-22-mod-lc}
\langle \lambda_2(\tau_1,u_{1-})\,\lambda_2^*(\tau_2,u_{2-})\rangle 
&=&
\langle \lambda^\ast_2(\tau_1,u_{1-})\,\lambda_2(\tau_2,u_{2-})\rangle 
=  
W(\tau_{12};u_{2-},u_{1-}) 
\\
\label{corr-mixed1-mod-lc}
\langle \lambda_1(\tau_1,u_{1+})\,\lambda_2^*(\tau_2,u_{2-})\rangle 
&=&
\overline{\langle \lambda_2(\tau_2,u_{2-})\,\lambda_1^*(\tau_1,u_{1+})\rangle}
= 
e^{\textrm{i} \alpha_{\textrm{\tiny v}}} \, W(\tau_{12};u_{1+},-u_{2-})
\\
\label{corr-mixed2-mod-lc}
\langle \lambda_1^\ast(\tau_1,u_{1+})\,\lambda_2(\tau_2,u_{2-})\rangle 
&=&
\overline{ \langle \lambda_2^\ast(\tau_2,u_{2-})\,\lambda_1(\tau_1,u_{1+})\rangle }
=
e^{-\textrm{i} \alpha_{\textrm{\tiny v}}} \,
W(\tau_{12};u_{1+},-u_{2-})\,.
\eea
In the axial phase, this analysis leads to
\bea
\label{corr-11-mod-ax-pm}
\langle \chi_1(\tau_1, u_{1+})\,\chi^\ast_1(\tau_2,u_{2+})\rangle =
\langle \chi^\ast_1(\tau_1,u_{1+})\,\chi_1(\tau_2,u_{2+})\rangle 
&=&
W(\tau_{12} ; u_{1+} , u_{2+})
\\
\label{corr-22-mod-ax-pm}
\langle \chi_2(\tau_1,u_{1-})\,\chi_2^*(\tau_2,u_{2-})\rangle 
=
\langle \chi^\ast_2(\tau_1,u_{1-})\,\chi_2(\tau_2,u_{2-})\rangle 
&=&  
W(\tau_{12} ; u_{2-} , u_{1-})
\\
\label{corr-mixed1-mod-ax-pm}
\langle \chi^\ast_1(\tau_1,u_{1+})\,\chi_2^*(\tau_2,u_{2-})\rangle 
=
\overline{\langle \chi_2(\tau_2,u_{2-})\,\chi_1(\tau_1,u_{1+})\rangle}
&=& 
e^{\textrm{i} \alpha_{\textrm{\tiny a}}} \,  W(\tau_{12} ; u_{1+} , -u_{2-})
\\
\label{corr-mixed2-mod-ax-pm}
\langle \chi_1(\tau_1,u_{1+})\,\chi_2(\tau_2,u_{2-})\rangle 
=
\overline{ \langle \chi_2^\ast(\tau_2,u_{2-})\,\chi_1^\ast(\tau_1,u_{1+})\rangle }
&=&
e^{-\textrm{i} \alpha_{\textrm{\tiny a}}} \,  W(\tau_{12} ; u_{1+} , -u_{2-})\,.
\eea

\section{Conclusions}
\label{sec_conclusions}

In this manuscript we have studied the modular Hamiltonians of an interval $A=[a,b]\subset \RR_+$
for the boundary conformal field theory defined by the massless Dirac fermion on the half-line,
in both the inequivalent phases that can be introduced by imposing the global conservation of energy,
where either the charge or the helicity is preserved. 
These modular Hamiltonians have been obtained in terms of the components of the Dirac field 
as the sum of the local term (\ref{K_A-local-def}), where the energy density (\ref{T00-lambda-def}) occurs, 
and the bi-local terms given by (\ref{K_A-bilocal-def}) and (\ref{T-bilocal-def}).
Each bi-local term involves fields that must be evaluated in two conjugate points 
$x$ and $\tilde{x} =ab/x$ within the interval $A$. Moreover, 
the modular Hamiltonians preserve the global symmetry characterising the phase, that is either 
the vector phase transformation (\ref{v1}) or the axial phase transformation (\ref{a1}).
In this respect we recall that
the bi-local terms in the modular Hamiltonians for the massless Dirac field defined on translationally invariant spaces \cite{Casini:2009vk, Klich:2015ina, Klich:2017qmt, Blanco:2019xwi,Fries:2019ozf}
preserve both vector and axial symmetry.

Another main result of this manuscript are the modular flows of the Dirac field,
given by (\ref{lambda-mod-flow-ab}) in the vector phase and by (\ref{chi-mod-flow-ab}) in the axial phase.
These flows mix two modular trajectories that start at conjugate points at $\tau=0$ (see Fig.\;\ref{figure-xi}) 
of fields having different chirality in the vector phase and different charge in the axial phase. 
By employing these results, we have found the correlation functions of the fields along the modular flow, 
which are (\ref{corr-11-mod})-(\ref{corr-mixed2-mod}) in the vector phase and 
(\ref{corr-11-mod-ax})-(\ref{corr-mixed2-mod-ax}) in the axial phase.
These correlators satisfy the KMS condition, 
which characterises the modular flow \cite{takesaki-book}.
Furthermore, these correlation functions 
satisfy the modular equation of motion determined by the modular flow (see e.g. (\ref{modeq})).  
The above results, first obtained at $t=0$ have been also extended in the diamond region shown in Fig.\,\ref{figure-diamond-bdy}.

As for the entanglement entropies of the interval on the half-line, the expression (\ref{ent-ent-bdy}) has been found
and, by considering the special case where the interval is adjacent to the boundary,  
the Affleck-Ludwig boundary entropy $g=1$ has been obtained, in agreement with \cite{Affleck:1991tk,Smith:2019jnh}.

In the model considered in this manuscript only reflection occurs at the boundary.
In a companion manuscript \cite{MT20def}, the results obtained here are employed
to study the modular Hamiltonians of equal disjoint intervals on the line placed symmetrically with respect to
a defect that allows both reflection and transmission.

It would be interesting to find the modular Hamiltonians of bipartitions involving many disjoint intervals
for other conformal field theory models 
(see \cite{Caraglio:2008pk,Furukawa:2008uk,  Calabrese:2009ez, Calabrese:2010he,Coser:2013qda, DeNobili:2015dla}
for the entanglement entropies).
In \cite{Arias:2016nip,Eisler:2019rnr,DiGiulio:2019cxv,Eisler:2020lyn}
some entanglement hamiltonians for free quantum field theories have been obtained 
as the continuum limit of the corresponding entanglement hamiltonians of the lattice model
\cite{Peschel:2003rdm,EislerPeschel:2009review,Casini:2009sr,Eisler:2017cqi}.
The contours of the entanglement entropies are also interesting related quantities to consider \cite{ChenVidal2014,Coser:2017dtb}.
It would be instructive to study also the modular Hamiltonians found in this manuscript through lattice calculations.
The modular Hamiltonians for free quantum field theories in higher dimensions and in the presence of boundaries are also natural operators to explore
(see e.g. \cite{Berthiere:2016ott} for the entanglement entropies).
It is important to study also the modular Hamiltonians in quantum field theories where the conformal symmetry does not occur
\cite{Arias:2016nip,Eisler:2020lyn}.
They could provide important insights to understand some features of the renormalisation group flows, 
even in the cases where the conformal symmetry is broken only by boundary terms \cite{Friedan:2003yc,Casini:2016fgb, Casini:2018nym}.
Finally, let us mention that it is important to address the above open problem also
in the context of the AdS/BCFT correspondence \cite{Bachas:2001vj, Azeyanagi:2007qj, Takayanagi:2011zk, Fujita:2011fp, Nozaki:2012qd,  Seminara:2017hhh, Seminara:2018pmr, Jafferis:2014lza}.

\vskip 20pt 
\centerline{\bf Acknowledgments} 
\vskip 5pt 

We are grateful to Paolo Acquistapace, Ra\'ul Arias, Stefano Bianchini, Viktor Eisler and Giuseppe Mussardo 
for insightful discussions and correspondence.

\vskip 30pt

\newpage
%%%%%%%%%%%%%%%%%%%%%%%%%%%%%%%%%%%%%%%%%%%%%%%%%%%%%

\appendix

\section{Spectral problem for two equal intervals on the line}
\label{app_2intervals}

In this appendix we report the explicit form of the solution of the spectral problem (\ref{sp1})
for the union  $A_{\textrm{\tiny sym}} \equiv [-b,-a] \cup [a,b] \subset \RR$
of two disjoint equal intervals  on the line \cite{Casini:2009vk}.

The eigenvalues in (\ref{sp1}) can be written in terms of the real parameter $s$ as follows
\be
\label{sigma-s-def}
\sigma_s = \frac{\tanh(\pi s) + 1}{2} 
\;\;\qquad\;\;
s \in \mathbb{R}\,.
\ee
The eigenfunctions in (\ref{sp1}) are
\be
\label{phi_s form disjoint}
\phi_{s,p}(x) \,=\, \Theta_A(x)\,m_p(x)\, e^{-\textrm{i}\, s\, w(x)} 
\;\;\qquad\;\;
p \in \{1,2\}
\;\;\qquad\;\;
x\in A_{\textrm{\tiny sym}}
\ee
where the function $w(x)$ occurring in the phase is  defined in (\ref{w-function-def}),
the functions $m_p(x)$ determining the amplitude are 
\be
\label{m-1-2-def}
m_1(x) = \sqrt{\frac{b(b-a)}{\pi (a+b)}} \;\frac{x-a}{\sqrt{(b^2 - x^2)(x^2 - a^2)}}
\;\qquad\;
m_2(x) = \sqrt{\frac{a(b-a)}{\pi (a+b)}} \; \frac{x+b}{\sqrt{(b^2 - x^2)(x^2 - a^2)}}
\ee
and $\Theta_A(x)$ can be written in terms of the Heaviside step function $\Theta$ as follows
\be
 \Theta_A(x)
 =
 \Theta(x+b)\, \Theta(-a-x) -  \Theta(x-a)\, \Theta(b-x) \,.
\ee
This expression is non vanishing only in $A_{\textrm{\tiny sym}}$.
In particular, $ \Theta_A(x)=+1$ when $x\in (-b,-a)$ 
and $ \Theta_A(x)=-1$ when $x\in (a,b)$.

%%%%%%%%%%%%%%%%%%%%%%%%%%%%%%%%%%%%%%%%%%%
%\newpage

\section{Modular flow from a shift operator}
\label{app-shift-operator}

In this Appendix we discuss the solution of the partial differential equation 
that allows to determine the modular flows reported in Sec.\,\ref{sec_mod_flow}.
In the Appendix\;\ref{app-shift-operator-general} the solution of the general case is derived
and in the Appendix\;\ref{app-flow-local} we describe
its application to some cases characterised by local modular Hamiltonians.

\subsection{General case}
\label{app-shift-operator-general}

The general form of the partial differential equation underlying the modular flows discussed in this manuscript reads
\be
\label{pde-gen-AB}
\partial_{\mathsf{t}} \psi(\mathsf{t},x)
=
V(x) \, \partial_x \psi(\mathsf{t},x) + Y(x)\, \psi(\mathsf{t},x) 
\;\;\qquad\;\;
\psi(0,x) = \psi(x)
\ee
where $\psi(x)$ corresponds to the given initial configuration of the field.

In order to solve (\ref{pde-gen-AB}), first one identifies 
a proper multiplicative factor 
that simplifies the form of the equation.
In particular, let us redefine the field $\psi(\mathsf{t},x) $ as 
\be
\label{psiVpsi-def}
\psi(\mathsf{t},x)  = A(x) \, \Psi(\mathsf{t},x) 
\ee
where $A(x)$ satisfies the ordinary differential equation $V  A' + Y A   = 0$,
which  can be also written as $\tfrac{d}{dx} \log A = -\,Y / V$,
whose solution reads
\be
\label{Vpsi-def}
A(x) \equiv A(x_0)\, e^{-\gamma(x;x_0)}
\;\;\qquad\;\;
\gamma(x;x_0) \equiv \int_{x_0}^x \frac{Y(y)}{V(y)}\, \rd y\,.
\ee
Plugging (\ref{psiVpsi-def}) into (\ref{pde-gen-AB}), 
the partial differential equation for $\Psi(\mathsf{t},x)$ becomes
\be
\label{pde-gen-AB-Psi}
\partial_{\mathsf{t}} \Psi(\mathsf{t},x)
\,=\,
V(x) \, \partial_x \Psi(\mathsf{t},x) 
\;\;\qquad\;\;
\Psi(0,x) = \frac{\psi(x)}{A(x)} \equiv \Psi(x)
\ee
where $\Psi(x)$ provides the known initial configuration. 
When $Y(x)$ vanishes identically, we have that $A(x) =1$ identically.

The unique solution of the partial differential equation (\ref{pde-gen-AB-Psi}) can be written in terms of the 
shift operator $e^{\mathsf{t}\, V(x)\, \partial_x}$ as follows \cite{hamermesh-book}
\be
\label{Psi-soln}
\Psi(\mathsf{t},x) 
=
e^{\mathsf{t}\, V(x)\, \partial_x}\, \Psi(x) 
=
e^{\mathsf{t} \,\partial_w} F(w)
=
F(w(x)+\mathsf{t})
=
\Psi \big(\xi(\mathsf{t},x)\big)
\ee
where we have introduced
\be
\label{w-prime-def}
w'(x) = \frac{1}{V(x)}
\;\;\qquad\;\;
\Psi(\zeta) = F(w(\zeta))
\ee
and
\be
\label{xi-def-general}
\xi(\mathsf{t},x)\,\equiv\, w^{-1}\big( w(x)+ \mathsf{t} \big)
\ee
which satisfies $\xi(0,x) = x$.

By employing the auxiliary variable $\zeta \equiv w(x)+\mathsf{t}$ and (\ref{w-prime-def}), 
one finds that
\be
\label{eqs-t-s-xi}
\partial_{\mathsf{t}} \xi(\mathsf{t},x) = V\big( \xi(\mathsf{t},x) \big)
\;\;\qquad\;\;
\partial_x \xi(\mathsf{t},x)  = \frac{V\big( \xi(\mathsf{t},x) \big)}{V(x)}\,.
\ee
We observe that (\ref{xi-def-general}) can be obtained also through the method of the characteristics.
This method is based on the first equation in (\ref{eqs-t-s-xi}), which gives $w(\xi) - w(\xi_0) = \mathsf{t}$,
once combined with (\ref{w-prime-def}), where $w(\xi_0)= w(\xi)|_{ \mathsf{t}=0}$ and $\xi|_{ \mathsf{t}=0} = x$.
Thus, $w(\xi)  = w(x) + \mathsf{t}$, which is equivalent to (\ref{xi-def-general}).

Finally, from (\ref{psiVpsi-def}), (\ref{Vpsi-def}) and (\ref{pde-gen-AB-Psi}), we can construct the solution of (\ref{pde-gen-AB}) as follows
\be
\psi(\mathsf{t},x)  
=
A(x_0)\, e^{-\gamma(x;x_0)}\,
\Psi \big(\xi(\mathsf{t},x)\big)
=
A(x_0)\, e^{-\gamma(x;x_0)}\,
\frac{\psi \big(\xi(\mathsf{t},x)\big)}{A\big(\xi(\mathsf{t},x)\big)}
\ee
where in the last expression we can employ (\ref{Vpsi-def}) to get
$A(\xi(\mathsf{t},x)) = A(x_0)\, e^{-\gamma(\xi(\mathsf{t},x);x_0)}$.
The final expression for  the solution of (\ref{pde-gen-AB}) can be written as 
\be
\label{gen-solution-mod-flow}
\psi(\mathsf{t},x)  
\,=\,
\frac{e^{\gamma(\xi(\mathsf{t},x);x_0)}}{e^{\gamma(x;x_0)}}\;
\psi \big(\xi(\mathsf{t},x)\big)
\,\equiv\,
e^{\Gamma(\mathsf{t},x)}\,
\psi \big(\xi(\mathsf{t},x)\big)
\ee
where
\be
\Gamma(\mathsf{t},x) = \int_{x}^{\xi(\mathsf{t},x)} \frac{Y(y)}{V(y)}\, \rd y
\ee
which is independent of $x_0$.
Since $\xi(0,x) = x$, it is straightforward to check that the solution 
(\ref{gen-solution-mod-flow}) satisfies the initial condition $ \psi(0 ,x)  = \psi(x) $,
as required in (\ref{pde-gen-AB}).

\subsection{Modular flows from local modular Hamiltonians}
\label{app-flow-local}

In various cases of physical interest 
\cite{Hislop:1981uh,Casini:2011kv,Wong:2013gua,Cardy:2016fqc},
the modular Hamiltonian of the interval $A=[a,b]$ at $t=0$ for the Dirac field reads
\be
\label{rho_a-def-norm}
K_A \,=\, 2\pi \int_A \beta_0(x) \, T_{tt}(0,x)\, \rd x
\ee
where $T_{tt}(t,x)$ is the energy density and $\beta_0(x)$ is a weight function
that characterises the underlying case.

The modular flow of the Dirac field generated by (\ref{rho_a-def-norm}) is given by (\ref{mod-evolution-lambda}).
It can be found by solving the following partial differential equation with a given initial configuration for the field
\be
\label{partial differential equation-1-beta0}
\textrm{i}\,\frac{\partial \psi(\tau,x)}{\partial \tau} 
= \big[ \, K_A\, , \,\psi(\tau,x)\,\big]_-
=\, 
\textrm{i}\, 2\pi 
\left(
\beta_0(x) \, \partial_x \psi(\tau,x) + \frac{1}{2} \, \partial_x \beta_0(x)\, \psi (\tau,x)
\right).
\ee

This differential equation corresponds to the special case of (\ref{pde-gen-AB}) where 
$V(x)$ and $Y(x)$ are obtained from weight function $\beta_0(x)$ occurring in (\ref{rho_a-def-norm}) as follows
\be
\label{AB-special-case-beta0}
\mathsf{t} = 2\pi \, \tau
\;\;\qquad\;\;
V(x) =  \beta_0(x)
\;\;\qquad\;\;
Y(x) = \frac{1}{2}\, \beta'_0(x)\,.
\ee
This implies that the modular flow $\psi(\tau,x)$ can be found by specialising to this case
the general solution (\ref{gen-solution-mod-flow}) derived in the Appendix\;\ref{app-shift-operator-general}.
The function $w(x)$ in (\ref{w-prime-def})
and therefore also $\xi(\tau,x)$ in (\ref{xi-def-general})
cannot be written without the explicit expression of $\beta_0(x)$.
Indeed, in this case the equations in (\ref{eqs-t-s-xi}) hold for $\xi(\tau,x)$ with $V(x)$ given in (\ref{AB-special-case-beta0}).

Specialising the function $\gamma(x;x_0)$ in (\ref{Vpsi-def}) to the case defined by (\ref{AB-special-case-beta0}), we obtain
\be
\gamma(x;x_0) = \frac{1}{2}\, \log\!\left( \frac{\beta_0(x)}{\beta_0(x_0)} \right)
\ee
hence the function multiplying the field in (\ref{gen-solution-mod-flow}) becomes
\be
e^{\Gamma(\tau,x)}
=
\left(
\frac{\beta_0(\xi(\tau,x))}{\beta_0(x)}
\right)^{1/2}.
\ee
Thus, the solution (\ref{gen-solution-mod-flow}) in the special case defined in (\ref{AB-special-case-beta0}) reads
\be
\label{ode-1-solution-bis-0}
\psi(\tau,x)  
\,=\,
\sqrt{ \frac{\beta_0\big( \xi(\tau,x) \big)}{\beta_0(x)} } \; \psi \big( \xi(\tau,x) \big)
 \,=\,
\sqrt{ \partial_x \xi(\tau,x) } \; \psi \big(\xi(\tau,x) \big)
\ee
where (\ref{eqs-t-s-xi}) has been used.
Since $\xi(0,x)=x$,
the initial condition $\psi(0,x)  = \psi(x)  $ is satisfied.

The solution (\ref{ode-1-solution-bis-0}) tells us that the modular flows generated by the modular Hamiltonians (\ref{rho_a-def-norm}) are local
because they do not mix fields localised in different points.

\subsubsection{Examples with translation invariance}
\label{app-examples-mf-trans}

We find worth enumerating some interesting configurations where the bipartition involves a single interval $A$ 
and whose modular Hamiltonian takes the local form (\ref{rho_a-def-norm}).
In these examples the underlying system is invariant under spatial translations.

Let us consider a configuration characterised by its weight function $\beta_0(x)$.
This weight function allows to find $w(x)$ through  (\ref{w-prime-def}) and (\ref{AB-special-case-beta0}).
Then, the function $\xi(\tau, x)$ is constructed through  (\ref{xi-def-general}) and (\ref{AB-special-case-beta0}).
%\textcolor{red}{\bf [come si fissa la costante di integrazione]}
%%
The modular flow for the massless Dirac field can be found by specialising 
the expression (\ref{ode-1-solution-bis-0}) to the case of interest. 
%%%%

In the following cases, where the underlying system is translations invariant, 
the solution (\ref{ode-1-solution-bis-0}) can be employed to find 
the correlators of the components $\psi_r$ (with $r\in \{1,2\}$) of the Dirac field along the modular flow
\be
\langle \psi_r(\tau_1,x_1)\,\psi_r^*(\tau_2,x_2)\rangle 
=
\sqrt{ \frac{\beta_0\big(\xi(\tau_1,x_1)\big)\,\beta_0\big(\xi(\tau_1,x_2)\big) }{\beta_0(x_1)\, \beta_0(x_2)} } 
\; 
\langle 
\psi \big(\xi(\tau_1,x_1)\big)\, \psi^\ast\!\big(\xi(\tau_2,x_2)\big)
\rangle\,.
\ee
More explicitly,  in the following cases one finds that this correlator takes the form (\ref{Wfunc-g})
\be
\label{Wfunc-g-local-app}
\langle \psi_r(\tau_1,x_1)\,\psi_r^*(\tau_2,x_2)\rangle 
=
W(\tau; x, y) 
=
\frac{G(x,y) \big( e^{w(x)} - e^{w(y)} \big)}{2\pi \textrm{i} \,\big(e^{w(x)+\pi \tau} - e^{w(y)-\pi \tau} - \textrm{i} \varepsilon\big)} 
\ee
where we have introduced the antisymmetric function $G(x,y)\equiv g(x,y) /[ e^{w(x)} - e^{w(y)}]$.
We observe that (\ref{Wfunc-g-local-app}) satisfies the following partial differential equation
\be
\label{Weq-local-app}
\left[\, \frac{1}{2\pi}\,\partial_{\tau} - B_\textrm{\tiny loc}(x) \right] W(\tau;x,y) =  0
\;\;\qquad\;\;
B_\textrm{\tiny loc}(x) \equiv
 \beta_0(x)\, \partial_x + \frac{1}{2}\, \partial_x \beta_0(x)
\ee
under the condition that $G(x,y)$ is a solution of (we recall that $w'(x) = 1/\beta_0(x)$)
\be
\bigg(
\partial_x 
+
\frac{1}{2} \, B_w(x,y)
\bigg)\,
G(x,y)
=0
\;\;\qquad\;\;
B_w(x,y) \equiv \frac{\beta'_0(x)}{\beta_0(x)} + \frac{e^{w(x)} + e^{w(y)}}{\big(e^{w(x)} - e^{w(y)}\big) \beta_0(x)}\,.
\ee 
In the following cases,  it turns out that
$B_w(x,y) $ is independent of the underlying bipartition 
and that $G(x,y)$ is proportional to the two-point correlator of the entire system with $x, y \in A$.

The first example that we consider is given by a system in its ground state on the line,
where the bipartition is defined by a single interval $A=[a,b] \in \RR$ and by its complement 
\cite{Hislop:1981uh, Brunetti:1992zf, Casini:2011kv}.
The modular Hamiltonian takes the local form (\ref{rho_a-def-norm}) with (see also (\ref{eigenfunc-1int}))
\be
\label{beta-loc-singleint-line}
 \beta_0(x) \equiv \frac{(b-x)(x-a)}{b-a}
 \;\;\qquad\;\;
  w(x) =  \log\!\bigg( \frac{x-a}{b-x} \bigg)\,.
\ee
The function $w(x)$ provides $\xi(\tau,x) $ as follows
\be
\label{xi-interval-line-gs}
 \xi(\tau,x) 
=
w^{-1}\big( w(x)+ 2\pi\,\tau\big)
 =
 \frac{a  + b\, e^{w(x)}\, e^{2\pi \tau}}{1 +e^{w(x)} \, e^{2\pi \tau}}
 =
 \frac{(b-x) \,a+(x-a)\, b\, e^{2\pi \tau}}{b-x +(x-a)\, e^{2\pi \tau}}
\ee
For any $x\in (a,b)$, we have that $\xi(\tau,x) \to a$ as $\tau\to -\infty$ and $ \xi(\tau,x) \to b$ as $\tau\to +\infty$ in (\ref{xi-interval-line-gs}).
Instead, $\xi(\tau,a) = a$ and $\xi(\tau,b) =b$ for any $\tau \in \RR$.
\\ 
The correlators along the modular flow are given (\ref{Wfunc-g-local-app}) with
\be
G(x,y) = \frac{1}{x-y}\,.
\ee

It is worth considering  the limiting regime of this first example where the subsystem becomes the semi-infinite line $x>0$
(i.e. $a=0$ and $b\to +\infty$), which corresponds to the case studied by Bisognano and Wichmann \cite{Bisognano:1975ih,Bisognano:1976za}.
In this limit, from  (\ref{beta-loc-singleint-line}), we have $ \beta_0(x) = x$ with $x>0$
and the expression (\ref{xi-interval-line-gs}) reduces to $ \xi(\tau,x) = x\, e^{2\pi \tau}$ with $x>0$,
which are the dilations of the semi-infinite line parameterised by $\tau$.

The expression (\ref{rho_a-def-norm}) describes also the modular Hamiltonian of an interval $A=[a,b]$ 
on a circle of finite length $L$, when the entire system is in its ground state. 
We remark that we consider anti-periodic boundary conditions.
Non local terms can occur in the modular Hamiltonian if other boundary conditions are imposed
\cite{Klich:2015ina, Klich:2017qmt, Erdmenger:2020nop}.
In this case, which has been already studied in \cite{Wong:2013gua, Cardy:2016fqc},
the weight function $\beta_0(x)$ and the  corresponding $w(x)$ read respectively
\be
 \beta_0(x) \equiv  \frac{L}{\pi}\; \frac{\sin[\pi (b-x)/L]\, \sin[\pi(x-a)/L]}{\sin[\pi(b-a)/L]}
 \qquad
 w(x) =  \log\!\bigg( \frac{\sin[\pi (x-a)/L]}{\sin[\pi (b-x)/L]} \bigg) \,.
\ee
The expression of $w(x)$ allows to write $\xi(\tau,x)$ as 
\be
 \xi(\tau,x) \,=\,
\frac{L}{2\pi\,  \textrm{i}}\, \log\! \bigg(
 \frac{e^{\textrm{i} \pi (b+a)/L} +  e^{\textrm{i} 2\pi b/L}\, e^{w(x)+2\pi \tau}
 }{ 
 e^{\textrm{i} \pi (b-a)/L}  +  e^{w(x)+2\pi \tau}}
\bigg)
\ee
and the correlators along the modular flow as (\ref{Wfunc-g-local-app}) with
\be
G(x,y) = \frac{1}{\sin[\pi(x-y)/L]}\,.
\ee

The last example is given by an interval $A=[a,b] \subset \RR$ on the infinite line 
and a system at finite temperature $1/\beta$.
In this case the modular Hamiltonian has the local form (\ref{rho_a-def-norm}) with \cite{Wong:2013gua, Cardy:2016fqc}
\be
 \beta_0(x) \equiv  \frac{\beta}{\pi}\; \frac{\sinh[\pi (b-x)/\beta]\, \sinh[\pi(x-a)/\beta]}{\sinh [\pi(b-a)/\beta]}
 \qquad
 w(x) =  \log\!\bigg( \frac{\sinh[\pi (x-a)/\beta]}{\sinh [\pi (b-x)/\beta ]} \bigg)\,.
\ee
This expression for $w(x)$ leads to the corresponding $\xi(\tau,x)$, which can be written as 
\be
 \xi(\tau,x) \,=\,
\frac{\beta}{2\pi }\, \log\! \bigg(
 \frac{e^{ \pi (b+a)/\beta} +  e^{2\pi b/\beta}\, e^{w(x)+2\pi \tau}
 }{ 
 e^{\pi (b-a)/\beta}  +  e^{w(x)+2\pi \tau}}
\bigg)
\ee
and  to the corresponding correlators along the modular flow, which are (\ref{Wfunc-g-local-app}) with
\be
G(x,y) = \frac{1}{\sinh[\pi(x-y)/\beta]}\,.
\ee

Notice that the first case can be obtained either from the second one in the limit $L \to \infty$
or from the last one in the limit $\beta \to \infty$.

\subsubsection{Modular flow of a primary field in CFT}
\label{app-examples-mf-trans-CFT}

A conformal field theory in two space-time dimensions contains the symmetric, 
conserved and traceless energy momentum tensor $\{T_{\mu \nu}(t,x) \, : \, \mu, \nu = t,x\}$. Let 
us consider
\be 
\label{chT}
T_\pm (t \pm x) = \frac{1}{2} \big[\,T_{x t}(t,x) \pm T_{tt}(t,x)\big] 
\ee 
the chiral right and left-moving combinations of the energy-momentum tensor. 
Introducing $A=[a,b]$, 
we are interested in the flows of a primary field of the theory
generated by the modular Hamiltonians 
\be
\label{rho_a-def-norm-cft}
K^\pm_A \,=\, 2\pi \int_A \beta_0(u_\pm) \, T_\pm (u_\pm )\, \rd u_\pm 
\;\;\qquad\;\; 
u_\pm = x\pm t \,.
\ee

Consider, for instance, any right-moving primary field $\phi (v_+)$ with conformal dimension $h$. 
By employing the transformation law \cite{luscher} 
\be
\big[\,T_+(u_+)\, ,\, \phi(v_+)\,\big]_{-}
\,=\,
\ri \, \Big( \delta(u_+-v_+) \,\partial_{v_+} \phi(v_+) - h\, \phi (v_+) \, \partial_{u_+} \delta (u_+-v_+) \Big)
\ee
and assuming that $\beta_0(a) = \beta_0(b) = 0$, we find  
\be
\label{diffeq}
\textrm{i}\,\partial_\tau \phi(\tau,v_+) = [ \, K^+_A\, , \,\phi(\tau,v_+)\,]_- 
=\, 
\textrm{i}\, 2\pi 
\Big(
\beta_0(v_+) \, \partial_{v_+} \psi(\tau,v_+) + h \, \, \psi (\tau,v_+) \,\partial_{v_+} \beta_0(v_+) 
\Big)
\ee 
which generalises (\ref{partial differential equation-1-beta0}) to an arbitrary conformal dimension $h$. 
Notice that the differential equation (\ref{diffeq}) determining the modular flow of $\phi$ is given by (\ref{pde-gen-AB}) with
\be
\label{AB-special-case-beta0-primary}
\mathsf{t} = 2\pi \, \tau
\;\;\qquad\;\;
V(x) =  \beta_0(x)
\;\;\qquad\;\;
Y(x) = h\, \beta'_0(x)\,.
\ee
Specifying the solution (\ref{gen-solution-mod-flow}) to this case, one obtains  
\be
\label{ode-1-solution-bis-0-primary}
\phi(\tau,v_+)  
=
\bigg[\, \frac{\beta_0\big( \xi(\tau,v_+) \big)}{\beta_0(v_+)} \,\bigg]^h  \phi \big( \xi(\tau,v_+) \big) 
=
\left [\partial_{v_+} \xi(\tau,v_+) \right ]^h \phi \big(\xi(\tau,v_+) \big) \, .
\ee
The modular evolution of a left-moving primary field of dimension $h$ is obtained from 
(\ref{ode-1-solution-bis-0-primary}) by replacing $v_+$ with $v_-$. 

We remark that the explicit form of the weight function $\beta_0(x)$ has not been specified in the above discussion. 
For the special case of a two dimensional conformal field theory in its ground state and 
the bipartition defined by the single interval $A=[a,b]$, the modular trajectory $\xi$ is given by (\ref{xi-interval-line-gs})
and the result (\ref{ode-1-solution-bis-0-primary}) has been already reported in \cite{Hislop:1981uh, Brunetti:1992zf, Hollands:2019hje}.

%%%%%%%%%%%%%%%%%%%%%%%%%%%%%%%%%%%%%%%%%%%
%\newpage

\section{Modular flows from bi-local modular Hamiltonians}
\label{app-bilocal-eh}

In this Appendix we describe the derivation 
of the modular flows of the massless Dirac field generated by two different bi-local modular Hamiltonians.
In the Appendix\;\ref{app-details-mod-flow} the case of the interval in the half-line is considered,
obtaining (\ref{psi-mod-flow-ab}),
while in Appendix\;\ref{app-details-mod-flow-2int} we discuss the case given by 
two disjoint equal intervals on the line, that has been solved in \cite{Casini:2009vk}.
In the Appendix\;\ref{app_pde_mc_2int} we report a partial differential equation 
satisfied by the correlators of the Dirac field along the modular flow generated by
the modular Hamiltonian of two disjoint intervals in a generic configuration on the line.

\subsection{Interval on the half-line}
\label{app-details-mod-flow}

In the following we discuss the derivation of (\ref{psi-mod-flow-ab}) as the solution of the 
system of the two coupled partial differential equations  (\ref{system-partial differential equation-Lambda-2comps})
with the initial condition (\ref{initialcond}).

By using (\ref{beta-loc-xtilde}) to find $- \beta_{\textrm{\tiny loc}}(\tilde{x})  \partial_{\tilde{x}} = \beta_{\textrm{\tiny loc}}(x) \,  \partial_{x} $,
the system (\ref{system-partial differential equation-Lambda-2comps}) can be written as
\be
\label{mod-flow-vec-2}
\bigg[ \,\frac{d}{d\tau} - 2\pi\, \beta_{\textrm{\tiny loc}}(x) \,  \partial_{x}  \bigg] \Psi(\tau,x) \,=\, 2\pi\, \boldsymbol{M}(x)\,\Psi(\tau,x)
\;\;\qquad\;\;
\Psi(\tau=0,x) \equiv \Psi(x)
\ee
where 
\be
\label{mixing-matrix-bdy}
\boldsymbol{M}(x) 
\equiv 
\left(\,
\begin{array}{cc}
\tfrac{1}{2}\,\partial_x\beta_{\textrm{\tiny loc}}(x)\;  & -\,e^{\textrm{i} \alpha}  \beta_{\textrm{\tiny bi-loc}}(x) 
\\
 \rule{0pt}{.5cm} 
e^{-\textrm{i} \alpha}  \beta_{\textrm{\tiny bi-loc}}(\tilde{x}) \; & -\tfrac{1}{2}\,\partial_{\tilde{x}} \beta_{\textrm{\tiny loc}}(\tilde{x}) 
\end{array} \, \right) 
\ee
which provides the coupling between the two equations. 

In order to write the system (\ref{mod-flow-vec-2}) in a simpler form, 
we redefine the fields in $\Psi(\tau,x)$ as
\be
\label{Mcal-def-bdy}
\Psi(\tau, x) = 
\boldsymbol{\mathcal{M}}(x)\, \tilde{\Psi}(\tau, x)
\;\;\qquad\;\;
\boldsymbol{\mathcal{M}}(x) 
\equiv
\bigg(
\begin{array}{cc}
c(x) & 0
\\
\rule{0pt}{.3cm}
0 & c(\tilde{x}) 
\end{array}  \bigg)
\ee
in terms of the function $c(x)$ satisfying the condition 
\be
\label{f-condition-der}
\beta_{\textrm{\tiny loc}}(x)\, c'(x)  = \, -\,\frac{1}{2}\,c(x)\,\partial_x\beta_{\textrm{\tiny loc}}(x) 
\ee
which leads to 
\be
\label{f-def-beta-loc}
 c(x) = \frac{c_0}{\sqrt{\beta_{\textrm{\tiny loc}}(x)}}
\ee
where $c_0$ is a  non vanishing constant. 
By plugging (\ref{Mcal-def-bdy}) into (\ref{mod-flow-vec-2}) and exploiting (\ref{f-condition-der}), one finds that
$\tilde{\Psi}(\tau, x)$ must solve the following system
\be
\label{mod-flow-vec-2-tilde}
\bigg[\, \frac{d}{d\tau} - 2\pi\, \beta_{\textrm{\tiny loc}}(x) \,  \partial_{x} \, \bigg] \tilde{\Psi}(\tau,x) 
\,=\, 
2\pi\, b(x)\, J_\alpha
\,\tilde{\Psi}(\tau,x)
%\;\;\qquad\;\;
%\tilde{\Psi}(\tau=0,x) = \tilde{\Psi}_0(x)
\ee
where we have introduced 
\be
\label{b-function-def}
b(x)
\,\equiv\,
\frac{c(\tilde{x})}{c(x)}\; \beta_{\textrm{\tiny bi-loc}}(x)
\,=\,
\frac{\sqrt{\beta_{\textrm{\tiny loc}}(x)\, \beta_{\textrm{\tiny loc}}(\tilde{x})}}{x+\tilde{x}}
\,=\,
b(\tilde{x})
\,=\,
\frac{\sqrt{a\,b} \, (b^2-x^2) \,(x^2-a^2)}{2\,(b-a)\, (a\,b + x^2)^2}
\ee
which is independent of the constant $c_0$ and 
the constant matrix $J_\alpha $ can be written in terms of the Pauli matrices $\boldsymbol{\sigma}_1$ and $\boldsymbol{\sigma}_2$ as follows
\be
J_\alpha \equiv
\bigg(
\begin{array}{cc}
0 &  -\,e^{\textrm{i} \alpha} 
\\
e^{-\textrm{i} \alpha} & 0
\end{array} \bigg)
=\,
-\,\textrm{i} \,\big[\,
\boldsymbol{\sigma}_1 \sin(\alpha) + \boldsymbol{\sigma}_2 \cos(\alpha) 
\big]\,.
\ee
The unitary matrix $J_\alpha$, that satisfies also $J_\alpha^{-1} = -J_\alpha$, 
provides the dependence on $\alpha$ in (\ref{mod-flow-vec-2-tilde}).
This matrix can be diagonalised through the unitary matrix $U_\alpha$, namely
\be
\label{J-alpha-diagonalised}
J_\alpha 
\,=\,
U_\alpha^{-1}
\bigg(\begin{array}{cc}
\textrm{i} \;\, & 0 \\
0 \;\, & -\textrm{i}
\end{array}\bigg)
\, U_\alpha
\;\;\qquad\;\;
U_\alpha 
\equiv
\frac{1}{\sqrt{2}}\,
\bigg(\begin{array}{cc}
- \textrm{i}\, e^{-\textrm{i} \alpha} \;\, & 1 \\
\textrm{i}\, e^{-\textrm{i} \alpha}  \;\, & 1
\end{array}\bigg)\,.
\ee

We find it convenient to introduce the following linear combinations of fields 
\be
\label{mu-tilde-fields-def}
\left(
\begin{array}{c}
\tilde{\mu}_1(\tau,x) 
\\  
\rule{0pt}{.45cm} 
\tilde{\mu}_2(\tau,\tilde{x}) 
\end{array}  
\right) 
\equiv\,
U_\alpha
\left(
\begin{array}{c}
\tilde{\psi}_1(\tau,x) 
\\  
\rule{0pt}{.45cm} 
\tilde{\psi}_2(\tau,\tilde{x}) 
\end{array}  
\right) \,.
\ee
Indeed, by employing (\ref{J-alpha-diagonalised}) and (\ref{mu-tilde-fields-def}) into (\ref{mod-flow-vec-2-tilde}), 
we observe that the partial differential equations for the fields $\tilde{\mu}_1(\tau,x) $ and $\tilde{\mu}_2(\tau,\tilde{x}) $ are decoupled.
They read
\be
\label{system-pde-mutilde}
\left\{
\begin{array}{l}
\displaystyle
\bigg[\, \frac{d}{d\tau} - 2\pi\, \beta_{\textrm{\tiny loc}}(x) \,  \partial_{x} \, \bigg] 
\tilde{\mu}_1(\tau,x) 
=\, 2\pi \, \textrm{i}\, b(x)\, \tilde{\mu}_1(\tau,x) 
\\
\rule{0pt}{.8cm}
\displaystyle
\bigg[\, \frac{d}{d\tau} + 2\pi\, \beta_{\textrm{\tiny loc}}(\tilde{x}) \,  \partial_{\tilde{x}} \, \bigg] 
\tilde{\mu}_2(\tau,\tilde{x}) 
=\, -\,2\pi \, \textrm{i}\, b(\tilde{x})\, \tilde{\mu}_2(\tau,\tilde{x}) 
\end{array}
\right.
\ee
where the second equation tells us that $\tilde{x}$ is a convenient spatial variable for $\tilde{\mu}_2$.

Since one of the equation in (\ref{system-pde-mutilde}) is obtained from the other one 
by exchanging $(\tau, x) \leftrightarrow (-\tau, \tilde{x})$ and $\tilde{\mu}_1 \leftrightarrow \tilde{\mu}_2$,
the solution of this system can be constructed from the solution of a single partial differential equation given by
\be
\label{pde-model-mu-tilde}
\bigg[\, \frac{d}{d\tau} - 2\pi\, \beta_{\textrm{\tiny loc}}(x) \,  \partial_{x} \, \bigg] 
\tilde{\mu}(\tau,x) 
=\, 2\pi \, \textrm{i}\, b(x)\, \tilde{\mu}(\tau,x) \,.
\ee

This partial differential equation belongs to the class defined by (\ref{pde-gen-AB}), whose solution has been 
derived in the Appendix\;\ref{app-shift-operator-general}.
In particular, (\ref{pde-model-mu-tilde}) corresponds to the case defined by 
\be
\label{int-disjoint-bdy-special-case}
\mathsf{t} = 2\pi \, \tau
\;\;\qquad\;\;
V(x) =  \beta_{\textrm{\tiny loc}}(x)
\;\;\qquad\;\;
Y(x) =  \textrm{i}\, b(x)\,.
\ee
Following the discussion reported in the Appendix\;\ref{app-shift-operator-general}, one first introduces $\xi(\tau,x)$ 
from (\ref{xi-def-general}), that in this case becomes
\be
\label{xi-def-w}
\xi(\tau,x) 
\,=\,
w^{-1}  \big(   w(x) + 2\pi \, \tau\big)
\ee
where $w(x)$ is given by (\ref{w-function-def}) and 
its explicit expression of $\xi(\tau,x) $ has been reported in (\ref{xi-def}).
Then, one specialises the integral in (\ref{Vpsi-def}) to the case defined by (\ref{int-disjoint-bdy-special-case}), finding
\be
\int \! \frac{b (\xi)}{\beta_{\textrm{\tiny loc}}(\xi)}\, d\xi
=  \omega(\xi)  +\textrm{const}
\;\;\qquad\;\;
\omega(\xi) 
\equiv 
 \textrm{arctan}\big(\xi/\sqrt{ab}\, \big)\,.
\ee
This allows to write the solution of (\ref{pde-model-mu-tilde}) by specifying (\ref{gen-solution-mod-flow}) to this case. 
The result reads
\be
\label{soln-mu-tilde}
\tilde{\mu}\big(\tau,\xi(\tau,x)\big)
=
\frac{e^{\textrm{i}  \,\omega(\xi(\tau, x)) }}{e^{\textrm{i}  \,\omega(\xi(0, x) ) }}
\, \tilde{\mu}\big(\xi(\tau,x)\big)
\ee
where $ \tilde{\mu}(x)$ corresponds to the initial configuration at $\tau=0$
and we have adapted the notation of (\ref{gen-solution-mod-flow}) to this case.

The result (\ref{soln-mu-tilde}) allows to find the solution of the system of decoupled partial differential equations in (\ref{system-pde-mutilde}) as
\be
\label{mu-tilde-vec-soln}
\left(\,
\begin{array}{c}
\tilde{\mu}_1(\tau,x) 
\\  
\rule{0pt}{.45cm} 
\tilde{\mu}_2(\tau,\tilde{x}) 
\end{array}  
\right) 
=
\left(
\begin{array}{c}
e^{\textrm{i}\,\Delta\omega (\xi(\tau, x) )} 
\, \tilde{\mu}_{1}\big(\xi(\tau,x)\big)
\\  
\rule{0pt}{.45cm} 
e^{\textrm{i}\,\Delta\omega (\xi(-\tau, \tilde{x}) )}
\, \tilde{\mu}_{2}\big(\xi(-\tau,\tilde{x})\big)
\end{array}  
\right) 
\ee
where the function $\Delta \omega (\xi(\tau_0, x))$, that must satisfy $\Delta \omega(\xi( 0 , x))=0$
to match the initial configuration, is defined as
\be
\label{Delta-omega-def}
\Delta \omega \big(\xi(\tau_0, x)\big)
\equiv 
\omega\big(\xi(\tau_0, x) \big)- \omega\big(\xi(0, x) \big)
=
\omega\big(\xi(\tau_0, x) \big)- \omega(x)\,.
\ee

By first combining (\ref{mu-tilde-fields-def}) and (\ref{mu-tilde-vec-soln}) and  then using (\ref{x-tilde-xi-tilde}), 
it is straightforward to obtain
\be
\label{lambda-tilde-soln-same-arg}
\left(
\begin{array}{c}
\tilde{\psi}_1(\tau,x) 
\\  
\rule{0pt}{.45cm} 
\tilde{\psi}_2(\tau,\tilde{x}) 
\end{array}  
\right) 
=\, 
U_\alpha^{-1}
\left(
\begin{array}{cc}
e^{\textrm{i}\,\Delta\omega (\xi(\tau, x) )} & 0
\\  
\rule{0pt}{.45cm} 
0 & e^{\textrm{i}\,\Delta\omega (\tilde{\xi}(\tau, x))}
\end{array}  
\right) 
U_\alpha
\left(
\begin{array}{c}
\tilde{\psi}_{1}\big(\xi(\tau,x)\big)
\\  
\rule{0pt}{.45cm} 
 \tilde{\psi}_{2}\big(\tilde{\xi}(\tau, x)\big)
\end{array}  
\right) .
\ee
Since $\textrm{arctan}(1/y) = \tfrac{\pi}{2} - \textrm{arctan}(y)$ for $y>0$, we have
\be
\label{xi-tilde-Delta-omega}
\Delta \omega \big(\tilde{\xi}(\tau_0, x)\big)
=
\omega\big(\tilde{\xi}(\tau_0, x) \big)- \omega(\tilde{x})
=
-\,
\Delta \omega \big(\xi(\tau_0, x)\big)
\ee
hence (\ref{lambda-tilde-soln-same-arg}) can be written as 
\be
\label{soln-lambda-tilde}
\left(
\begin{array}{c}
\tilde{\psi}_1(\tau,x) 
\\  
\rule{0pt}{.45cm} 
\tilde{\psi}_2(\tau,\tilde{x}) 
\end{array}  
\right) 
=
\left(\,
\begin{array}{cc}
 \cos\!\big[\Delta\omega \big(\xi(\tau, x) \big)\big]   &  - \,e^{\textrm{i} \alpha}  \sin\!\big[\Delta\omega \big(\xi(\tau, x) \big)\big] 
\\
\rule{0pt}{.5cm} 
e^{- \textrm{i} \alpha}  \sin\!\big[\Delta\omega \big(\xi(\tau, x) \big)\big]    &  \cos\!\big[\Delta\omega \big(\xi(\tau, x) \big)\big] 
\end{array} \,\right)
\,
\left(
\begin{array}{c}
\tilde{\psi}_{1}\big(\xi(\tau,x)\big)
\\  
\rule{0pt}{.45cm} 
 \tilde{\psi}_{2}\big(\tilde{\xi}(\tau,x)\big)
\end{array}  
\right) .
\ee

The unitary $2\times 2$ matrix in the r.h.s. provides the mixing of different fields along the modular flow.
Setting $\theta = \Delta\omega (\xi(\tau, x) )$, this matrix can be written in terms of the Pauli matrices $\boldsymbol{\sigma}_1$ 
and $\boldsymbol{\sigma}_2$ and of the $2 \times 2$ identity matrix $\mathbb{I}$ as follows
\be
\label{exp-J-alpha}
e^{-J_\alpha \theta}
\,=\,
 \cos (\theta)\, \mathbb{I}
-
\textrm{i}\, \sin (\theta) \big[  \boldsymbol{\sigma}_1 \sin(\alpha) + \boldsymbol{\sigma}_2 \cos(\alpha) \big] 
\,=\,
\bigg(
\begin{array}{cc}
 \cos (\theta)   &  - \,e^{\textrm{i} \alpha} \sin (\theta) 
\\
e^{- \textrm{i} \alpha} \sin (\theta)    &  \cos (\theta)
\end{array} \bigg)\,.
\ee
Notice that $e^{-J_\alpha \theta}$ becomes orthogonal when $\alpha =0$.

The modular evolution of the doublet (\ref{Lambda-doublet-def}) is obtained by 
combining  the fields redefinition (\ref{Mcal-def-bdy}) and (\ref{soln-lambda-tilde}).
The result reads
\be
\label{soln-lambda}
\left(
\begin{array}{c}
\psi_1(\tau,x) 
\\  
\rule{0pt}{.45cm} 
\psi_2(\tau,\tilde{x}) 
\end{array}  
\right) 
=\,
\boldsymbol{\mathfrak{M}}(x)
\left(
\begin{array}{c}
\psi_{1}\big(\xi(\tau,x)\big)
\\  
\rule{0pt}{.45cm} 
 \psi_{2}\big(\tilde{\xi}(\tau,x)\big)
\end{array}  
\right) 
\ee
where
\be
\label{Mfrak-def-alpha}
\boldsymbol{\mathfrak{M}}(x)
\equiv\,
\boldsymbol{\mathcal{M}}(x)
\left(\,
\begin{array}{cc}
 \cos\!\big[\Delta\omega \big(\xi(\tau, x) \big)\big]   &  - \,e^{\textrm{i} \alpha}  \sin\!\big[\Delta\omega \big(\xi(\tau, x) \big)\big] 
\\
\rule{0pt}{.5cm} 
e^{- \textrm{i} \alpha}  \sin\!\big[\Delta\omega \big(\xi(\tau, x) \big)\big]    &  \cos\!\big[\Delta\omega \big(\xi(\tau, x) \big)\big] 
\end{array} \,\right)
\,\boldsymbol{\mathcal{M}} \big(\xi(\tau, x) \big)^{-1}\,.
\ee

We can rewrite (\ref{soln-lambda}) considering $\psi_2(\tau,x)  $ instead of $\psi_2(\tau,\tilde{x})  $ in the second component of (\ref{soln-lambda}).
By first using (\ref{x-tilde-xi-tilde}) and then (\ref{xi-tilde-Delta-omega}), one obtains
\be
\left\{ 
\begin{array}{l}
\displaystyle
\psi_1(\tau,x)  
=
c(x) \!
\left(
 \cos\!\big[\Delta\omega \big(\xi(\tau, x) \big)\big]  \,
  \frac{\psi_{1}\big(\xi(\tau,x)\big)}{c\big(\xi(\tau,x)\big)}
-  
e^{\textrm{i} \alpha} 
\sin\!\big[\Delta\omega \big( \xi(\tau, x) \big)\big] \,
   \frac{\psi_{2}\big(\tilde{\xi} (\tau, x)\big)}{ c\big(\tilde{\xi}(\tau, x)\big)}
\right)
\\
\rule{0pt}{1cm}
\displaystyle
\psi_2(\tau,x)  
=
c(x) \!
\left(
 \cos\!\big[\Delta\omega \big(\xi(-\tau, x) \big)\big]  \,
   \frac{\psi_{2}\big(\xi (-\tau, x)\big)}{ c\big(\xi(-\tau, x)\big)}
-
e^{-\textrm{i} \alpha} 
\sin\!\big[\Delta\omega \big( \xi(-\tau, x) \big)\big] \,
   \frac{\psi_{1}\big(\tilde{\xi} (-\tau, x)\big)}{ c\big(\tilde{\xi}(-\tau, x)\big)}
 \right)
\end{array}\right.
\ee
which can be rewritten by observing that (\ref{beta-loc-xtilde}) and (\ref{f-def-beta-loc}) give $c(\tilde{\xi}) = \tfrac{\xi}{\sqrt{ab}}\, c(\xi)$, and that
\be
\label{cos-sin-Delta}
 \cos\!\big[\Delta\omega \big(\xi(\pm\tau, x) \big)\big]
=
\frac{a\,b+\xi\, x}{\sqrt{(a\,b+\xi^2)\,(a\,b+x^2)}}
\qquad
 \sin\!\big[\Delta\omega \big(\xi(\pm\tau, x) \big)\big]
=
\frac{\sqrt{a\,b}\, (\xi - x)}{\sqrt{(a\,b+\xi^2)\,(a\,b+x^2)}} 
\ee
where $\xi = \xi(\pm\tau, x)$ in the r.h.s.'s.
This rewriting leads to the final result reported in (\ref{psi-mod-flow-ab}).

Finally, let us remark that (\ref{soln-lambda}) and (\ref{Mfrak-def-alpha}) lead to
\be
\label{soln-lambda-psi2-rescaled}
\left(
\begin{array}{c}
\psi_1(\tau,x) 
\\  
\rule{0pt}{.45cm} 
e^{\textrm{i} \alpha} \,\psi_2(\tau,\tilde{x}) 
\end{array}  
\right) 
=\,
\boldsymbol{\mathfrak{M}}(x)\big|_{\alpha=0}
\left(
\begin{array}{c}
\psi_{1}\big(\xi(\tau,x)\big)
\\  
\rule{0pt}{.45cm} 
e^{\textrm{i} \alpha} \, \psi_{2}\big(\tilde{\xi}(\tau,x)\big)
\end{array}  
\right) 
\ee
where $\boldsymbol{\mathfrak{M}}(x)|_{\alpha=0}$ 
is the mixing matrix (\ref{Mfrak-deef-2int}) determining the modular flow of the Dirac field on the line 
generated by the modular Hamiltonian  of two disjoint equal intervals
(see the  Appendix\;\ref{app-details-mod-flow-2int}).

%\newpage
%%%%%%%%%%%%%%%%%%%%%%%%%%%%%%%%%%%%%%%%%%%%%%%%%%%%%
\subsection{Two disjoint equal intervals on the line}
\label{app-details-mod-flow-2int}

In order to make this manuscript self-consistent, in the following we find it worth deriving 
the modular flow of the massless Dirac field in the ground state 
and the correlators of the resulting field
when the subsystem is given by 
two disjoint equal intervals in the infinite line $A_{\textrm{\tiny sym}} \equiv [-b,-a] \cup [a,b] \subset \RR$,
by adapting to this case the procedure discussed in the Appendix\;\ref{app-details-mod-flow}
for the interval in the half-line. 
This analysis corresponds to a special case of the results obtained in 
 \cite{Casini:2009vk, Longo:2009mn} for the union of a generic number of disjoint intervals with arbitrary lengths.

The modular flow of the Dirac field is the solution of the following system of partial differential equations 
\be
\label{mod-flow-pde-4eq-2int}
\frac{d}{d\tau}
\left(
\begin{array}{c}
\psi_1(x,\tau) \\ \psi_1(-\tilde{x},\tau) \\ \psi_2(x,\tau)  \\ \psi_2(-\tilde{x},\tau) 
\end{array} \right) 
=\, 
2\pi \,\big[\, \boldsymbol{B}_{\textrm{\tiny sym}}(x)  \oplus \big( \! - \!\boldsymbol{B}_{\textrm{\tiny sym}}(x) \big)\,\big] 
\left(
\begin{array}{c}
\psi_1(x,\tau) \\ \psi_1(-\tilde{x},\tau) \\ \psi_2(x,\tau)  \\ \psi_2(-\tilde{x},\tau) 
\end{array}  \right) 
\ee
where the $2\times 2$ matrix differential operator $\boldsymbol{B}_{\textrm{\tiny sym}}(x) $ can be expressed in terms of 
$B_{\textrm{\tiny loc}}$ defined in (\ref{B-func-def}) as follows
\be
\boldsymbol{B}_{\textrm{\tiny sym}}(x) 
\equiv
\bigg(\!
\begin{array}{cc}
B_{\textrm{\tiny loc}}(x) \; \, & -\beta_{\textrm{\tiny bi-loc}}(x) 
\\
 \rule{0pt}{.3cm} 
 -\beta_{\textrm{\tiny bi-loc}}(-\tilde{x}) \;\, & B_{\textrm{\tiny loc}}(-\tilde{x}) 
\end{array}  \bigg) \,.
\ee

The block diagonal structure in (\ref{mod-flow-pde-4eq-2int}) implies that we can focus on the doublet given by 
\be
\label{Psi-doublet-2int}
\Psi(\tau,x)
\equiv
\bigg(
\begin{array}{c}
\psi(\tau,x) \\  \rule{0pt}{.3cm} \psi(\tau,-\tilde{x}) 
\end{array}  \bigg) 
\ee
that must solve the following system of partial differential equations
\be
\label{mod-flow-2-matrix-2int}
\hspace{-.7cm}
\frac{d}{d\tau} \,\Psi(\tau,x)
\,=\,
 2\pi \,
 \boldsymbol{B}_{\textrm{\tiny sym}}(x) 
\,\Psi(\tau,x)\,.
\ee
Since $\beta_{\textrm{\tiny loc}}(-\tilde{x}) \,  \partial_{(-\tilde{x})} = \beta_{\textrm{\tiny loc}}(x) \,  \partial_{x} $,
the system (\ref{mod-flow-2-matrix-2int}) can be also written as 
\be
\label{mod-flow-vec-2-2int}
\bigg[ \,\frac{d}{d\tau} - 2\pi\, \beta_{\textrm{\tiny loc}}(x) \,  \partial_{x}  \bigg] \Psi(\tau,x) \,=\, \boldsymbol{M}_{\textrm{\tiny sym}}(x)\,\Psi(\tau,x)
\;\;\qquad\;\;
\Psi(\tau=0,x) \equiv \Psi(x)
\ee
where 
\be
\label{Msymm-matrix-def}
\boldsymbol{M}_{\textrm{\tiny sym}}(x)
\equiv \,
2\pi
\left(
\begin{array}{cc}
\tfrac{1}{2}\,\partial_x\beta_{\textrm{\tiny loc}}(x)\; \; &  - \beta_{\textrm{\tiny bi-loc}}(x) 
\\
 \rule{0pt}{.5cm} 
-\! \beta_{\textrm{\tiny bi-loc}}(-\tilde{x}) \;\; & \tfrac{1}{2}\,\partial_{(-\tilde{x})} \beta_{\textrm{\tiny loc}}(-\tilde{x}) 
\end{array}  \right) .
\ee
From (\ref{xtilde}), (\ref{beta-loc-def}) and (\ref{beta-biloc-def}), one observes that
\be
\label{beta-parity-relations}
\beta_{\textrm{\tiny loc}}(-\tilde{x}) = \beta_{\textrm{\tiny loc}}(\tilde{x})
\;\;\qquad\;\;
\beta_{\textrm{\tiny bi-loc}}(-\tilde{x})  = -\, \beta_{\textrm{\tiny bi-loc}}(\tilde{x}) 
\;\;\qquad\;\;
\partial_{(-\tilde{x})} \beta_{\textrm{\tiny loc}}(-\tilde{x}) = -\partial_{\tilde{x}} \beta_{\textrm{\tiny loc}}(\tilde{x}) 
\ee
which allow to compare (\ref{Msymm-matrix-def}) and (\ref{mixing-matrix-bdy}), finding
\be
\boldsymbol{M}_{\textrm{\tiny sym}}(x)
=
\boldsymbol{M}(x) \big|_{\alpha=0}\,.
\ee

In order to write the system (\ref{mod-flow-vec-2-2int}) in a simpler form, one introduces
the doublet $\tilde{\Psi}(\tau, x)$ through the function
$c(x)$ defined in (\ref{f-def-beta-loc}) as follows
\be
\label{psi-tilde-def-2int}
\Psi(\tau, x) = 
\boldsymbol{\mathcal{M}}(x)\, \tilde{\Psi}(\tau, x)
\;\;\qquad\;\;
\boldsymbol{\mathcal{M}}(x) 
\equiv
\bigg(
\begin{array}{cc}
c(x) & 0
\\
\rule{0pt}{.3cm}
0 & c(-\tilde{x}) 
\end{array}  \bigg)\,.
\ee
Since for $c(x)$ the condition (\ref{f-condition-der}) holds,
the doublet $\tilde{\Psi}(\tau, x)$ is a solution of the following system
\be
\label{mod-flow-vec-2int}
\bigg[\, \frac{d}{d\tau} - 2\pi\, \beta_{\textrm{\tiny loc}}(x) \,  \partial_{x} \, \bigg] \tilde{\Psi}(\tau,x) 
\,=\, 
2\pi\, b(x)\, J
\,\tilde{\Psi}(\tau,x)
\;\;\qquad\;\;
J \equiv J_{\alpha}\big|_{\alpha=0} = 
\bigg(\,
\begin{array}{cc}
0 \; &  -1
\\
1 \;& 0
\end{array} \bigg)
\ee
where $b(x)$ is the function introduced in (\ref{b-function-def}), that satisfies
\be
b(x)
=
\frac{c(-\tilde{x})}{c(x)}\; \beta_{\textrm{\tiny bi-loc}}(x)
\,=\,
-\,\frac{c(x)}{c(-\tilde{x})}\; \beta_{\textrm{\tiny bi-loc}}(-\tilde{x})
\,=\,
b(-\tilde{x})\,.
\ee
Comparing (\ref{mod-flow-vec-2int}) with (\ref{mod-flow-vec-2-tilde}), 
we observe that the partial differential equation to solve for two equal disjoint intervals on the line corresponds 
to the partial differential equation underlying the interval on the half-line in the special case given by $\alpha = 0$.

The diagonalisation in (\ref{J-alpha-diagonalised}) tells us that $J = U^{-1} \,\textrm{diag}(\textrm{i}\, , -\textrm{i}) \,U$, 
where $U \equiv U_{\alpha=0} $;
hence it is natural to introduce
\be
\label{mu-tilde-def-2int}
\bigg(
\begin{array}{c}
\tilde{\mu}_-(\tau,x) 
\\  
\rule{0pt}{.3cm} 
\tilde{\mu}_+(\tau,-\tilde{x}) 
\end{array}  
\bigg) 
\equiv\,
U\,
\bigg(
\begin{array}{c}
\tilde{\psi}(\tau,x) 
\\  
\rule{0pt}{.3cm} 
\tilde{\psi}(\tau,-\tilde{x}) 
\end{array}  
\bigg) 
\;\;\qquad\;\;
U \equiv U_\alpha \big|_{\alpha=0}
=
\frac{1}{\sqrt{2}}\,
\bigg(\begin{array}{cc}
- \textrm{i}\; \;  & 1 \\
\textrm{i}  \;\; & 1
\end{array}\bigg)\,.
\ee
Combining (\ref{mod-flow-vec-2int}) and (\ref{mu-tilde-def-2int}), one finds that 
 $\tilde{\mu}_-(\tau,x) $ and $\tilde{\mu}_+(\tau,\tilde{x}) $ satisfy the following 
decoupled partial differeential equations
\be
\label{system-pde-mutilde-2int}
\left\{
\begin{array}{l}
\displaystyle
\bigg[\, \frac{d}{d\tau} - 2\pi\, \beta_{\textrm{\tiny loc}}(x) \,  \partial_{x} \, \bigg] 
\tilde{\mu}_-(\tau,x) 
=\, 2\pi \, \textrm{i}\, b(x)\, \tilde{\mu}_-(\tau,x) 
\\
\rule{0pt}{.8cm}
\displaystyle
\bigg[\, \frac{d}{d\tau} - 2\pi\, \beta_{\textrm{\tiny loc}}(-\tilde{x}) \,  \partial_{(-\tilde{x})} \, \bigg] 
\tilde{\mu}_+(\tau,-\tilde{x}) 
=\, - \,2\pi \, \textrm{i}\, b(x)\, \tilde{\mu}_+(\tau,-\tilde{x}) 
=\, -\,2\pi \, \textrm{i}\, b(-\tilde{x})\, \tilde{\mu}_+(\tau,-\tilde{x}) 
\end{array}
\right.
\ee
where one equation can be obtained from the other one by exchanging
$(\tau,x) \leftrightarrow (\tau,- \tilde{x})$ and $\tilde{\mu}_- \leftrightarrow  \tilde{\mu}_+$.
Notice that the sign of $\tau$ remains unchanged in this case.
From (\ref{system-pde-mutilde-2int}), we observe that the underlying partial differential to solve is again (\ref{pde-model-mu-tilde}),
whose solution (\ref{soln-mu-tilde}) can be employed also in this analysis.

In this case, by inverting the function $w(x)$ in (\ref{w-function-def}), one finds that 
we need to introduce 
\be
\label{xi-pm-def}
\xi_\pm(\tau,x) 
\,=\,
\frac{ (b-a)\big(e^{2\pi \tau + w(x)}-1\big) 
\pm
\sqrt{(b-a)^2\big(e^{2\pi \tau + w(x)}-1\big)^2 + 4 ab \,\big(e^{2\pi \tau + w(x)}+1\big)^2}
}{
2\,\big(1+e^{2\pi \tau + w(x)}\big)}\,.
\ee
It is worth combining these expressions and the Heavyside step function $\Theta(x)$ to define
\be
\label{xi-2int}
\xi(\tau, x) \equiv  \Theta(x) \, \xi_+(\tau,x) +  \Theta(-x) \, \xi_-(\tau,x)\,.
\ee

In terms of this $\xi(\tau, x) $ and of (\ref{Delta-omega-def}), the solution of (\ref{system-pde-mutilde-2int}) reads
\be
\left(\,
\begin{array}{c}
\tilde{\mu}_-(\tau,x) 
\\  
\rule{0pt}{.45cm} 
\tilde{\mu}_+(\tau,-\tilde{x}) 
\end{array}  
\right) 
=
\left(
\begin{array}{c}
e^{\textrm{i}\,\Delta\omega (\xi(\tau, x) )} 
\, \tilde{\mu}_{-}\big(\xi(\tau,x)\big)
\\  
\rule{0pt}{.45cm} 
e^{-\textrm{i}\,\Delta\omega (\xi(\tau, -\tilde{x}) )}
\, \tilde{\mu}_{+}\big(\xi(\tau,-\tilde{x})\big)
\end{array}  
\right) .
\ee
By using this result and (\ref{mu-tilde-def-2int}), one obtains 
\be
\label{psi-tilde-soln-2int}
\left(
\begin{array}{c}
\tilde{\psi}_(\tau,x) 
\\  
\rule{0pt}{.45cm} 
\tilde{\psi}(\tau,-\tilde{x}) 
\end{array}  
\right) 
=\, 
U^{-1}
\left(
\begin{array}{cc}
e^{\textrm{i}\,\Delta\omega (\xi(\tau, x) )} & 0
\\  
\rule{0pt}{.45cm} 
0 & e^{-\textrm{i}\,\Delta\omega (\xi(\tau, -\tilde{x}) )}
\end{array}  
\right) 
U\,
\left(
\begin{array}{c}
\tilde{\psi}\big(\xi(\tau,x)\big)
\\  
\rule{0pt}{.45cm} 
\tilde{\psi}\big(\xi(\tau,-\tilde{x})\big)
\end{array}  
\right) .
\ee
Now it is useful to observe that 
\be
\xi(\tau,-\tilde{x}) = - \frac{a\, b}{\xi(\tau,x)} \equiv - \,\tilde{\xi}(\tau,x)
\ee
that allows to write (\ref{psi-tilde-soln-2int}) as
\be
\label{pde-system-2int-psi-tilde}
\left(
\begin{array}{c}
\tilde{\psi}_(\tau,x) 
\\  
\rule{0pt}{.45cm} 
\tilde{\psi}(\tau,-\tilde{x}) 
\end{array}  
\right) 
=\, 
U^{-1}
\left(
\begin{array}{cc}
e^{\textrm{i}\,\Delta\omega (\xi(\tau, x) )} & 0
\\  
\rule{0pt}{.45cm} 
0 & e^{-\textrm{i}\,\Delta\omega (-\tilde{\xi}(\tau, x) )}
\end{array}  
\right) 
U\,
\left(
\begin{array}{c}
\tilde{\psi}\big(\xi(\tau,x)\big)
\\  
\rule{0pt}{.45cm} 
\tilde{\psi}\big(\!-\! \tilde{\xi}(\tau,x)\big)
\end{array}  
\right) .
\ee
Since $\textrm{arctan}(-1/y) = \textrm{arctan}(y) - \tfrac{\pi}{2} \,\textrm{sign}(y)$ for $y\in \RR$, in this expression we can use that
\be
\label{xi-tilde-Delta-omega-2int}
\Delta \omega (- \,\tilde{\xi}(\tau, x))
=
\omega(-\,\tilde{\xi}(\tau, x))- \omega(-\tilde{x})
=
\Delta \omega \big(\xi(\tau, x)\big)\,.
\ee

By employing this observation in (\ref{pde-system-2int-psi-tilde}) first and then the fields redefinition (\ref{psi-tilde-def-2int}),
we find that the modular flow of the doublet (\ref{Psi-doublet-2int}) reads
\be
\label{soln-lambda-2int}
\left(
\begin{array}{c}
\psi(\tau,x) 
\\  
\rule{0pt}{.45cm} 
\psi(\tau,-\tilde{x}) 
\end{array}  
\right) 
=\,
\boldsymbol{\mathfrak{M}}(x)
\left(
\begin{array}{c}
\psi\big(\xi(\tau,x)\big)
\\  
\rule{0pt}{.45cm} 
 \psi\big(\!-\!\tilde{\xi}(\tau,x)\big)
\end{array}  
\right) 
\ee
where the mixing matrix is 
\be
\label{Mfrak-deef-2int}
\boldsymbol{\mathfrak{M}}(x)
\equiv\,
\boldsymbol{\mathcal{M}}(x)
\left(\,
\begin{array}{cc}
 \cos\!\big[\Delta\omega \big(\xi(\tau, x) \big)\big]  \; \;&  -  \sin\!\big[\Delta\omega \big(\xi(\tau, x) \big)\big] 
\\
\rule{0pt}{.5cm} 
 \sin\!\big[\Delta\omega \big(\xi(\tau, x) \big)\big]    \; \; &  \cos\!\big[\Delta\omega \big(\xi(\tau, x) \big)\big] 
\end{array} \,\right)
\,\boldsymbol{\mathcal{M}} \big(\xi(\tau, x) \big)^{-1}\,.
\ee
By using (\ref{psi-tilde-def-2int}), the result (\ref{soln-lambda-2int}) can be written as
\be
\left\{ 
\begin{array}{l}
\displaystyle
\psi(\tau,x)  
\,=\,
c(x) \left(
 \cos\!\big[\Delta\omega \big(\xi(\tau, x) \big)\big]  \,
  \frac{\psi\big(\xi(\tau,x)\big)}{c\big(\xi(\tau,x)\big)}
-  
\sin\!\big[\Delta\omega \big( \xi(\tau, x) \big)\big] \,
   \frac{\psi \big(\!-\! \tilde{\xi}(\tau, x)\big)}{ c\big(\!-\!\tilde{\xi}(\tau, x)\big)}
\right)
\\
\rule{0pt}{.9cm}
\displaystyle
\psi(\tau,-\tilde{x})  
\,=\,
c(-\tilde{x}) \left(
 \cos\!\big[\Delta\omega \big(\xi(\tau, x) \big)\big]  \,
   \frac{\psi \big(\!-\! \tilde{\xi} (\tau, x)\big)}{ c\big(\!-\!\tilde{\xi}(\tau, x)\big)}
+
\sin\!\big[\Delta\omega \big( \xi(\tau, x) \big)\big] \,
   \frac{\psi\big(\xi (\tau, x)\big)}{ c\big(\xi(\tau, x)\big)}
 \right)
\end{array}\right.
\ee
where we remark that the second equation is equivalent to the first one;
indeed, by renaming the spatial variable, it becomes
\be
\psi(\tau,x)  
\,=\,
c(x) \left(
 \cos\!\big[\Delta\omega \big(\!-\!\tilde{\xi}(\tau, x) \big)\big]  \,
   \frac{\psi \big(\xi (\tau, x)\big)}{ c\big(\xi(\tau, x)\big)}
+
\sin\!\big[\Delta\omega \big( \!-\!\tilde{\xi}(\tau, x) \big)\big] \,
   \frac{\psi\big(\!-\!\tilde{\xi} (\tau, x)\big)}{ c\big(\!-\!\tilde{\xi}(\tau, x)\big)}
 \right)
\ee
where
\be
\cos\!\big[\Delta\omega \big( \!-\! \tilde{\xi}(\tau, x) \big)\big] 
=
\cos\!\big[\Delta\omega \big( \xi(\tau, x) \big)\big] 
\;\;\qquad\;\;
\sin\!\big[\Delta\omega \big( \!-\! \tilde{\xi}(\tau, x) \big)\big] 
=
- \sin\!\big[\Delta\omega \big( \xi(\tau, x) \big)\big] \,.
\ee

Summarising, 
when the bipartition of the infinite line is determined by $A_{\textrm{\tiny sym}}$, 
the modular flow of both the components of the Dirac field reads
\be
\psi(\tau,x)  
\,=\,
c(x) \left(
 \cos\!\big[\Delta\omega \big(\xi(\tau, x) \big)\big]  \,
  \frac{\psi\big(\xi(\tau,x)\big)}{c\big(\xi(\tau,x)\big)}
-  
\sin\!\big[\Delta\omega \big( \xi(\tau, x) \big)\big] \,
   \frac{\psi \big(\!-\! \tilde{\xi}(\tau, x)\big)}{ c\big(\!-\!\tilde{\xi}(\tau, x)\big)}
\right) .
\ee
By using (\ref{beta-loc-xtilde}), (\ref{f-def-beta-loc}) and (\ref{beta-parity-relations}),
we find it worth writing this expression also as 
\be
\label{psi-mod-flow-2int}
\psi(\tau,x)
=
\left[
P(\xi;  x) 
\left(
 \big( a \,b + x \,\xi \big) \,\psi(\xi) 
- \frac{a\,b}{\xi }\, \big(\xi - x\big)\, \psi(-ab/\xi)
\right)
\right]
\! \bigg|_{\xi = \xi(\tau,x)}
\ee
where $P(\xi;  x) $ has been defined in (\ref{P-function-def}) and $ \xi(\tau,x)$ is given by (\ref{xi-2int}).

Employing the modular flow (\ref{psi-mod-flow-2int}) and the correlators 
for the Fock vacuum of the fields (\ref{psi1-line}) and (\ref{psi2-line}) on the line,
one finds that the correlators of the two components of the Dirac field along the modular flow read
\cite{Longo:2009mn, Hollands:2019hje}
\bea
\label{mod-corr-2int-sym}
\langle \psi_1(\tau_1,x_1)\,\psi_1^*(\tau_2,x_2)\rangle =
W(\tau_{12}; x_1, x_2)
\\
\langle \psi_2(\tau_1,x_1)\,\psi_2^*(\tau_2,x_2)\rangle 
=
W(\tau_{12}; x_2, x_1) 
\eea
where $W$ is the function defined in (\ref{Wfunc}) with $w(x)$ given by (\ref{w-function-def}).

\subsection{Two disjoint intervals on the line: Modular equation of motion}
\label{app_pde_mc_2int}

The modular Hamiltonian of a subregion made by the union of a generic number 
of disjoint intervals on the line for the massless Dirac field 
and the corresponding modular flow have been studied by Casini and Huerta in \cite{Casini:2009vk},
while the correlators the Dirac field along the modular flow have been found in \cite{Longo:2009mn}.

The modular Hamiltonian of two disjoint intervals $A \equiv [a_1, b_1] \cup [a_2, b_2]$
on the line can be written as the sum $K_A  = K_A^{\textrm{\tiny loc}} + K_A^{\textrm{\tiny bi-loc}} $,
where the local term $K_A^{\textrm{\tiny loc}} $ and the bi-local term $K_A^{\textrm{\tiny bi-loc}}$
are defined respectively as  \cite{Casini:2009vk}
\be
\label{K_A-2int-terms}
K_A^{\textrm{\tiny loc}} 
=
2\pi 
\int_A 
\beta_{\textrm{\tiny loc}}(x) \, T_{tt}(0,x)\, \rd x 
\;\;\;\qquad\;\;\;
K_A^{\textrm{\tiny bi-loc}} 
=
2\pi 
\int_A 
\beta_{\textrm{\tiny bi-loc}}(x) \, T_{\textrm{\tiny bi-loc}}(0,x, x_{\textrm{\tiny c}} ) \, \rd x 
\ee
where $T_{tt}(t,x)$ is the local operator (\ref{T00-lambda-def}), while
$T_{\textrm{\tiny bi-loc}}(t,x, x_{\textrm{\tiny c}} )$ is the bi-local operator (\ref{T-bilocal-2int}).
The weight functions in (\ref{K_A-2int-terms}) can be written as follows
\be
\label{betas-2int-gen}
\beta_{\textrm{\tiny loc}}(x)  = \frac{1}{w'(x)} 
\;\;\qquad\;\;
\beta_{\textrm{\tiny bi-loc}}(x)  = \frac{\beta_{\textrm{\tiny loc}}(x_{\textrm{\tiny c}}(x))}{x - x_{\textrm{\tiny c}}(x)} 
\ee
where
\be
\label{w-function-2int}
w(x) \,=\,  
\log \!\left [ - \frac{(x-a_1)(x-a_2)}{(x-b_1)(x- b_2)} \right ]
\ee
and $x_{\textrm{\tiny c}}(x)$ is the point conjugate to $x\in A$ satisfying the condition $w(x_{\textrm{\tiny c}}(x))=w(x)$, namely
\be
\label{x-conjugate-2int}
x_{\textrm{\tiny c}}(x)
\equiv 
\frac{(b_1 b_2 - a_1 a_2)\, x +(b_1 + b_2)\, a_1 a_2 - (a_1 + a_2)\, b_1 b_2}{(b_1 - a_1 + b_2 - a_2)\, x + a_1 a_2 - b_1 b_2 }\,.
\ee
We remark that $x$ and its conjugate point $x_{\textrm{\tiny c}}(x) $ belong to different intervals in $A$.

For the symmetric configuration $A = A_{\textrm{\tiny sym}}$, 
we have that (\ref{w-function-2int}) becomes (\ref{w-function-def}) 
and (\ref{x-conjugate-2int}) simplifies to $x_{\textrm{\tiny c}}(x) = -\tilde{x} = -a b/x$;
hence the two expressions in (\ref{betas-2int-gen}) reduce to (\ref{beta-loc-def}) and (\ref{beta-biloc-def});
hence the final result is the modular Hamiltonian 
reported in (\ref{K_Asym-decomposition}), (\ref{K_Asym-terms}) and (\ref{T-bilocal-2int}).

The correlators of the components of the Dirac field along the modular flow 
found in \cite{Longo:2009mn,Hollands:2019hje} read 
\bea
\label{Wfunc-g-local-app-2int-1}
\langle \psi_1(\tau_1,x_1)\,\psi_1^*(\tau_2,x_2)\rangle =
W(\tau_{12}; x_1, x_2)
\\
\label{Wfunc-g-local-app-2int-2}
\langle \psi_2(\tau_1,x_1)\,\psi_2^*(\tau_2,x_2)\rangle 
=
W(\tau_{12}; x_2, x_1) 
\eea
with 
\be
\label{Wnew}
W(\tau; x, y) 
=
\frac{e^{w(x)} - e^{w(y)}}{2\pi \textrm{i} (x - y)}\;
\frac{1}{e^{w(x)+\pi \tau} - e^{w(y)-\pi \tau} - \textrm{i} \varepsilon}
\ee
where the function $w(x)$ is given by (\ref{beta-loc-singleint-line}) for a single interval 
and by (\ref{w-function-2int}) for the union of two disjoint intervals on the line.

We remark that (\ref{Wnew}) satisfies the properties given in (\ref{Widentity})
and this implies that the modular correlators obey the KMS condition \cite{Longo:2009mn}.

For two disjoint intervals, the differential equations (\ref{mod-flow-2-matrix-2int}), 
defining the modular flow of the massless Dirac field, 
imply that in the limit of $\varepsilon \to 0$ (\ref{Wnew}) satisfies the following modular equation of motion
\be
\label{pde-W-2int}
\left(\,
\frac{1}{2\pi}\, \partial_\tau - \beta_{\textrm{\tiny loc}}(x)  \, \partial_x - \frac{1}{2} \, \partial_x \beta_{\textrm{\tiny loc}}(x) 
\right)
W(\tau; x, y) 
+ 
\beta_{\textrm{\tiny bi-loc}}(x) \, W(\tau; x_{\textrm{\tiny c}}  , y) 
\,=\,
0
\ee
where $\beta_{\textrm{\tiny loc}}(x) $ and $\beta_{\textrm{\tiny bi-loc}}(x) $ are defined in (\ref{betas-2int-gen}) 
in terms of the function $w(x)$ in (\ref{w-function-2int}).

We remark that
the occurrence of the last term in the l.h.s. of (\ref{pde-W-2int}) is due to the fact that the modular Hamiltonian
contains also a bi-local term. 
Indeed, for a single interval $(a,b)$ on the line, the modular correlators
(which are (\ref{Wfunc-g-local-app-2int-1},\ref{Wfunc-g-local-app-2int-2}) with $w(x)$ given by (\ref{beta-loc-singleint-line}))
satisfy a modular equation of motion like (\ref{pde-W-2int}) with $\beta_{\textrm{\tiny bi-loc}}(x) = 0$ identically 
and $\beta_{\textrm{\tiny loc}}(x) = \beta_0(x)$ defined in (\ref{beta-loc-singleint-line}).

%%%%%%%%%%%%%%%%%%%%%%%%%%%%%%%%%%%%%%%%%%%%%%%%%%%%%
%\newpage

\bibliographystyle{nb}

\bibliography{refsbdy}

\end{document}